\documentclass[iop]{emulateapj}
\pdfoutput=1 
\usepackage{amsmath, amssymb}
\usepackage{graphicx}
\usepackage{natbib}
\usepackage{color}
\usepackage{tablefootnote}

\begin{document}

\title{The Lyman Alpha Reference Sample. VIII. Characterizing Lyman-Alpha Scattering in Nearby Galaxies}

\author{Joanna S. Bridge\altaffilmark{1, 2, 3, 4}, Matthew Hayes\altaffilmark{3}, Jens Melinder\altaffilmark{3}, G\"oran \"Ostlin\altaffilmark{3}, Caryl Gronwall\altaffilmark{1, 2}, Robin Ciardullo\altaffilmark{1, 2}, Hakim Atek\altaffilmark{5}, John M. Cannon\altaffilmark{6}, Max Gronke\altaffilmark{7,8}, Lucia Guaita\altaffilmark{9}, Alex Hagen\altaffilmark{1, 2, 10}, Edmund Christian Herenz\altaffilmark{3}, Daniel Kunth\altaffilmark{5}, Peter Laursen\altaffilmark{7}, J. Miguel Mas-Hesse\altaffilmark{11}, Stephen A. Pardy\altaffilmark{12}}

\email{Email: joanna.bridge@louisville.edu}
\altaffiltext{1}{Department of Astronomy \& Astrophysics, The Pennsylvania State University, University
Park, PA 16802}
\altaffiltext{2}{Institute for Gravitation and the Cosmos, The Pennsylvania State University, University Park, PA 16802}
\altaffiltext{3}{Department of Astronomy, Oskar Klein Centre, Stockholm University, AlbaNova University Centre, Stockholm, Sweden}
\altaffiltext{4}{Department of Physics and Astronomy, 102 Natural Science Building, University of Louisville, Louisville KY 40292}
\altaffiltext{5}{Institut dÕAstrophysique de Paris, 98 bis Bd Arago, 75014 Paris, France}
\altaffiltext{6}{Department of Physics \& Astronomy, Macalester College, Saint Paul, MN 55105}
\altaffiltext{7}{Institute of Theoretical Astrophysics, University of Oslo, Postboks 1029 Blindern, 0315 Oslo, Norway}
\altaffiltext{8}{Department of Physics, University of California, Santa Barbara, CA 93106, USA}
\altaffiltext{9}{N\'ucleo de Astronom\'ia, Facultad de Ingenier\'ia, Universidad Diego Portales, Av.\ Ej\'ercito 441, Santiago, Chile}
\altaffiltext{10}{vizual.ai, 3600 O'Donnell Street, Suite 250, Baltimore, MD 21224}
\altaffiltext{11}{Centro de Astrobiolog\'{\i}a (CSIC--INTA), ESA--ESAC, 28692 Villanueva de la Ca\~nada, Spain}
\altaffiltext{12}{Department of Astronomy, University of Wisconsin - Madison, Madison, WI 53706}

\begin{abstract}
We examine the dust geometry and Ly$\alpha$ scattering in the galaxies of the Lyman Alpha Reference Sample (LARS), a set of 14 nearby ($0.02 < z < 0.2$) Ly$\alpha$ emitting and starbursting systems with \emph{Hubble Space Telescope} Ly$\alpha$, H$\alpha$, and H$\beta$ imaging. We find that the global dust properties determined by line ratios are consistent with other studies, with some of the LARS galaxies exhibiting clumpy dust media while others of them show significantly lower Ly$\alpha$ emission compared to their Balmer decrement. With the LARS imaging, we present Ly$\alpha$/H$\alpha$ and H$\alpha$/H$\beta$ maps with spatial resolutions as low as $\sim 40$~pc, and use these data to show that in most galaxies, the dust geometry is best modeled by three distinct regions: a central core where dust acts as a screen, an annulus where dust is distributed in clumps, and an outer envelope where Ly$\alpha$ photons only scatter. We show that the dust that affects the escape of Ly$\alpha$ is more restricted to the galaxies' central regions, while the larger Ly$\alpha$ halos are generated by scattering at large radii. We present an empirical modeling technique to quantify how much Ly$\alpha$ scatters in the halo, and find that this ``characteristic" scattering distance correlates with the measured size of the Ly$\alpha$ halo. We note that there exists a slight anti-correlation between the scattering distance of Ly$\alpha$ and global dust properties.
\end{abstract} 

\section{Introduction}
Ever since the question was posed of how distant galaxies could be detected, the Ly$\alpha$ emission line resulting from the n = 2 to n = 1 transition of hydrogen at 1216~\AA~has been recognized as key to studying high redshift star-forming galaxies \citep{partridge1967}. Ly$\alpha$ reprocesses approximately two-thirds of the ionizing photons from hot, massive, short-lived stars \citep{spitzer1978}. It is therefore expected to be the brightest emission line from young stellar populations \cite[e.g.,][]{raiter2010}.  

However, due to the high absorption cross-section between the n = 1 and n = 2 state, the neutral gas reservoirs in a galaxy lead to a complicated resonant radiative transfer problem.  This process increases the path lengths that Ly$\alpha$ photons have to travel before escaping the galaxy and therefore also increases the chance of being absorbed by interstellar dust grains \cite[e.g.,][]{neufeld1990, verhamme2006, laursen2013, behrens2014,dijkstra2014}. The result is that Ly$\alpha$ is not observed in all star-forming galaxies, and in galaxies where it is observed, the sites of escaping emission and intrinsic production are different. Many factors govern the visibility of Ly$\alpha$, such as kinematics, the porosity and clumpiness of the interstellar medium (ISM), and scattering \citep{kunth1994, kunth1998, mashesse2003, lequeux1995, heckman2011, wofford2013}.

Nevertheless, Ly$\alpha$ has been used to find tens of thousands of high-redshift galaxies and, with the advent of experiments such as the Hobby Eberly Telescope Dark Energy Experiment (HETDEX; \citealt{hill2008}) and \emph{James Webb Space Telescope} (\emph{JWST}; \citealt{gardner2006}), it will remain the prime tracer of galaxy formation and evolution in the next decade. Therefore, it is of paramount importance that the processes by which Ly$\alpha$ is emitted from galaxies are very well understood. This requires understanding Ly$\alpha$ radiative transfer on small scales.

One of the best samples in which to study resolved Ly$\alpha$ is the Lyman Alpha Reference Sample (LARS; \citealt{hayes2013, hayes2014, ostlin2014, pardy2014, riverathorsen2015, herenz2016, duval2016}). The LARS sample comprises 14 nearby ($0.02 < z < 0.2$) Ly$\alpha$ emitting and/or absorbing galaxies with extensive \emph{Hubble Space Telescope} (\emph{HST}) broadband and narrowband imaging, supplemented by observations from numerous other telescopes. The goal of these observations is to probe exactly how the Ly$\alpha$ photons travel and by what mechanism they escape. The work presented here will use this unique sample of galaxies to model the behavior of dust and Ly$\alpha$ scattering in star-forming galaxies.

One way to understand what Ly$\alpha$ tells us about a galaxy is to compare it to the very well understood transition of H$\alpha$, the H~I Balmer transition from n = 3 to n = 2 at 6563~\AA~\citep[e.g.,][]{hayes2007, ostlin2009, herenz2016}. In star-forming galaxies, Balmer emission serves as a proxy for the total ionizing flux of the stars embedded in the galaxies, and is therefore one of the more direct ways (when corrected for reddening) to quantify a system's current star formation \citep{kennicutt1998}. The same recombination events that generate Ly$\alpha$ photons also generate H$\alpha$, meaning that H$\alpha$ traces Ly$\alpha$ but without the complicated radiative transfer of resonance. The comparison of Ly$\alpha$ and H$\alpha$ is therefore a way of understanding what we are missing when we focus only on the Ly$\alpha$ emission. This is particularly useful because the reason that high-redshift galaxies are studied in Ly$\alpha$ is because their H$\alpha$ emission is shifted out of the optical (and near infrared for $z > 2.5$) and is therefore often unavailable. If we are able to quantify what information is missed by the complicated radiative transfer of Ly$\alpha$ compared to H$\alpha$, we can then use that knowledge in high-redshift galaxy studies.

In this work, we use the H$\alpha$ and H$\beta$ emission of the LARS galaxies to simulate their Ly$\alpha$ emission to obtain a \emph{characteristic scattering distance} for Ly$\alpha$ photons, and show how dust geometry creates deviations from a simple scattering model. We summarize the LARS data set and reductions in \S 2.  In \S 3, we perform spatially resolved analysis of H$\alpha$/H$\beta$ and Ly$\alpha$/H$\alpha$ emission line ratios to pinpoint mechanisms for Ly$\alpha$ wscape. We describe how H$\alpha$ observations can be used to create simulated Ly$\alpha$ distributions to obtain a characteristic scattering distance for each galaxy in \S 4. We discuss the implications of this study in \S5, and conclude with \S 6, describing how this work can be expanded upon in the future.

In this work, we adopt the standard $\Lambda$CDM cosmology, with $H_0=70$ km s$^{-1}$ Mpc$^{-1}$, $\Omega_M=0.3$, and $\Omega_\Lambda=0.7$ \citep{komatsu2011}.

\section{Data}
The 14 LARS galaxies were chosen from the GALEX General Release 2 and SDSS Data Release 6 catalogs. The galaxies were chosen to have H$\alpha$ equivalent widths of $EW_{\textrm{H}\alpha}> 100$~\AA, ensuring a sample with significant active star formation and Ly$\alpha$ photon production. 
The sample is bright in both the far UV ($16.6<m_{FUV} <19.2$) and near UV ($16.2<m_{NUV}<19.1$) and has a far UV luminosity  range of $9.2 \, L_{\odot} < \textrm{log}(L_{FUV}) < 10.7\,L_{\odot}$. These luminosities are on par with higher-redshift Ly$\alpha$ emitters and Lyman break galaxies (LBGs), making the LARS sample a set of reasonable high-redshift analog galaxies. For further sample details, see \cite{ostlin2014} (Tables 1 -- 3 therein.) While the details of the dataset and data reduction can also be found in \cite{hayes2013} and \cite{ostlin2014}, we summarize the procedures here. 

\subsection{Observations}
The LARS galaxies were imaged by the \emph{HST} using a synthetic narrowband centered on Ly$\alpha$. The  technique used to measure the Ly$\alpha$ emission line is described in detail in \cite{ostlin2014} and \cite{hayes2009}. Briefly, a synthetic narrowband image was created by subtracting the F140LP and F150LP or the F125LP and F140LP long-pass filters (depending on the galaxy redshift) on the Solar Blind Channel to observe the Ly$\alpha$ line. Either the Wide Field Camera 3 (WFC3) or the ACS were then used to obtain H$\alpha$ and H$\beta$ emission maps. These three emission maps serve as the basis for our following analysis.

\subsection{Data Processing}
The images were reduced using the \texttt{Drizzlepac}\footnote{http://drizzlepac.stsci.edu/} package in the standard \emph{HST} reduction algorithm. We used custom built software called \texttt{Lyman-Alpha Extraction Software} (\texttt{LaXs}; \citealt{hayes2009}; \citealt{ostlin2014}) to subtract the continuum from the Ly$\alpha$ images. To find the underlying continuum flux in Ly$\alpha$ (and to lesser extent H$\alpha$ and H$\beta$), we performed spectral energy distribution fitting using observations from five continuum filters on \emph{HST} (two in the FUV and three in the optical). To briefly summarize:  we fit each pixel with a 3-component model, comprised of a young, star-forming population, a stellar continuum from older stars, and line+continuum produced by photoionization. These were fit to the data with a $\chi^2$ fitting method and population spectra from \texttt{Starburst99} templates \citep{leitherer1999}. The free parameters in this procedure are the equivalent width of the Ly$\alpha$ line, the age (of the young population), E(B-V)$_{\mathrm{s}}$, and the masses of the young and old populations. The fitting routine also accounts for flux from the nebular continuum by utilizing the H$\alpha$ and H$\beta$ observations. For more details see \cite{hayes2009}. The images' point spread functions were also matched using custom-made software (\citealt{hayes2016}; Melinder et al.\ \emph{in prep}).

\subsection{Photometric Analysis}
Using the appropriate aperture for each galaxy is important, as we need to determine a size that encompasses as much of the Ly$\alpha$ emission as possible without including too much background. The \citet{petrosian1976} radius determined from the UV continuum does not encompass the outer regions of the Ly$\alpha$ halo for every galaxy. We therefore created an aperture by convolving the UV continuum image with a $\sigma=50$~pixel Gaussian kernel, creating a larger aperture. Using this convolved image, we calculated an isophotal radius at which the local intensity is 20\% of the average flux contained within the radius, and used this radius to define the aperture for each galaxy. However, in cases where the Ly$\alpha$ halo morphology is particularly extended or more complex, even this radius is not large enough to encompass all of the Ly$\alpha$ emission.  For LARS01, LARS02, LARS05, LARS07, and LARS08, we therefore used 1.5 times this radius to define the aperture to include all of the Ly$\alpha$ emission from the galaxy. Using an isophotal measurement of the radius, rather than a fixed circular aperture, maximizes the signal while decreasing the background. The aperture is large enough to contain the extended Ly$\alpha$ emission while excluding excess background. An average of 75\% of the Ly$\alpha$ emission is captured within the calculated apertures.

\section{Emission Line Ratios and Dust Geometry}
Intrinsically, the ratio of Ly$\alpha$ to H$\alpha$ in Case B nebulae with $T = 5,000 - 20,000$ K and $n_e = 10^2 -10^4$ cm$^{-3}$ varies from 8.1 to 11.6 \citep{hummer1987}. Convention dictates that, roughly, Ly$\alpha$/H$\alpha\sim8.7$ \citep{hu1998, hayes2015}. Deviations from this ratio capture the physics of Ly$\alpha$ radiative transfer as well as information about the dust extinction, while departures from H$\alpha$/H$\beta$ = 2.86 can be used to probe the dust content of a galaxy \citep{pengelly1964,brocklehurst1971}. Therefore we use the Ly$\alpha$ and H$\alpha$ emission maps to calculate these ratios for the LARS galaxies, giving insight into the radiation physics, dust geometry, and reddening in galaxies. 

\begin{figure}[t]
\centering
\scalebox{0.6}
{\includegraphics{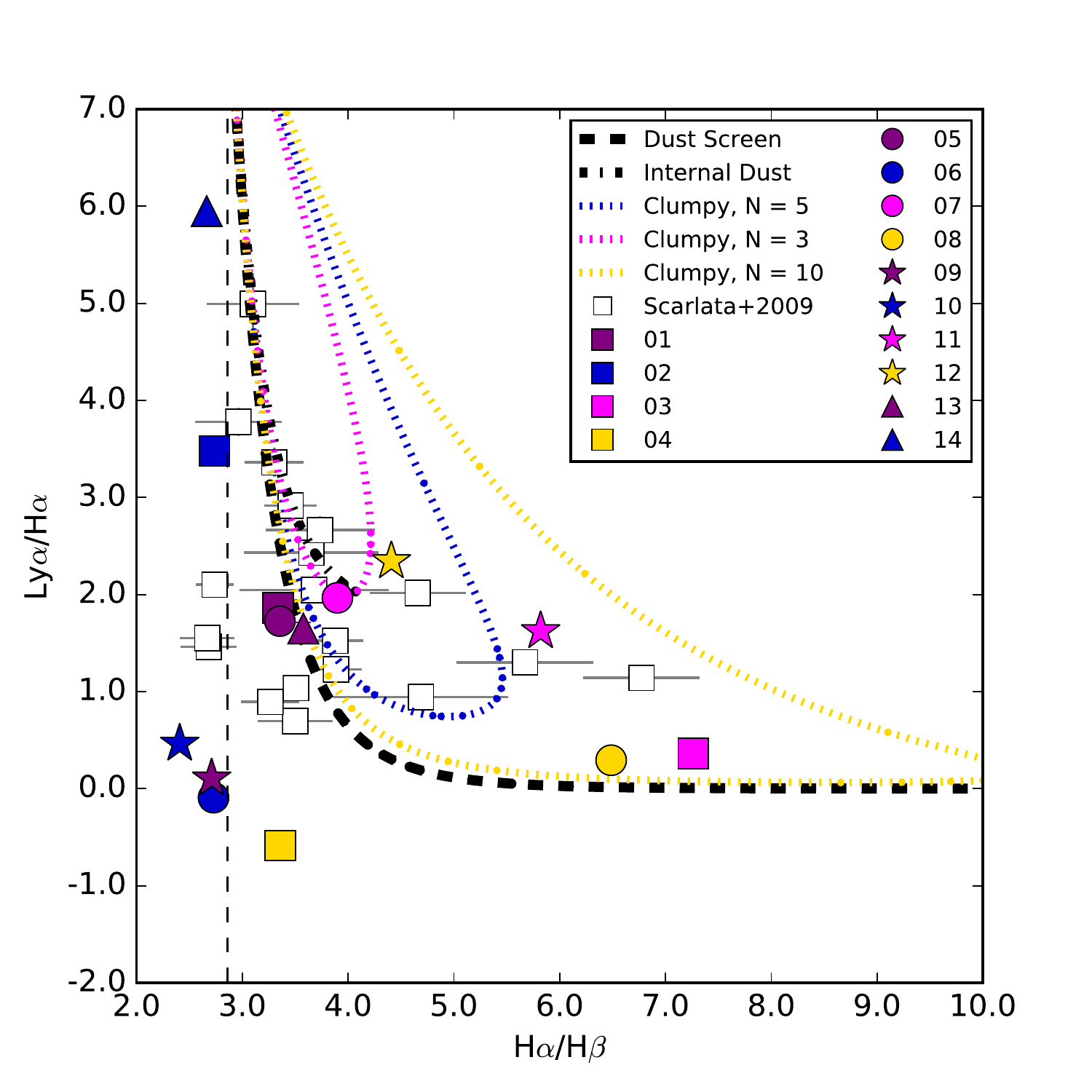}}
\caption{Ly$\alpha$/H$\alpha$ versus H$\alpha$/H$\beta$ for the 14 LARS galaxies. The various curves denote different dust geometry models. A simple uniform dust screen is shown as a black dashed line, while a smooth internal dust model is shown with a black dashed-dotted line. The blue, magenta, and yellow dotted lines show a clumpy dust model for different numbers of clumps along the line of sight ($N$). Moving counter-clockwise along the curves corresponds to higher optical depths of the dust. For most of the galaxies, the error bars ($\pm1\sigma$ assuming a normal distribution) are within the plot markers. Shown for comparison are the $z\sim0.3$ \emph{GALEX} LAEs from \citet{scarlata2009}. The vertical dashed black line shows the intrinsic Balmer decrement value of H$\alpha$/H$\beta$=2.86.}
\label{all_ratios}
\end{figure}

\begin{deluxetable*}{ccccc}[t]
\tablecolumns{5}
\tablecaption{LARS Integrated Fluxes\label{fluxtable}}
\tablehead{\colhead{LARS ID} & \colhead{$z$} & \colhead{Ly$\alpha$ (erg s$^{-1}$ cm$^{-2}$)} & \colhead{H$\alpha$ (erg s$^{-1}$ cm$^{-2}$)} & \colhead{~H$\beta$ (erg s$^{-1}$ cm$^{-2}$)}}
\startdata
01 & 0.028 &$(5.25\pm0.03)\times10^{-13}$ & $2.82\times10^{-13}$ & $8.44\times10^{-14}$ \\[5pt]
02 & 0.030 &$(2.56\pm0.01)\times10^{-13}$ &  $7.36\times10^{-14}$ & $2.69\times10^{-14}$ \\[5pt]
03 & 0.031 &$(9.87\pm0.07)\times10^{-14}$ &  $2.72\times10^{-13}$ & $3.75\times10^{-14}$ \\[5pt]
04 & 0.033 &$(-1.12\pm0.01)\times10^{-13}$ &  $1.91\times10^{-13}$ & $5.67\times10^{-14}$ \\[5pt]
05 & 0.034 &$(2.75\pm0.02)\times10^{-13}$ &  $1.60\times10^{-13}$ & $5.76\times10^{-14}$ \\[5pt]
06 & 0.034 &$(-2.30\pm0.45)\times10^{-15}$ &  $2.48\times10^{-14}$ & $9.11\times10^{-15}$\\[5pt]
07 & 0.038 &$(2.30\pm0.02)\times10^{-13}$ & $1.17\times10^{-13}$ & $3.00\times10^{-14}$ \\[5pt]
08 & 0.038 &$(1.01\pm0.01)\times10^{-13}$& $3.44\times10^{-13}$ & $5.31\times10^{-14}$ \\[5pt]
09 & 0.047 &$(4.58\pm0.16)\times10^{-14}$ & $4.33\times10^{-13}$ & $1.60\times10^{-13}$ \\[5pt]
10 & 0.057 &$(1.56\pm0.05)\times10^{-14}$ & $3.37\times10^{-14}$ & $1.40\times10^{-14}$ \\[5pt]
11 & 0.084 &$(1.47\pm0.00)\times10^{-13}$ & $9.06\times10^{-14}$& $1.56\times10^{-14}$ \\[5pt]
12 & 0.102 &$(1.71\pm0.01)\times10^{-13}$ &  $7.28\times10^{-14}$ & $1.65\times10^{-14}$ \\[5pt]
13 & 0.147 &$(6.78\pm2.68)\times10^{-14}$ & $4.11\times10^{-14}$ & $1.15\times10^{-14}$ \\[5pt]
14 & 0.181 &$(1.63\pm0.00)\times10^{-13}$ & $2.75\times10^{-14}$ & $1.03\times10^{-14}$
\enddata
\end{deluxetable*}

\subsection{Integrated Emission Lines}
In Figure~\ref{all_ratios}, we show the integrated line ratios for the 14 LARS galaxies, with the fluxes given in Table~\ref{fluxtable}. Following \citet{scarlata2009}, we show several models describing dust geometry. A simple dust screen model with $R_V = 3.07$  \citep{cardelli1989} is denoted by the black dashed line. The black dash-dotted line demonstrates a smooth internal dust model \citep{mathis1972}, where the stars, gas, and dust are evenly mixed. This model is described as $I_\lambda/I_{\lambda}^0 = (1 - e^{-\tau_\lambda})/\tau_\lambda$, where $I_\lambda$ is the observed intensity of the source, $I_\lambda^0$ is the assumed intrinsic intensity, and $\tau_\lambda$ is the optical depth of the medium. Finally, we show models for various clumpy media following the formalism of \cite{natta1984},
\begin{equation}
\label{natta}
\frac{I_\lambda}{I_\lambda^0} = \textrm{exp}(-N(1-e^{-\tau_{c, \lambda}}))
\end{equation}
where $I_\lambda$ and $I_\lambda^0$ are defined as before, $N$ is the mean number of clumps along the line of sight, and $\tau_{c, \lambda}$ is the optical depth of each clump. The number of clumps per sightline follow a Poisson distribution and individually obey the Cardelli model, with each clump having the same optical depth. These models do not include Ly$\alpha$ radiative transfer effects. \citet{scarlata2009} found that a majority of $z\sim0.3$ Ly$\alpha$ emitting galaxies were best described by the clumpy dust model. Meanwhile, for the galaxies that lie below the uniform dust screen model, they noted that a mechanism that would preferentially dampen the Ly$\alpha$ photons but leave the Balmer photons unaffected would be required, such as a neutral medium through which the Ly$\alpha$ photons would pass without scattering.  

We find that while several of the LARS galaxies lie in the region of Figure~\ref{all_ratios} described by a clumpy dust model, others show Ly$\alpha$ to be significantly lower than that inferred from the Balmer decrement. Of particular note are the galaxies for which the H$\alpha$/H$\beta$ ratios are consistent with the intrinsic value, but have Ly$\alpha$/H$\alpha$ values anywhere from $\sim75$\% (LARS14) to nearly 100\% (LARS09) less than than the intrinsic value of 8.7. This is in agreement with previous studies that have found that integrated Ly$\alpha$ emission is often much weaker than expected from recombination theory \citep[e.g.,][]{terlevich1993, giavalisco1996, atek2008, scarlata2009}. Overall, the LARS galaxies show Ly$\alpha$/H$\alpha$ ratios that span the whole range from no additional destruction of photons beyond that done by dust to the formation of damped Ly$\alpha$ absorption profiles.

The internal dust model is difficult to differentiate from a clumpy dust model at low optical depth and a small number of clumps. However, given the global values shown in Figure~\ref{all_ratios}, it is unlikely that the internal dust best explains the ISM in the LARS galaxies.

\begin{figure*}[!t]
\centering
\scalebox{0.5}
{\includegraphics{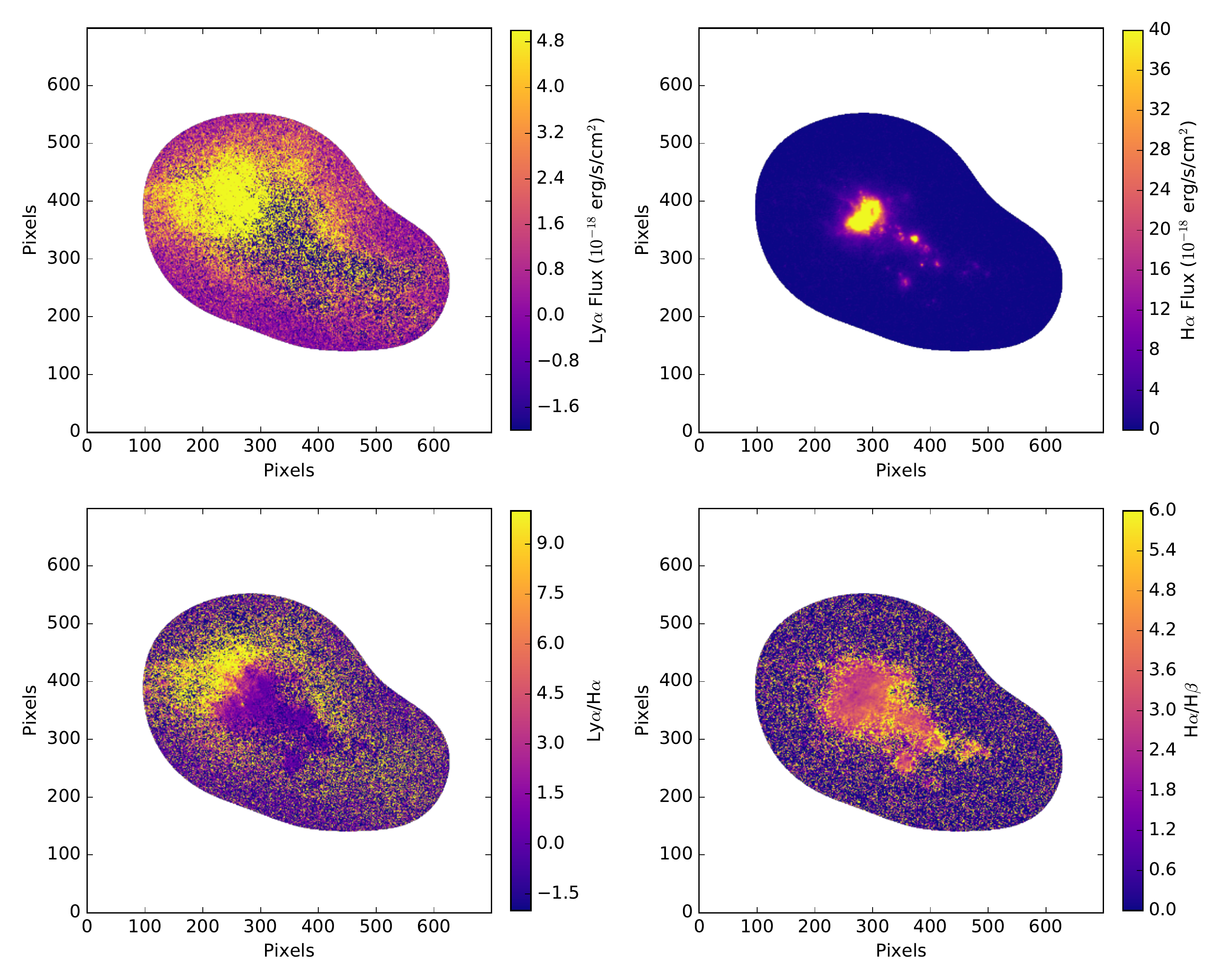}}
\caption{Images of LARS01. The calculated aperture has been applied. \emph{Top left}: The Ly$\alpha$ emission map. \emph{Top right}: The H$\alpha$ emission map. \emph{Bottom left}: The Ly$\alpha$/H$\alpha$ ratio, with an average uncertainly of 0.01. \emph{Bottom right}: The H$\alpha$/H$\beta$ ratio with an average uncertainly of 0.001.}
\label{images01}
\end{figure*}

\begin{figure*}[!t]
\centering
\scalebox{0.5}
{\includegraphics{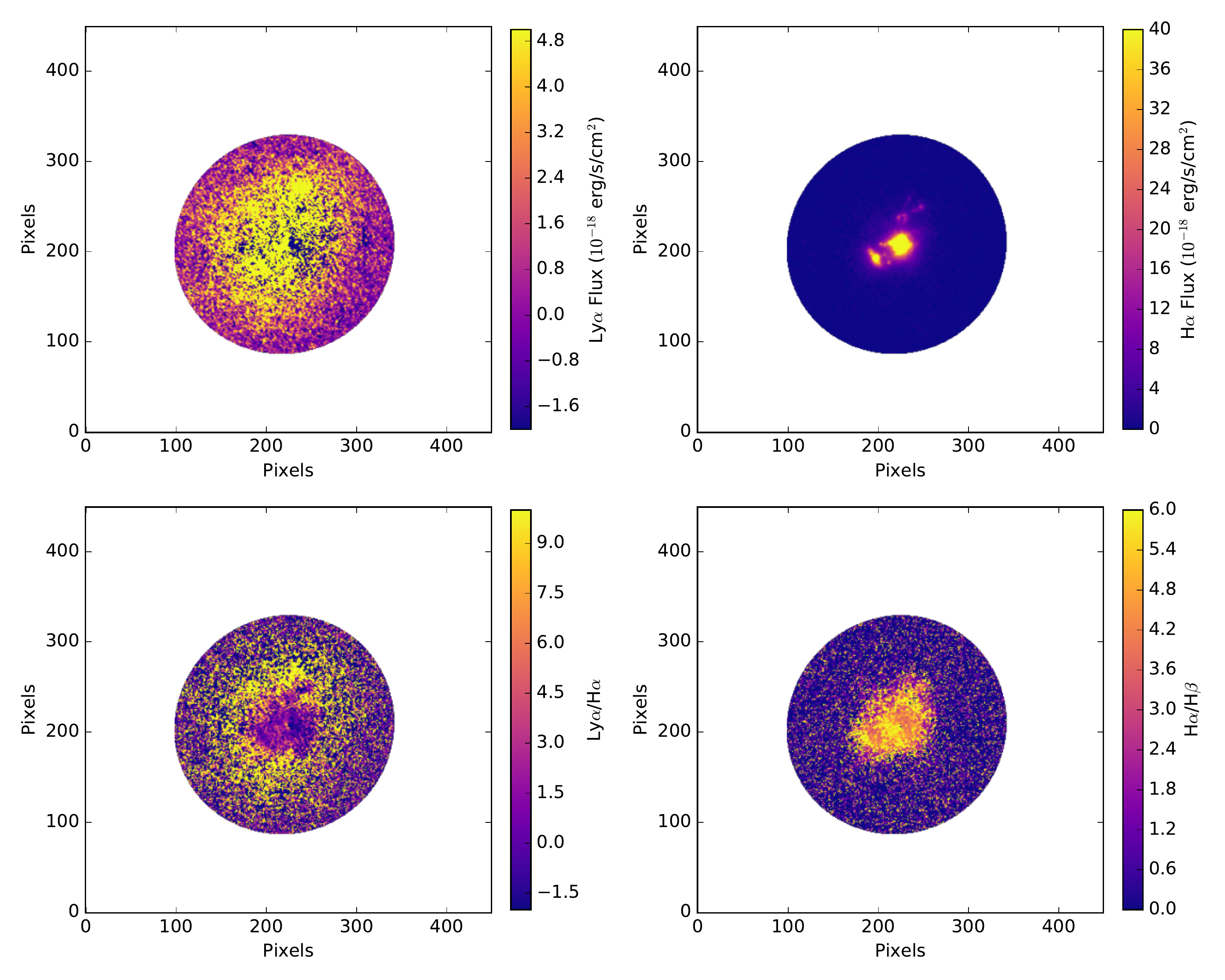}}
\caption{Images of LARS12. The calculated aperture has been applied. \emph{Top left}: The Ly$\alpha$ emission map. \emph{Top right}: The H$\alpha$ emission map. \emph{Bottom left}: The Ly$\alpha$/H$\alpha$ ratio, with an average uncertainty of 0.02. \emph{Bottom right}: The H$\alpha$/H$\beta$ ratio, with an average uncertainly of 0.005.}
\label{images12}
\end{figure*}

\subsection{Resolved Emission Line Ratios}\label{resolved}
Using the resolved line maps of the LARS galaxies, we are able to probe the dust geometry down to the minimum physical scales allowed by the \emph{HST} point spread function (see Melinder et al.\ \emph{in prep}), which for these galaxies is approximately 40 to 130 parsecs, depending on redshift. By using the emission line ratios and knowing where exactly in the galaxies different line ratios occur, we can trace what regions in the galaxies are affecting the integrated line ratios. In this section, we will compare two galaxies that are each representative of differing dust models, LARS01 and LARS12. The \emph{HST} Ly$\alpha$ and H$\alpha$ emission maps are shown for these two galaxies in top panels of Figures~\ref{images01} and~\ref{images12}, respectively. Because the H$\beta$ images have low signal-to-noise in the outer regions of the galaxies we have implemented a cut in the H$\beta$ line maps of $f_{\textrm{H}\beta} = 1\times10^{-18}$ erg/s/cm$^2$. We then use only the pixels from all three emission maps that correspond to H$\beta$ values greater than this cut.

\begin{figure}
\centering
\scalebox{0.41}
{\includegraphics{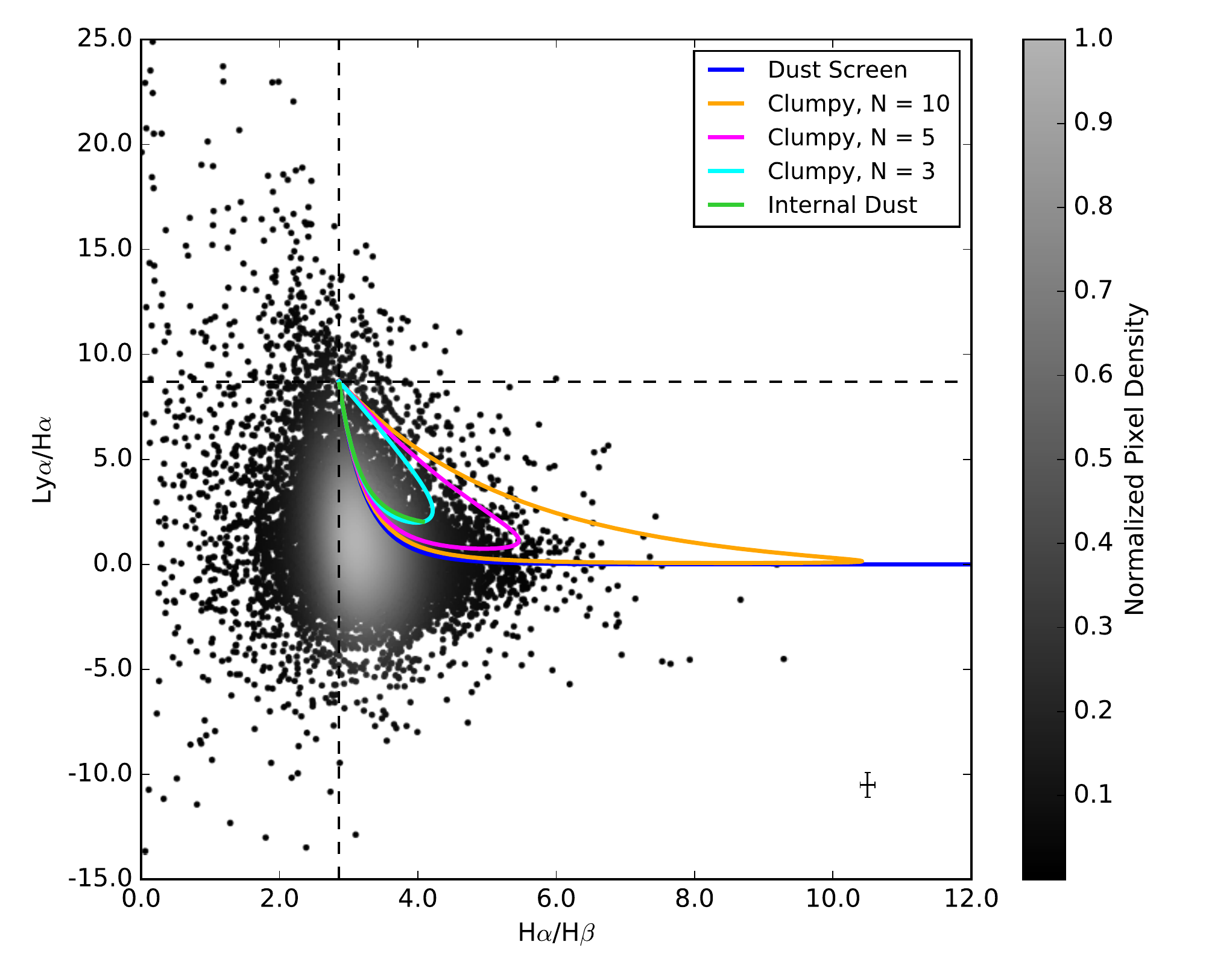}}
\caption{Ly$\alpha$/H$\alpha$ versus H$\alpha$/H$\beta$ for LARS01. The uniform dust screen model is plotted as a blue line, and three different clumpy dust models for $N=3,5,10$ are shown in orange, magenta, and cyan, respectively. The smooth internal dust model is plotted with a green line. The horizontal black dashed line indicates the intrinsic ratio of Ly$\alpha$/H$\alpha\sim8.7$, and the vertical dashed line shows H$\alpha$/H$\beta$ = 2.86. The normalized density of points is shown in the gray scale (determined using a Gaussian kernel density estimator). The median error bar is shown in the bottom right corner.}
\label{ratios}
\end{figure}

LARS01, which is displayed in Figure~\ref{all_ratios} as a purple square, is an example of a galaxy that is not well described by a clumpy dust model. We show the Ly$\alpha$/H$\alpha$ versus H$\alpha$/H$\beta$ values for the galaxy in Figure~\ref{ratios}.  As in Figure~\ref{all_ratios}, we have plotted the various dust models.  While some pixels fall in the clumpy model region, most lie below the plotted dust models. \cite{scarlata2009} noted that the dust models shift slightly downwards if the scattering of Ly$\alpha$ photons \emph{within} the clumps is included. This effectively increases the optical depth of these photons. Further dampening of Ly$\alpha$ may be due to these photons traveling through the neutral hydrogen in the ISM. Additionally, a number of pixels correspond to regions where the Ly$\alpha$ is the same or brighter than predicted by recombination physics. In the lower panels of Figure~\ref{images01}, the Ly$\alpha$/H$\alpha$ and H$\alpha$/H$\beta$ ratios are shown for the galaxy.

In addition to the bright Ly$\alpha$, there are also a significant number of pixels that have Ly$\alpha$/H$\alpha < 0$. Some of these may be due to noise, producing artificially negative Ly$\alpha$ values. Nevertheless, our cut in H$\beta$ minimizes the number of pixels affected by noise. Therefore, the negative Ly$\alpha$/H$\alpha$ values are more likely to be where Ly$\alpha$ is seen in absorption. This is due to continuum light being either absorbed or scattered by the n = 2 to n = 1 resonance. The Ly$\alpha$ emission maps shown in the top right panels of Figures~\ref{images01} and~\ref{images12} display the negative Ly$\alpha$ pixels in dark purple. The fact that the negative pixels are spatially correlated, particularly in the more central regions of the galaxies, indicates that these are real absorption features.

\begin{figure}
\centering
\scalebox{0.41}
{\includegraphics{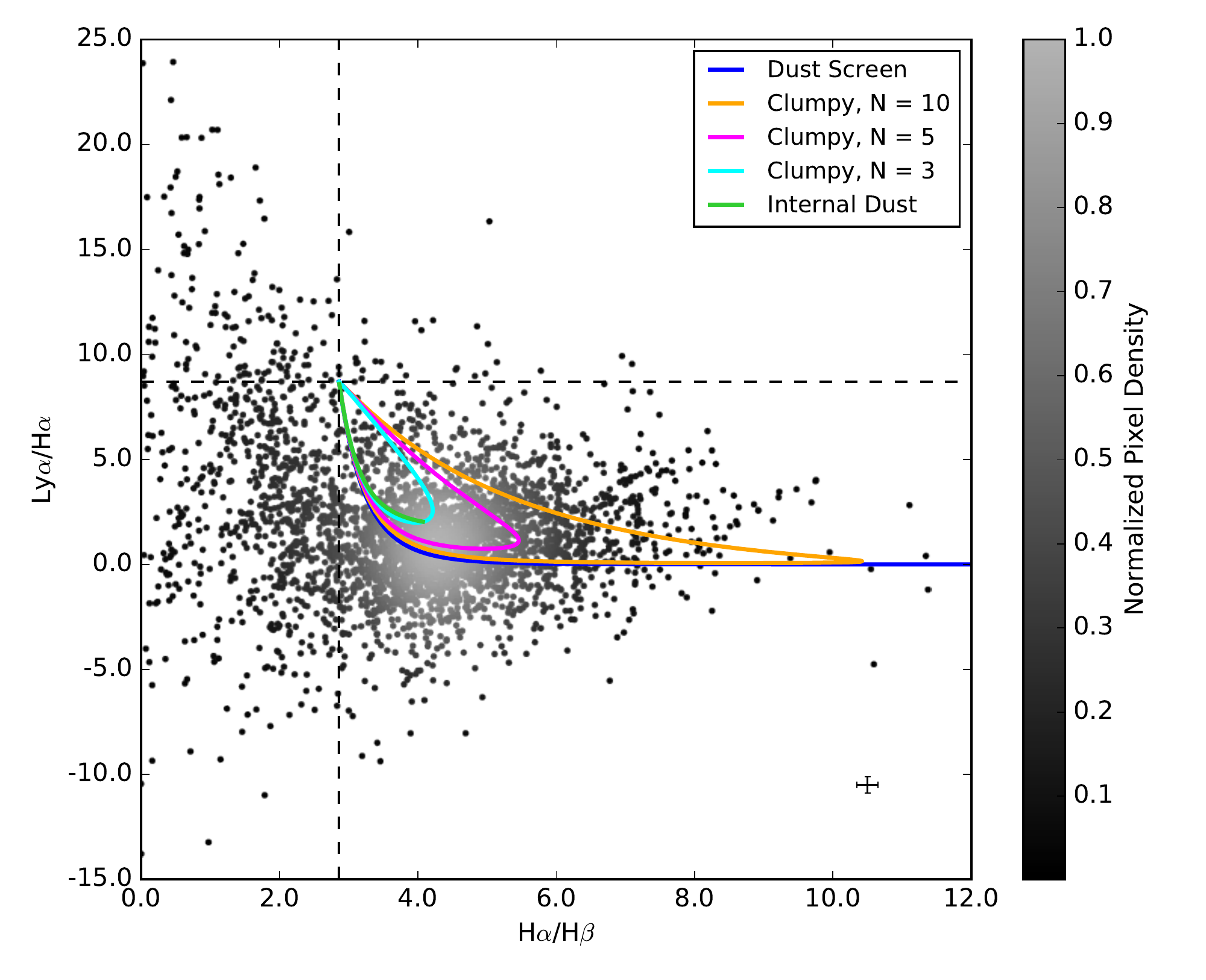}}
\caption{Same as Figure~\ref{ratios} but for LARS12. LARS12 is generally a redder galaxy, and many more pixels lie in the clumpy dust model regions than for LARS01.}
\label{ratios12}
\end{figure}

Unlike LARS01, LARS12, denoted in Figure~\ref{all_ratios} by a yellow star, falls directly in the region of the clumpy dust models. We show the Ly$\alpha$/H$\alpha$ versus H$\alpha$/H$\beta$ values for the pixels in LARS12, along with the same dust models, in Figure~\ref{ratios12}. As would be expected from the global Ly$\alpha$/H$\alpha$ and H$\alpha$/H$\beta$ ratios, a majority of the pixels fall in the regions of the plot well-described by the clumpy dust models with varying $N$ (numbers of clumps along the line of sight) and $\tau$ (optical depth). Overall, LARS12 has much higher H$\alpha$/H$\beta$ values than LARS01, indicating that the galaxy is significantly dustier. We show the Ly$\alpha$/H$\alpha$ and H$\alpha$/H$\beta$ ratios of LARS12 in the bottom panels of Figure~\ref{images12}.

\begin{figure*}
\centering
\scalebox{0.32}
{\includegraphics{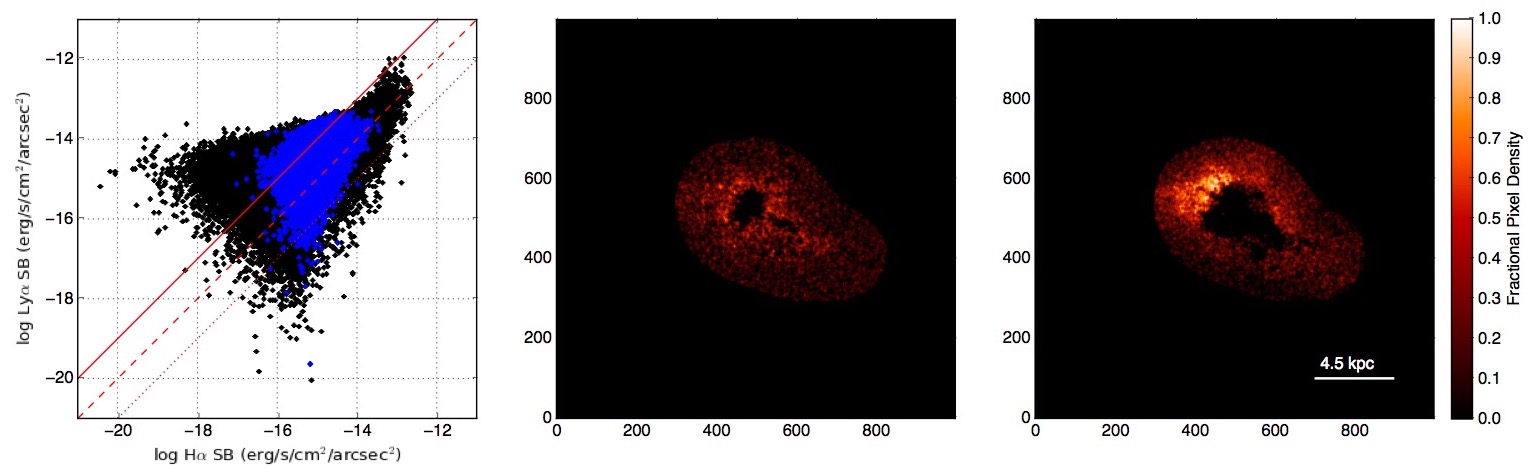}}
\caption{\emph{Left}: Log Ly$\alpha$ surface brightness vs. log H$\alpha$ surface brightness for LARS01. Overplotted in blue are the pixels in the galaxy that lie above the simple dust screen model  (Figure~\ref{ratios}). Note that since the values are plotted on a logarithmic scale, only positive values are included. The solid red line indicates the intrinsic value of Ly$\alpha$ versus H$\alpha$ surface brightnesses under Case B recombination, and the dashed and dotted red lines show a one-to-one relation and ten times below that, respectively. \emph{Center}: The blue pixels in the left panel are traced back to their origin in the Ly$\alpha$ image. The calculated aperture has been applied to each galaxy. The color scale of the image corresponds to how densely the pixels are packed, with the lighter colors showing a greater number of pixels in the region and the dark colors showing no pixels that correspond to the blue pixels on the left. The color scale is normalized to one. \emph{Right}: The Ly$\alpha$ image showing only the pixels that lie above the intrinsic Ly$\alpha$/H$\alpha$ value denoted by the solid red line in the left panel. These are the pixels that indicate scattering of Ly$\alpha$ photons into the line of sight, and largely trace the diffuse Ly$\alpha$ emission.}
\label{sbplotcardscatt01}
\end{figure*}

To quantify the comparison of Ly$\alpha$ and H$\alpha$ in a galaxy, we examine their respective surface brightnesses \citep{ostlin2009}. We plot the surface photometry for LARS01 in the left panel of Figure~\ref{sbplotcardscatt01}. Shown in black are the logarithmic Ly$\alpha$ surface brightness plotted against the logarithmic H$\alpha$ surface brightness.  The solid red line indicates the intrinsic Ly$\alpha$/H$\alpha\sim 8.7$ expected under the assumption of Case B recombination. Lower Ly$\alpha$/H$\alpha$ ratios indicate either dust absorption or scattering of the Ly$\alpha$ photons out of the line of sight, while Ly$\alpha$/H$\alpha$ values greater than $\sim8.7$ show where Ly$\alpha$ has likely been scattered \emph{into} the line of sight.

Overplotted in blue in the left panel of Figure~\ref{sbplotcardscatt01} are the individual pixels of LARS01 that fall in the region corresponding to the clumpy dust models of \citet{natta1984} (defined as any pixel lying above the dust screen curve in Figure~\ref{ratios}). Most of the blue pixels show Ly$\alpha$/H$\alpha$ ratios well below the intrinsic relationship, indicating that the Ly$\alpha$ photons are either absorbed by dust or scattered out of the line of sight by neutral hydrogen \citep{hayes2015}. Some pixels indicate that Ly$\alpha$ photons have also been scattered into the line of sight, as they are overluminous for the corresponding H$\alpha$ surface brightnesses. Both of these scenarios can be explained by the presence of a clumpy medium that can both dampen as well as scatter the Ly$\alpha$. The brightest Ly$\alpha$ pixels (in the upper right region of the left panel of Figure~\ref{sbplotcardscatt01}) are not affected by the clumpy medium.

In the center panel of Figure~\ref{sbplotcardscatt01}, we trace the pixels in the clumpy dust region of Figure~\ref{ratios} back to the LARS01 Ly$\alpha$ image. What is immediately obvious is that the Ly$\alpha$ photons that are escaping from the center of the galaxy are \emph{not} doing so through a clumpy dust medium. Not only is the central $\sim 0.65$~kpc diameter region the source of the brightest Ly$\alpha$ emission, but its H$\alpha$/H$\beta$ ratio is very close to the intrinsic value of 2.86.  Whatever minimal extinction exists in the region is consistent with that expected from a dust screen model. This suggests that the region has been cleared of static neutral hydrogen, either by the winds from a preponderance of supernovae or by a bubble blown out by the regionÕs active star formation.  The resulting outflows then allow Ly$\alpha$ photons in the wings of the line to escape \citep{mashesse2003, riverathorsen2015, herenz2016}. Moreover, the concentration of ionized hydrogen in the galaxy's central region (see Figure~\ref{images01}) indicates that most of the medium has been ionized, and is therefore transparent to Ly$\alpha$ photons. \cite{laursen2009} showed that the dust model used in Ly$\alpha$ radiative transfer calculations is relatively unimportant since the photons either travel through fairly dust-free regions, or they travel through dusty regions and are absorbed. This is in line with the general lack of Ly$\alpha$ attenuation in the center of LARS01. 

\begin{figure*}
\centering
\scalebox{0.32}
{\includegraphics{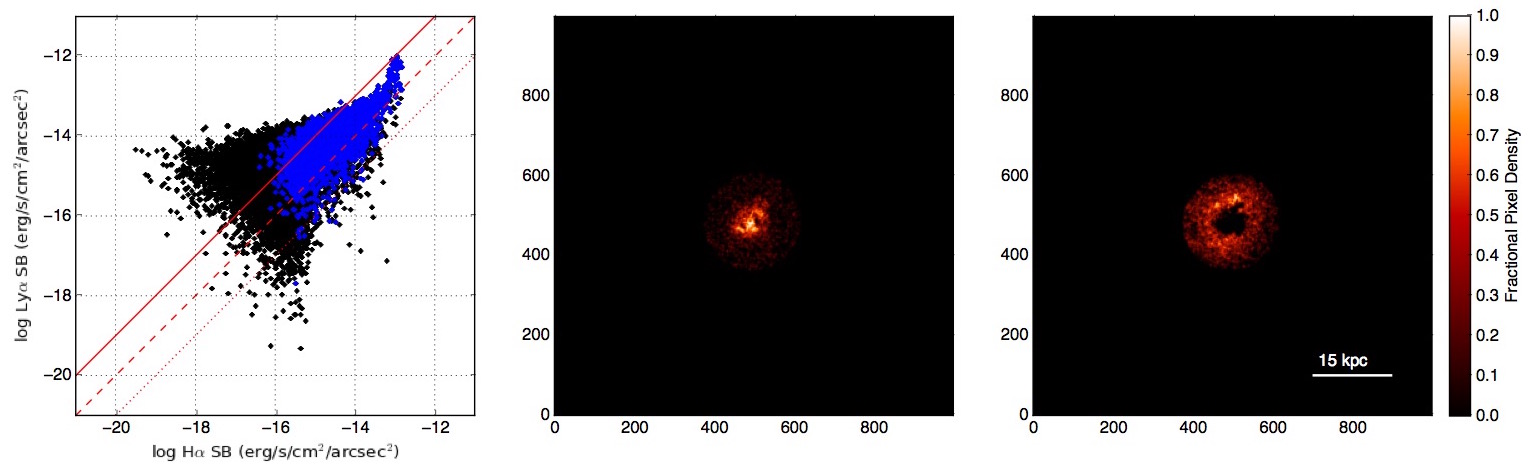}}
\caption{Same as Figure~\ref{sbplotcardscatt01} but for LARS12. LARS12 shows evidence of clumpy dust almost all the way to the center of the galaxy, save for a small region (just several pixels wide) at its center.}
\label{sbplotcardscatt12}
\end{figure*}

Most of the pixels that lie above the \cite{cardelli1989} dust screen model in Figure~\ref{ratios} come from a concentrated annular region around the center of the galaxy. This indicates that an annulus of clumpy dust exists outside the central star-forming region.

\cite{riverathorsen2015} observed the LARS galaxies using the Cosmic Origin Spectrograph (COS) on the \emph{HST}. They found that, in general, gas outflows with velocities greater than 50 km s$^{-1}$ occurred in all the galaxies where Ly$\alpha$ is observed. The model used to explain this behavior \citep{tagle1999, mashesse2003} posits that feedback from the actively star-forming regions in the center of the galaxy creates Rayleigh-Taylor instabilities at the border between the hot bubble and the neutral phase of the ISM. This results in a clumpy medium through which Ly$\alpha$ photons can either escape or be absorbed. This model could explain the regions we observe in LARS01 and the other LARS galaxies.

In contrast to LARS01, the integrated ratios of LARS12 place it solidly in the region of Figure~\ref{all_ratios} that is described by the clumpy dust models.~LARS12 is consistent with the clumpy dust model almost all the way to the nucleus (see Figure~\ref{sbplotcardscatt12}).~While Ly$\alpha$ escapes freely from the inner $\sim 0.65$ kpc of LARS01, the same cavity in LARS12 is only $\sim 0.4$~kpc in diameter. This regions is denoted by several pixels in the center of the galaxy, shown in the center panel of Figure~\ref{sbplotcardscatt12}. This difference can also be seen in the curve of growth analysis performed by \citet[][Figure 4]{hayes2014}: Êin LARS01, the integrated Ly$\alpha$ flux peaks close to the galactic center, while in LARS12, this peak is 5~kpc from the galaxy's center. In other words, in the center of LARS12, dust attenuation is suppressing Ly$\alpha$ emission in the center of LARS12 compared to what would be expected from a clumpy dust medium. 

We next isolate the pixels of the LARS01 image that fall \emph{above} the intrinsic Ly$\alpha$/H$\alpha$ value. The regions of the galaxy in which these pixels are located are of particular interest as they provide insight into the mechanisms that can transport Ly$\alpha$ to large radii. Previous studies have found that in some galaxies, as much as 70\% of the Ly$\alpha$ emission comes from the diffuse regions that are outside the more central areas of star formation \citep{hayes2005, atek2008}. These halos often reach galactocentric radii of 10 kpc. Several processes can theoretically modify the Ly$\alpha$/H$\alpha$ ratio, such as extreme high or low densities, shocks, and other non-equilibrium conditions, but the more likely explanation of an upward deviation from Ly$\alpha$/H$\alpha\sim8.7$ is the physics of Ly$\alpha$ radiative transfer \cite[e.g.,][]{hayes2015}. The fact that Ly$\alpha$/H$\alpha \gg 10$ is frequently observed tells us that scattering into the line of sight not only happens, but that it can dominate the integrated luminosity of a galaxy.

In the right panel of Figure~\ref{sbplotcardscatt01}, we show only the regions of LARS01 where Ly$\alpha$ photons are scattered into the line of sight. The pixels that contain overly bright Ly$\alpha$ make up the Ly$\alpha$ halo of LARS01. Figure~\ref{sbplotcardscatt12} shows the same quantity but for LARS12.  This galaxy has the same characteristics as LARS01, with the highest Ly$\alpha$/H$\alpha$ pixels being the most significant contributor to the emission from the outer halo. For Ly$\alpha$ photons to be this bright relative to H$\alpha$, the photons have to be \emph{scattered} into the line of sight. It has been shown that the size of extended Ly$\alpha$ halos around low-redshift Ly$\alpha$ emitters is inversely correlated with dust content, indicating that low dust abundance is necessary for Ly$\alpha$ photons to resonantly scatter to large radii \citep{hayes2013}. The fact that the high Ly$\alpha$/H$\alpha$ pixels correspond directly to the Ly$\alpha$ halo indicates that the halo is generated by a scattering process where little dust exists, rather than escaping directly through holes blown through the ISM by galactic winds.

We present figures similar to Figures~\ref{sbplotcardscatt01} and~\ref{sbplotcardscatt12} for the other LARS galaxies in the Appendix.  As might be expected from the results for LARS12, all of the galaxies that lie on or above the Cardelli model in Figure~\ref{all_ratios} (LARS03, 08, 11, and 13) show little to no cavity in the center of the galaxies. Rather than forming an annulus, the pixels corresponding to the clumpy dust model exist all the way to the center. In contrast, most of the galaxies of this study possess a scattering halo where Ly$\alpha$ is overly bright compared to H$\alpha$.

\section{An Empirical Model for Lyman-Alpha Scattering}
Here, we present a model to quantify how much Ly$\alpha$ photons scatter in galaxies. This model is based on the observations of dust geometries present in the LARS galaxies. To model the emission of Ly$\alpha$, we begin with an examination of the relative intensities of Ly$\alpha$ and H$\alpha$ in the LARS galaxies. Deviations from the intrinsic ratio of Ly$\alpha$/H$\alpha\sim8.7$ capture the physics of Ly$\alpha$ radiative transfer. At gas densities that are relevant in typical interstellar media, the opacity of the H$\alpha$ line is negligible. Our goal is to quantify this difference between H$\alpha$ and Ly$\alpha$, and thereby probe the effect that environment and dust geometry have on Ly$\alpha$ scattering. There is a stark difference between the Ly$\alpha$ and H$\alpha$ emission shown in Figures~\ref{images01} and~\ref{images12}, highlighting the scattering that Ly$\alpha$ photons undergo. By using the H$\alpha$ (with H$\beta$ to correct for reddening) images and our knowledge of the intrinsic Ly$\alpha$/H$\alpha$ ratio, we can calculate the intrinsic Ly$\alpha$ emission that should exist in comparison to what is actually observed.

\begin{figure*}[!t]
\centering
\scalebox{0.7}
{\includegraphics{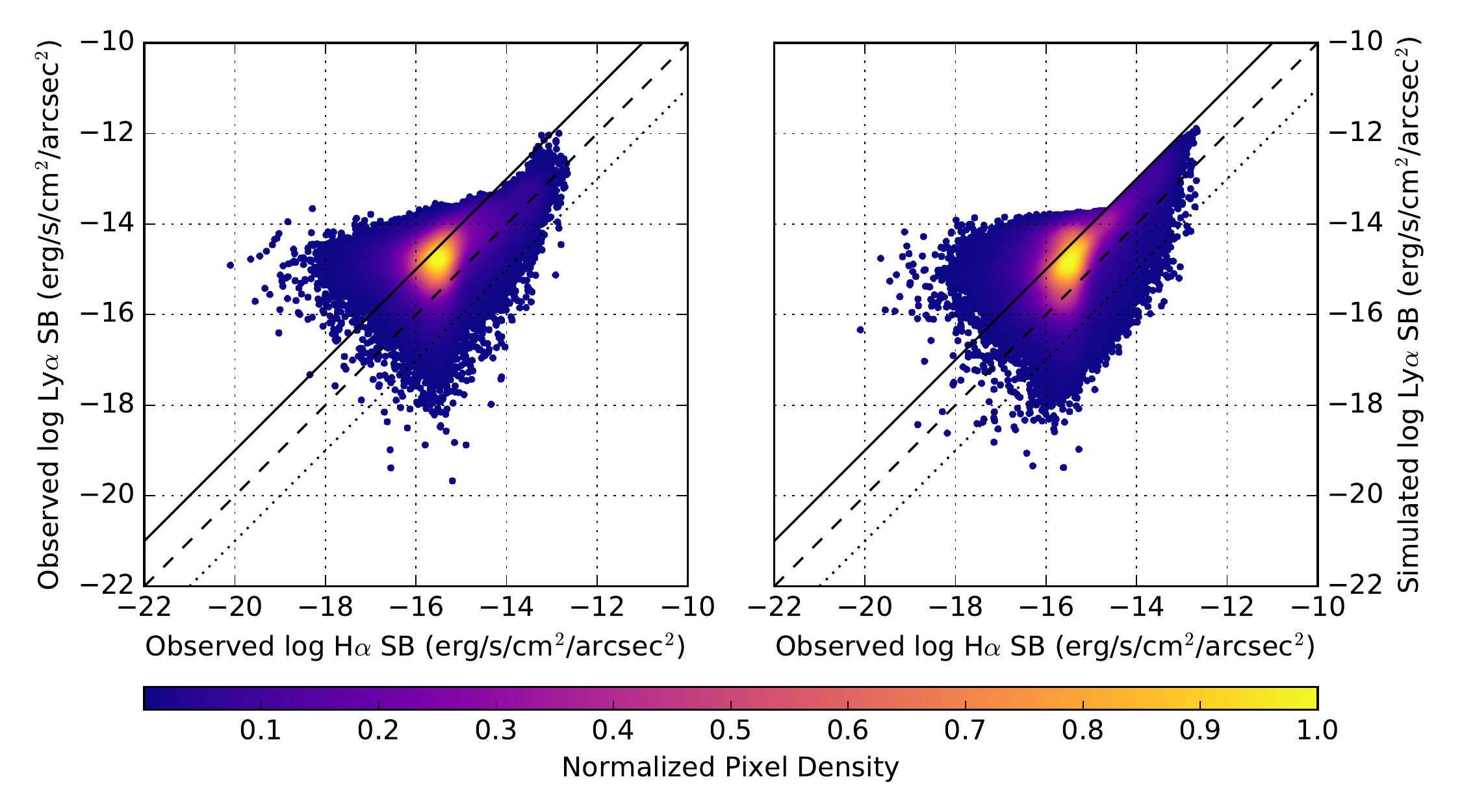}}
\caption{Surface brightness plots for LARS01. Shown in the left panel is the observed Ly$\alpha$ surface brightness versus H$\alpha$ surface brightness. Note that since the values are plotted on a logarithmic scale, only positive measurements are included. The solid black line denotes the Case B recombination value, while the dashed line represents a one-to-one ratio of Ly$\alpha$ and H$\alpha$ emission and the dotted line a dex below that. The right panel shows the same plot but for the simulated Ly$\alpha$ surface brightness from the model versus H$\alpha$ surface brightness. The normalized density of points is shown in the color scale (determined using a Gaussian kernel density estimator), and the density values shown in the color bar.}
\label{sbplot01}
\end{figure*}

\begin{figure*}[!t]
\centering
\scalebox{0.7}
{\includegraphics{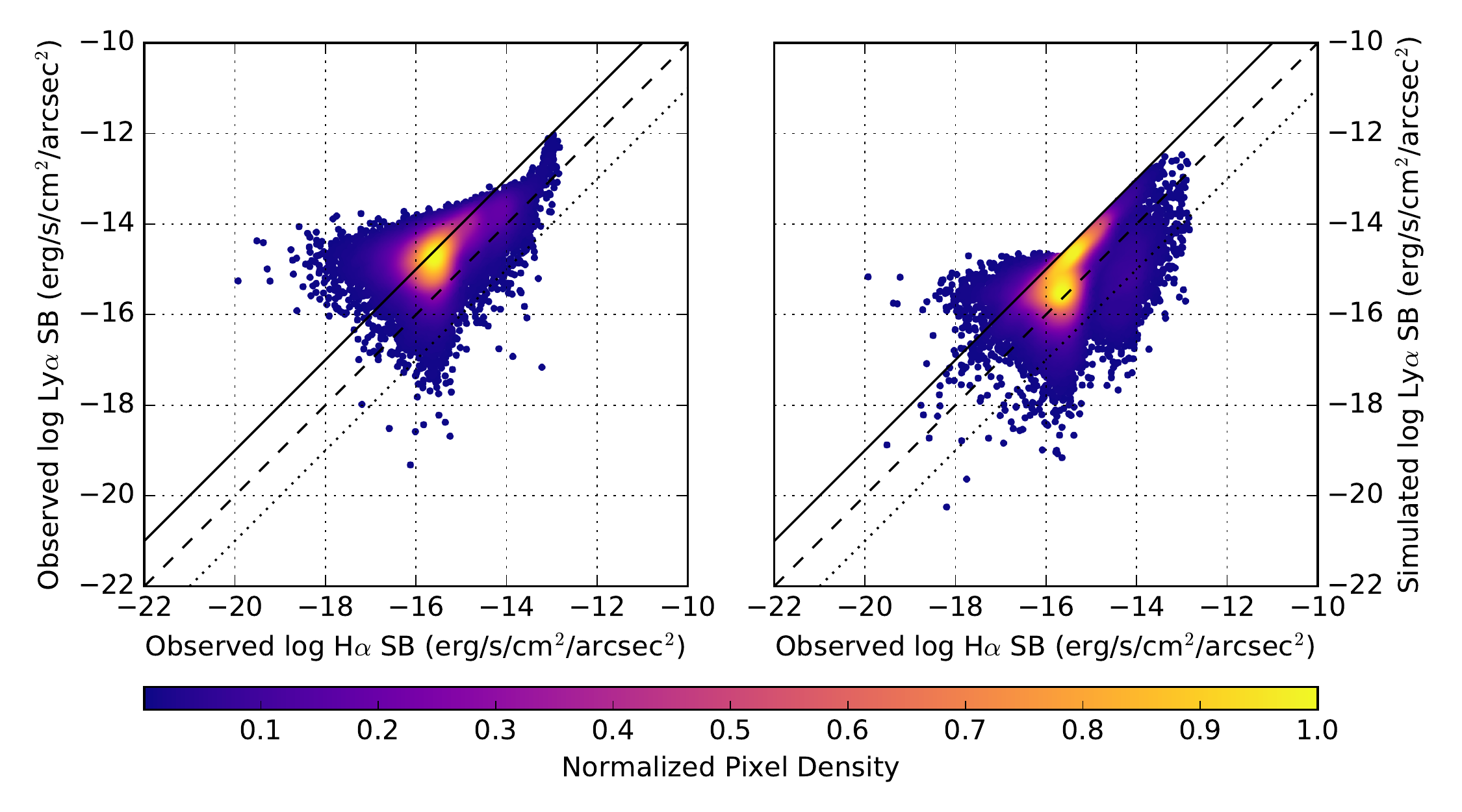}}
\caption{Same as Figure~\ref{sbplot01}, but for LARS12.}
\label{sbplot12}
\end{figure*}

\begin{figure}
\centering
\scalebox{0.45}
{\includegraphics{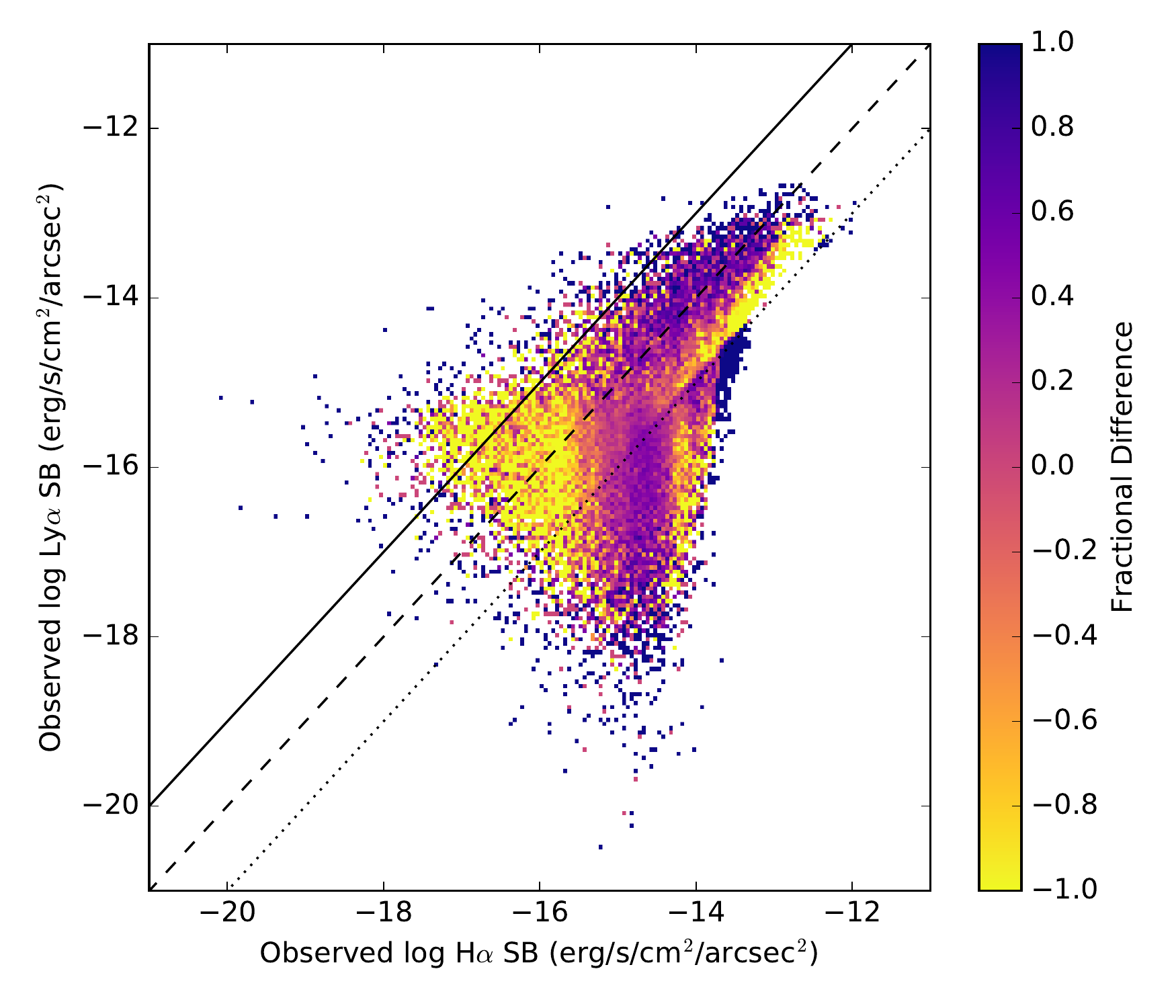}}
\caption{Fractional difference between the model and observed surface brightnesses for LARS01. The data have been binned and a positive difference (purple) indicates that there are more points in the observed bin than the model bin. Negative differences (yellow) occur where there are more points in the model bin than the observed data.}
\label{res01}
\end{figure}

\begin{figure}
\centering
\scalebox{0.45}
{\includegraphics{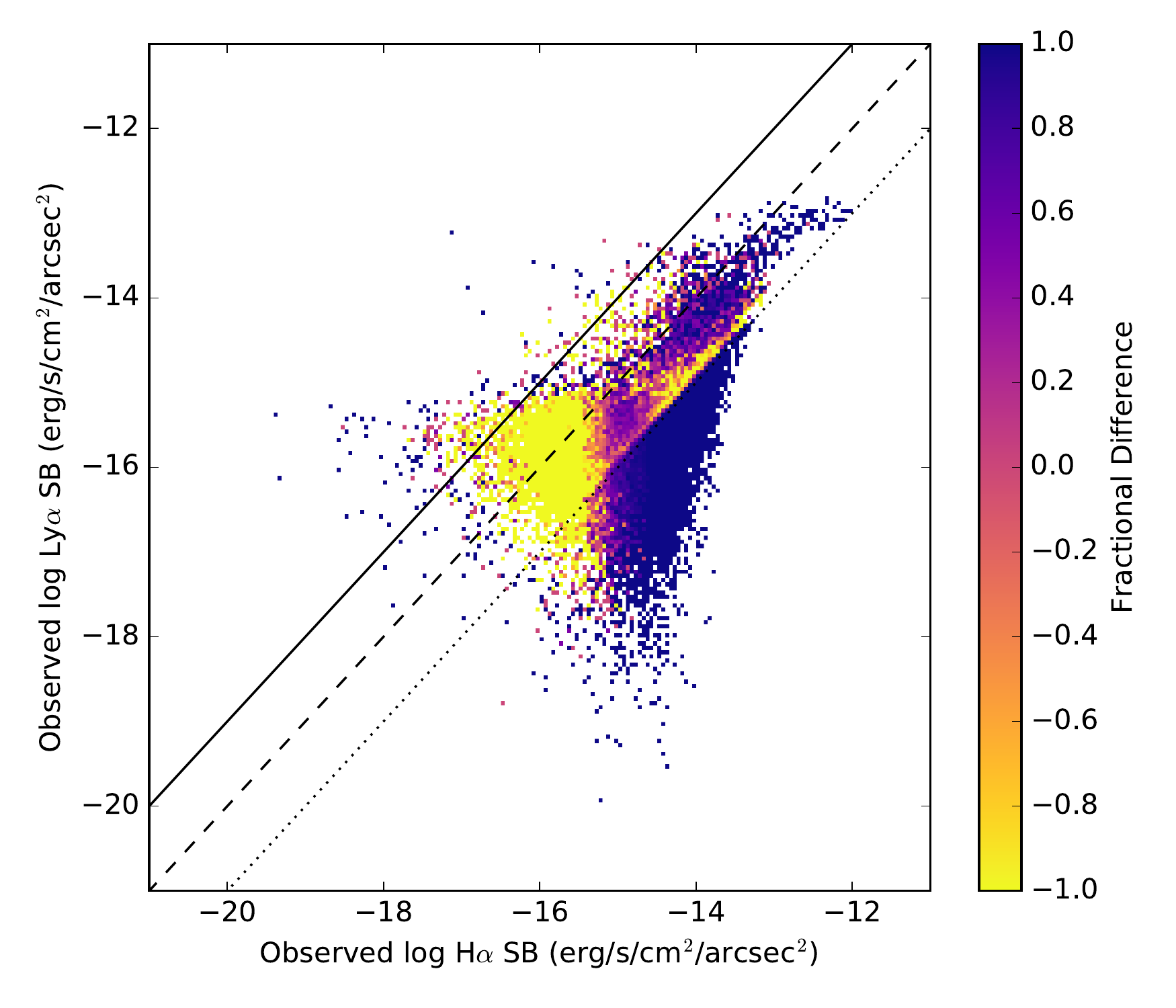}}
\caption{Same as Figure~\ref{res01}, but for LARS12.}
\label{res12}
\end{figure}

\subsection{Model Description}
Our scattering model is based on the observed properties of the dust absorption and photon scattering discussed in Section~\ref{resolved}. LARS01 shows that the pixels corresponding to a clumpy dust model congregate in an annulus around a central region within which the Ly$\alpha$ emission is brightest and only slightly extinguished. Meanwhile, the pixels that are overbright in Ly$\alpha$ compared to H$\alpha$, which is an indication of scattering back into the line of sight, form the Ly$\alpha$ halo. Following this, we posit a three-component galaxy model: a central region in which the Ly$\alpha$ photons are attenuated by only a uniform dust screen, an annular region that contains a clumpy dust geometry, and finally an outer halo region with no dust where the Ly$\alpha$ photons can scatter out to large distances. Of particular interest are how and how much the Ly$\alpha$ photons scatter in the halo region, as one of the big questions about Ly$\alpha$ escape is how the dust affects the escape of Ly$\alpha$ photons.

We begin multiplying the 2D H$\alpha$ emission line map by the intrinsic ratio of 8.7 to obtain an estimate of the intrinsic Ly$\alpha$ distribution that is unaffected by dust or scattering. We then create three separate images: one derived using a dust screen model, another using a clumpy dust model, and a third where the scattering of Ly$\alpha$ photons is the only physical process.  After each image is created, we  piece them together, using the dust screen model for the central region, the clumpy dust model in an annulus around the center, and the scattering model for the outer regions. Below, we describe the details of each segment, and how the sizes of each region are determined.

We modulate the central region Ly$\alpha$ emission with a simple dust screen model with
\begin{equation}
E(B-V) = \frac{2.5}{k_{\textrm{H}\beta} - k_{\textrm{H}\alpha}} \textrm{log}_{10}\bigg(\frac{\textrm{H}\alpha/\textrm{H}\beta}{2.86}\bigg)
\end{equation}
where $k_{\textrm{H}\alpha}$ = 2.455 and $k_{\textrm{H}\beta}$ = 3.520, as described by \citet{cardelli1989}. Here, the H$\alpha$/H$\beta$ values used are drawn randomly from the Balmer decrement distribution for the galaxy being modeled and applied to each pixel in the central region. The reason we draw randomly from a distributions of Balmer decrements rather than the measured value for each pixel is because the H$\beta$ is not always well measured, making a reliable value at each pixel impossible to determine. This extinction is then applied as
\begin{equation}
L_{\textrm{Ly}\alpha} = 8.7 \times L_{\textrm{H}\alpha}\times 10^{-0.4E(B-V)k_{\textrm{Ly}\alpha}}
\end{equation}
The size of the central region is a free parameter that is fit for in the model.

The annulus between the central region and the Ly$\alpha$ halo has a more complex dust component than the central region as it is assumed to contain clumpy media. This changes the effect that the dust has on the Ly$\alpha$ photons, resulting in a different absorption cross-section. We model this effect with Equation~\ref{natta}. Using the intrinsic H$\alpha$/H$\beta$ ratio of 2.86, we calculate the number of clumps along the line of sight for each pixel in the annular region. From this set of clump values we then draw randomly to create a distribution of clumps along the line of sight in the annulus and apply the random values to all of the pixels in the region. As with the central region, the size of the annulus is determined during the model fitting. Additionally, we assume a normal distribution of $\tau$ centered around 1.0 with a standard deviation of 0.25. This has the effect of magnifying the absorption slightly as compared to the dust screen model applied to the central region. Using other optical depth values would slightly increase or decrease the absorption in the annular clumpy dust region. The choice of this optical depth distribution was largely empirical, chosen because any higher optical depth and it was difficult for the model to reproduce any of the surface brightness distributions from the annular region. Using a global distribution of $\tau$ centered at 1 is a simplifying assumption, but its use does not change qualitatively the overall conclusions drawn from the model outputs. 

To model the Ly$\alpha$ halo at large radii, we ``scatter" the photons by convolving the simulated Ly$\alpha$ (where Ly$\alpha$ = 8.7 $\times$ H$\alpha$) emission with a symmetric two-dimensional Gaussian kernel. The convolution operation empirically mimics the effect of scattering the Ly$\alpha$ from the centralized emission that is seen in the H$\alpha$ maps toward larger angular distances seen in the Ly$\alpha$ halo. That a Gaussian two-dimensional kernel is appropriate can be understood from the radiative transfer point of view: once in the wings of the line, a Ly$\alpha$ photon performs a random walk \citep{adams1972}. An ensemble of random ``walkers" in an inhomogenous medium will result in a Gaussian distribution after sufficient steps \citep{gronke2016,gronke2017}. Other kernels were tested, such as a Lorentzian kernel, with similar results. 

The resulting model has three input parameters: the average scattering distance of Ly$\alpha$ photons in the halo region, represented by the Gaussian kernel width, the isophotal size of the central region, and the isophotal size of the annulus. We parameterize the sizes of the two regions as a fraction of the aperture calculated in Section 2.3.

To find the best fit model for each galaxy, we use a Markov Chain Monte Carlo implementation called \texttt{emcee} \citep{foreman2013}. This technique ensures that the parameter spaces for each input are fully explored without being excessively computationally expensive. For each iteration of input parameters, we compare the simulated Ly$\alpha$ vs. H$\alpha$ surface brightness distribution to the observed surface brightness distribution to determine which set of parameters best fits the data. We do this by first creating a 2D histogram of the original Ly$\alpha$ versus H$\alpha$ surface brightness distribution. After each iteration in the calculation, a new surface brightness distribution is created and a 2D histogram of that distribution is made. To compare the observed and modeled surface brightnesses, we test the fit by minimizing the quantity
\begin{equation}
\sum_{i,j=0}^{n} \frac{(\textrm{Observation}[i, j] - \textrm{Model}[i, j])}{\textrm{Observation}[i, j]}
\end{equation}
where $n$ is the number of bins, and $(i,j)$ represents the bins of the 2D histogram. The output of \texttt{emcee} is the kernel width and region sizes that best match the observed surface brightness distributions. 

\subsection{Model Results}
In Figures~\ref{sbplot01} and~\ref{sbplot12}, we compare the observed Ly$\alpha$ versus H$\alpha$ surface brightnesses to our simulated values. For LARS01, the resulting scattering kernel is $\sigma=0.72^{+0.38}_{-0.27}$ kpc, where the errors give the 16th and 84th percentiles of the walkers from the MCMC algorithm. This kernel encodes the average scattering distance that each Ly$\alpha$ photon undergoes in any given galaxy. In contrast, the halo surrounding LARS12 has a Ly$\alpha$ photon scattering distance of $\sigma=1.03^{+1.74}_{-0.69}$ kpc. The best fit scattering model for all of the LARS galaxies is given in Table~\ref{table}. 

\begin{deluxetable*}{cccc}
\tablecolumns{4}
\tablewidth{0pt}
\tablecaption{LARS Scattering Properties}
\tablehead{\colhead{LARS ID} & \colhead{\shortstack{Halo Scattering \\Distance (kpc)}} & \colhead{\shortstack{Central Region \\Radius (kpc)}} & \colhead{\shortstack{Annular Region\\ Radius (kpc)}}}
\startdata
01 & $0.72^{+0.38}_{-0.27}$ & $0.36^{+1.35}_{-0.26}$ & $2.28^{+0.16}_{-0.26}$ \\[5pt]
02 & $0.36^{+0.04}_{-0.034}$ &  $1.52^{+0.59}_{-1.3}$ & $3.50^{+0.49}_{-1.30}$ \\[5pt]
03 & $0.35^{+0.05}_{-0.05}$ &  $3.08^{+0.31}_{-0.24}$ & $3.71^{+0.36}_{-0.24}$ \\[5pt]
04 & $0.70^{+0.54}_{-0.42}$ &  $2.17^{+0.14}_{-0.54}$ & $2.92^{+2.63}_{-0.54}$ \\[5pt]
05 & $0.40^{+0.08}_{-0.09}$ &  $1.59^{+0.46}_{-0.32}$ & $2.67^{+0.58}_{-0.32}$ \\[5pt]
06 & $0.41^{+0.06}_{-0.09}$ &  $1.94^{+0.73}_{-1.07}$ & $4.39^{+0.58}_{-1.07}$ \\[5pt]
07 & $0.37^{+0.14}_{-0.26}$ & $1.43^{+0.32}_{-0.33}$ & $2.22^{+0.47}_{-0.33}$ \\[5pt]
08 & $0.44^{+0.07}_{-0.08}$ & $2.71^{+1.09}_{-1.46}$ & $5.58^{+1.75}_{-1.46}$ \\[5pt]
09 & $0.56^{+0.11}_{-0.11}$ & $0.40^{+0.13}_{-0.68}$ & $7.18^{+1.55}_{-0.68}$ \\[5pt]
10 & $0.52^{+0.08}_{-0.06}$ & $1.35^{+0.23}_{-0.11}$ & $2.30^{+0.27}_{-0.11}$ \\[5pt]
11 & $1.56^{+0.03}_{-0.49}$ & $6.41^{+0.50}_{-1.51}$ & $6.92^{+0.21}_{-1.51}$ \\[5pt]
12 & $1.03^{+1.74}_{-0.69}$ &  $1.23^{+2.46}_{-1.06}$ & $4.77^{+1.04}_{-1.06}$ \\[5pt]
13 & $2.24^{+0.18}_{-0.29}$ & $3.29^{+0.52}_{-0.33}$ & $6.62^{+0.31}_{-0.33}$ \\[5pt]
14 & $1.68^{+2.83}_{-1.13}$ & $1.96^{+3.91}_{-1.68}$ & $7.58^{+1.65}_{-1.68}$
\enddata
\label{table}
\end{deluxetable*}

The right panels of Figures~\ref{sbplot01} and~\ref{sbplot12} show the models for the Ly$\alpha$ versus H$\alpha$ surface brightnesses. Note that the pixels in the brightest Ly$\alpha$ region lie directly on the Case B recombination line or just below it - this is because in the central regions, only a dust screen model is applied, meaning there is no way for the photons to scatter to \emph{brighter} Ly$\alpha$ values. The clumpy dust annulus corresponds to the region further down and to the left in the surface brightness distribution, where there is more reddening due to the clumpy nature of the dust, so H$\alpha$ is dampened compared to the Ly$\alpha$. The region of the figure with the overbright Ly$\alpha$ compared to H$\alpha$ is from the Ly$\alpha$ halo. Note that in our model, the borders between the three zones are sharp, whereas in the data, the transitions are more gradual. For example, both Figures~\ref{sbplot01} and~\ref{sbplot12} show a distinct corner in the simulated surface brightness distributions where the annular clumpy dust butts up agains the brighter Ly$\alpha$ pixels that are above the intrinsic Ly$\alpha$/H$\alpha$ line. The natural conclusion is that there is a continuum to the region edges as they bleed into each other, and/or that some radiative transfer physics is not being captured by the model. This difference is apparent in the models for both LARS01 and LARS12. For the model comparisons for all LARS galaxies, see the Appendix.

In Figures~\ref{res01} and~\ref{res12}, we show the fractional differences between the model and the observed Ly$\alpha$ and H$\alpha$ surface brightnesses. While the model itself is performed on individual pixels, in order to determine how well the model matches the observed distribution, we binned both the observed and modeled distributions, and compared how many pixels exist in each bin. We chose a large number of bins ($60\times60$) to be able to compare on a fairly fine scale, while still maintaining enough points in each bin. The differences were then calculated by finding the observed minus predicted differences in each bin. A positive difference (purple) indicates that there are more points in the observed bin than the model bin. Negative differences (yellow) occur where there are more points in the model bin than the observed. From the figures, it is apparent that our model matches the surface brightness distribution for LARS01 very well.  For LARS12, the differences are more stark, with the model distributing points that should be in the Ly$\alpha$ halo (shown in yellow in Figure~\ref{res12}) to the area of the surface brightness distribution that corresponds to the clumpy annulus (shown in purple in the same figure.) 

\begin{figure}[!t]
\centering
\scalebox{0.58}
{\includegraphics{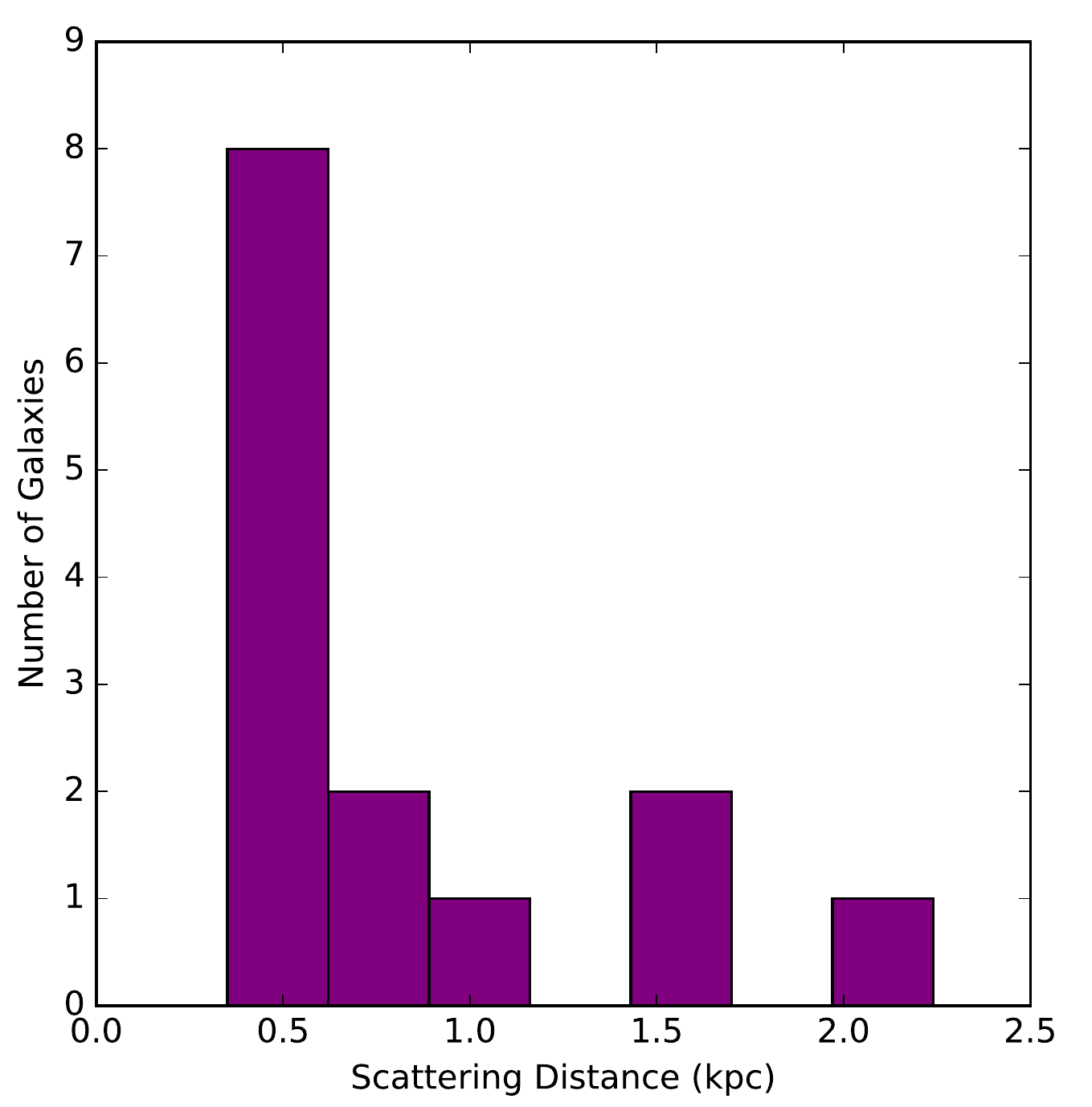}}
\caption{The distribution of the characteristic scattering distances resulting from the model.  A majority of the galaxies have a scattering distance of one kiloparsec or less.} 
\label{hist}
\end{figure}

The characteristic scattering distances for the LARS galaxies range from a fraction of a kiloparsec to several kiloparsecs, and are shown in Figure~\ref{hist}. We discuss the implications of these results in the following section.

\begin{figure}[!t]
\centering
\scalebox{0.58}
{\includegraphics{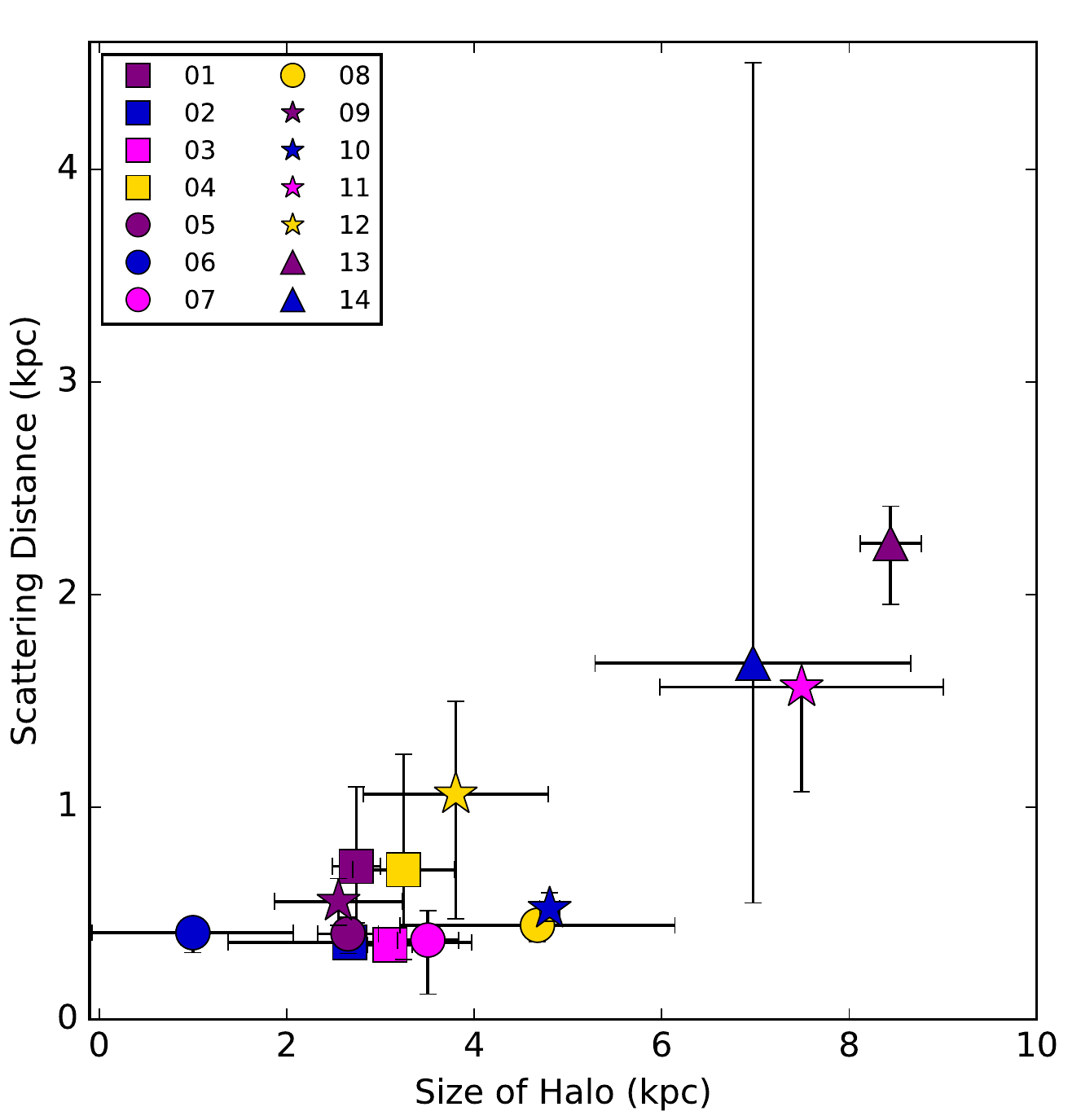}}
\caption{The characteristic scattering distances of the LARS galaxies versus the sizes of the Ly$\alpha$ halos. The Ly$\alpha$ halo size corresponds to the size of the outer galaxy regions as determined by the MCMC fitting algorithm. The Spearman's correlation coefficient is $\rho=0.62$ with a probability of a correlation arising by chance of $p=0.019$. The errors give the 16th and 84th percentiles of the distribution of the walkers from the MCMC algorithm that determined scattering distance.}
\label{scattvhalo}
\end{figure}

\begin{figure}[!t]
\centering
\scalebox{0.58}
{\includegraphics{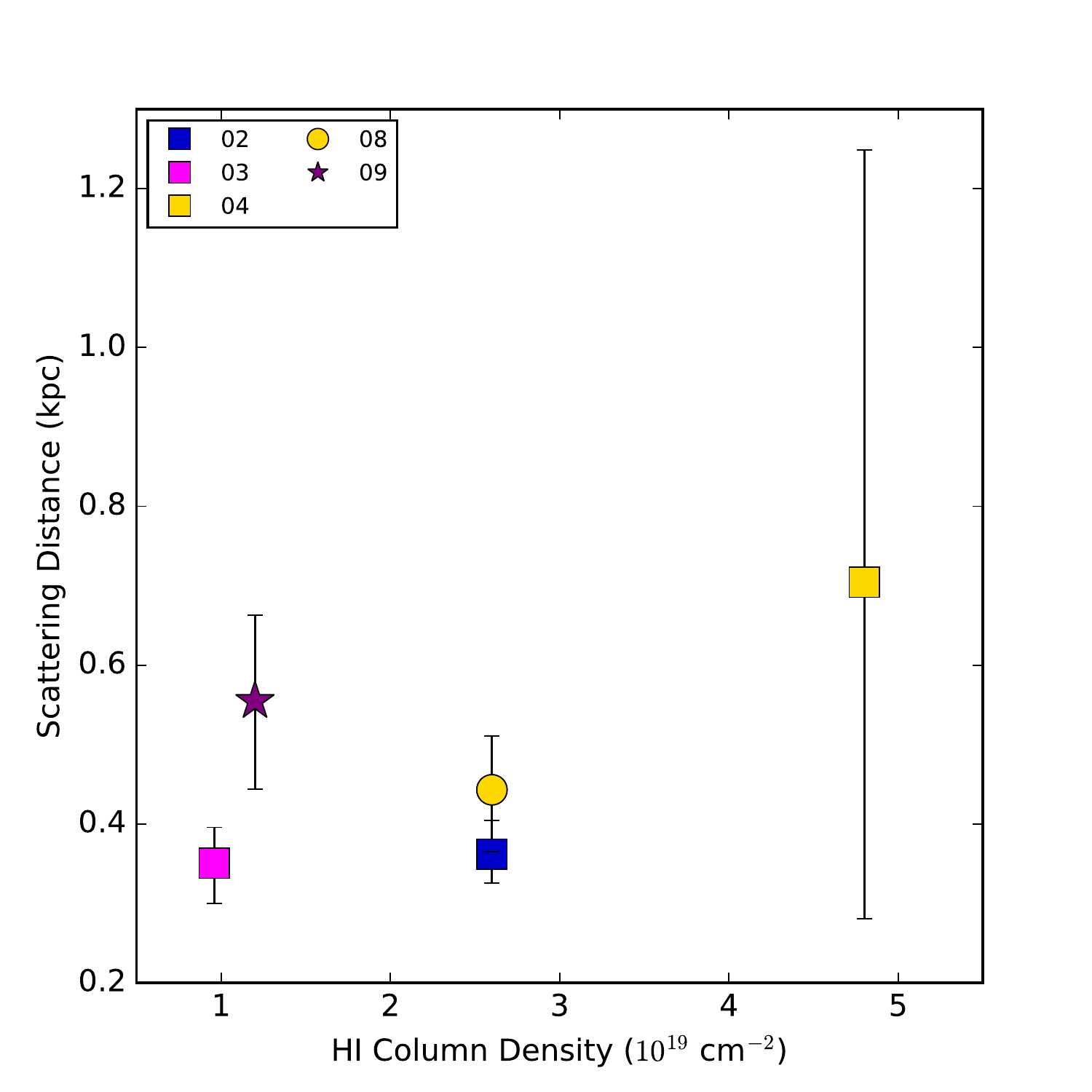}}
\caption{The characteristic scattering distance versus the HI column density for the LARS galaxies for which these data exist \citep{pardy2014}. The Spearman correlation coefficient is $\rho=0.67$ with a probability of a correlation arising by chance of $p=0.22$. The errors in the scattering distance represent the 16th and 84th percentiles of the distribution of the walkers from the MCMC algorithm that determined scattering distance.}
\label{scattvnHI}
\end{figure}

\begin{figure}[!h]
\centering
\scalebox{0.58}
{\includegraphics{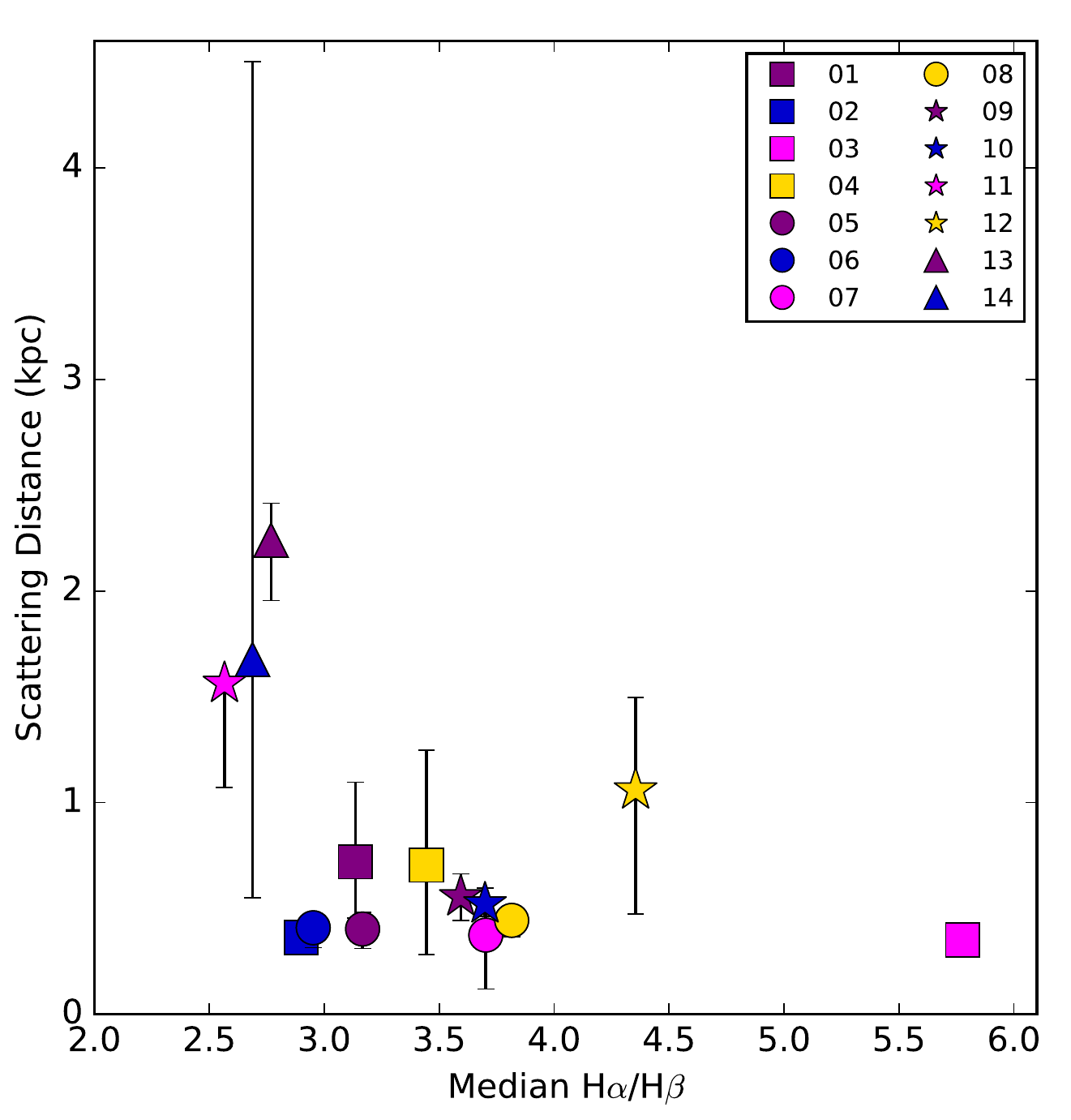}}
\caption{The characteristic scattering distance versus the global Balmer decrement of the LARS galaxies. The Spearman correlation coefficient is $\rho=-0.49$ with a probability of a correlation arising by chance of $p=0.078$. The errors in the scattering distance represent the 16th and 84th percentiles of the distribution of the walkers from the MCMC algorithm that determined scattering distance.}
\label{scattvred}
\end{figure}

\section{Discussion}
The Ly$\alpha$ scattering model presented here works very well for galaxies that exhibit the dust geometry for which it was designed - namely, those galaxies that have a central region with a non-clumpy dust distribution at the center, an annulus of clumpy dust, and a Ly$\alpha$ halo generated by scattering of photons. The model has difficulty reproducing the pixel distributions that indicate clumpy dust all the way to the center of the galaxies, which would result in little or no central region at all. Table~\ref{table} shows central region sizes that do not correspond well with what is observed. LARS12, for example, shows a small cleared central region a few pixels wide (0.4 kpc), while the model indicates a central region of $1.23^{+2.46}_{-1.06}$ kpc. As can be inferred from the error bars, this value is not very well-determined.

We note also that this model works best on the galaxies with symmetric reddening distributions. For example, the Balmer decrement distribution of LARS03 is skewed to the red and has a median value of H$\alpha$/H$\beta$ = 5.78. Our best-fit model for LARS03 gives a Ly$\alpha$ halo scattering distance of $\sigma$ = 0.351 kpc, but the resulting Ly$\alpha$ versus H$\alpha$ surface brightness distribution deviates rather significantly from the observed data. LARS12 is another red galaxy, with a median Balmer decrement value of H$\alpha$/H$\beta$ = 4.35. Our difficulty in matching its surface brightness distribution can be seen in in the right panel of Figure~\ref{sbplot12}, where Ly$\alpha$ is in places under-predicted relative to H$\alpha$. This may be due in part to the fact that we determine the reddening by drawing from a distribution, rather than using the actual H$\alpha$/H$\beta$ values for each pixel. If the galaxy is skewed very red, the Ly$\alpha$ pixels from the more central regions are pulled downwards in the surface brightness distribution. However, for galaxies with reddening distributions that are more Gaussian and with lower median H$\alpha$/H$\beta$ values, our model does quite well predicting the shape of the observed surface brightness distributions. 

Another issue is that some galaxies show what appears to be two populations in the simulated surface brightness distribution (e.g., see Figure~\ref{sbplot12}). This is an artifact of separating the galaxies into three discrete regions.  In some galaxies, such as LARS01, the resulting distribution of pixels centers in the same place as the observations (see Figure~\ref{sbplot01}) and the result is a good fit. But for other galaxies, the pixels do not get distributed as evenly, leaving some evidence of the different dust models imprinted on the Ly$\alpha$ vs.\ H$\alpha$ surface brightness. For the same reason, the pixel density is occasionally too high along the intrinsic Ly$\alpha$ vs.\ H$\alpha$ line, as the best-fitting dust model does not distributed the pixels evenly.

The part of the model that best reproduces the surface brightness distributions is that associated with the Ly$\alpha$ halo. The Gaussian smoothing kernel does well at mimicking how the Ly$\alpha$ photons scatter in the halo. This result can be seen in Figures~\ref{sbplot01} and~\ref{sbplot12} as well as the similar figures for the other LARS galaxies in the Appendix. Therefore, while the model may have difficulty reproducing some dust geometries, the results of the characteristic scattering distance of Ly$\alpha$ photons in the halo are robust. The one exception to this is LARS06 (see Figure~\ref{sbplot06}), which, as a net Ly$\alpha$ absorber, has very little Ly$\alpha$ halo.

As shown in Figure~\ref{scattvhalo}, the scattering distances scale linearly with the sizes of the Ly$\alpha$ halo for each galaxy, with a \cite{spearman1904} correlation coefficient of $\rho=0.62$ and a probability of a correlation arising by chance of 1.9\%. Here, the distance of the outer edge of the annulus was determined by the fitting algorithm, and since the radius of the overall aperture was fixed, the size of the outer halo region follows directly. The relationship between the characteristic scattering distance and the size of the halo is intuitive, as a galaxy with a longer characteristic Ly$\alpha$ photon scattering distance will experience more Ly$\alpha$ escape, resulting in a larger halo.

The amount of Ly$\alpha$ scattering should be strongly related to the density of the neutral scattering medium in a galaxy. We explored this relationship using data for the LARS galaxies from the Very Large Array (VLA) \citep{pardy2014}. At present, data are published for only five galaxies in the sample: LARS02, LARS03, LARS04, LARS08, and LARS09. We note that there is a possible positive correlation between the scattering distance and the HI column density as shown in Figure~\ref{scattvnHI}. However, this correlation is pinned only by LARS04, for which the scattering distance has large error bars. More extensive HI observations will be able to determine whether or not this correlation actually exists. 

The fact that the size of the Ly$\alpha$ halo correlates with the scattering distance means that the halo is produced mainly by scattering of HI within the galaxy.  Therefore, no other phenomena in addition to scattering is required to explain the halos. This conclusion was also reached work based on spectroscopic data of $2<z<4$ Ly$\alpha$ emitters in the VIMOS Ultra Deep Survey \citep[VUDS; ][]{guaita2017}. Relating the characteristic scattering distance with other physical parameters that are linked to the HI content (e.g., galaxy mass), would indicate that the extent of the Ly$\alpha$ emission depends on galaxy physical properties. This would be further proof that Ly$\alpha$ photons are generated inside the galaxy and are simply scattered by the neutral hydrogen.

Figure~\ref{scattvred} shows that there is a tentative inverse correlation between the Ly$\alpha$ scattering distance and the average reddening values of the LARS galaxies. The Spearman correlation coefficient for this relationship is $\rho=-0.49$ with a probability of a correlation arising by chance of 7.8\%. This behavior suggests that Ly$\alpha$ photons do not scatter as far in galaxies with higher dust content. This is consistent with studies that show that the Ly$\alpha$ escape fraction decreases with dust content \citep{hayes2013, atek2014}.

For our scattering model, we assumed certain dust geometries based on the pixel-by-pixel distribution for each galaxy. Since LARS galaxies are bright, with vigorous star formation, one might wonder whether our results can be extrapolated to the general population of Ly$\alpha$ emitting galaxies. For example, the dust distribution in LBGs is best represented with a clumpy shell configuration \citep[e.g.,][]{vijh2003}. That this same geometry of a clumpy dust annulus is found in a sample of LBG analogs at low redshift indicates that the dust behaviors evidenced by the LARS galaxies can be useful for determining higher redshift galaxy properties (e.g., Ly$\alpha$ escape fraction, UV output, etc.)

The results of this scattering model indicate that some specific geometries would be useful to study in full 3D radiative transfer models, such as the 3D Monte Carlo code MCLy$\alpha$ of \cite{verhamme2006,verhamme2008}. Simulations that use spherically symmetric shells of neutral gas that scatter photons to model Ly$\alpha$ radiative transfer have been very effective at reproducing observed Ly$\alpha$ spectra (in some cases even better than multiphase geometries \citep{gronke2015}). However, \cite{gronke2016} showed that using clumpy geometries of neutral gas with a large number of clumps along the line of sight was the best way to model the observed Ly$\alpha$ spectra. The observation-based model we have presented marries more complicated gas and dust geometries with the shell model, and will provide new avenues to pursue in 3D radiative transfer models.
\\
\section{Summary and Conclusions}
We have presented a study of the dust geometry of the 14 LARS galaxies, and how it affects Ly$\alpha$ escape from galaxies. Informed by this dust characterization, we have also developed a modeling technique to characterize how far, on average, Ly$\alpha$ photons scatter into the outer halo before escaping.  

The LARS galaxies provide a unique dataset with which to use pixel-by-pixel photometry to probe the galaxies' dust geometry down scales of $\sim40$ parsecs. This has allowed us to move from looking only at the global dust properties of the galaxies to understanding which galaxy regions are affecting the global properties the most.

Using LARS01 and LARS12 as examples of galaxies with quite different dust distributions, we probed the properties of each galaxy, including which regions correspond to a clumpy ISM, and which regions contain the most scattering. We found that for LARS01, whose global Ly$\alpha$/H$\alpha$ value places the galaxy in a region not well described by a clumpy medium, there are three distinct regions that determine the way in which Ly$\alpha$ escapes. At the center of the galaxy, Ly$\alpha$ escapes at or slightly below the intrinsic Ly$\alpha$/H$\alpha$ ratio of 8.7, implying little dust or neutral hydrogen along the line of sight. A clumpy medium surrounds the center in an annulus of pixels that are well-described with a large number of small dust clouds.  Finally, there exists a Ly$\alpha$ halo from which the overluminous (Ly$\alpha$/H$\alpha > 8.7$) Ly$\alpha$ photons scatter out to great distances before leaving the galaxy.

In contrast, the sizes of the regions of LARS12 are significantly different than for LARS01. For this galaxy, the clumpy ISM extends almost to the very center of the galaxy with only a small central region where the dust screen model is applicable. Otherwise, LARS12 has the typical Ly$\alpha$ halo seen in many of the LARS galaxies.  

Based upon these findings, we developed a model to quantify what effect these different dust regions have on Ly$\alpha$ escape. Our three-parameter model uses an MCMC algorithm to find the average scattering distance that Ly$\alpha$ undergoes in the halo of the galaxies, as well as the sizes of the areas with smooth and clumpy dust geometries. The resulting halo scattering distances correlate with the Ly$\alpha$ halo sizes and are slightly inversely correlated with the median Balmer decrement of the galaxies. 

Observations of the HI in the remaining LARS galaxies (P.I. Cannon; ID VLA/17A-240) will reveal whether or not there is a correlation between the HI column density and the modeled scattering distances.  Additionally, with an angular resolution of $\sim6''$, the VLA observations will be capable of determining the HI morphologies, which will allow for a more thorough exploration of how the Ly$\alpha$ scattering is affected by the presence of neutral hydrogen. Other future work will involve expanding our modeling algorithm to the full extended LARS (eLARS; P.I. \"Ostlin; ID 13483) sample, which comprises a further 28 nearby Ly$\alpha$ emitters with more disk-like morphologies and lower $H\alpha$ equivalent width cuts. The applicability of this scattering model to a broader sample of galaxies will determine how robust these initial results are.
\\
\\
\indent We thank the referee for their very useful comment on this manuscript. J\@.S\@.B\@.~acknowledges support from the Swedish Research Council (VR) through the National Science Foundation (NSF) Graduate Research Opportunities Worldwide Fellowship as well as NSF grant AST-1615526. M\@.H\@.~and G\@.\"O\@.~acknowledge the support of the VR and the Swedish National Space Board (SNSB).  M\@.H\@.~is a Fellow of the Knut and Alice Wallenberg Foundation. D\@.K\@.~is supported by the Centre National d'Etudes Spatiales (CNES)/Centre National de la Recherche Scientifique (CNRS) convention 131425. J\@.M\@.M\@.H\@.~is funded by the Spanish MINECO grant ESP2015-65712-C5-1-R. The Institute for Gravitation and the Cosmos is supported by the Eberly College of Science and the Office of the Senior Vice President for Research at the Pennsylvania State University. This work makes use of NASA's Astrophysics Data System and the \texttt{AstroPy} Python package \citep{astropy2013, muna2016}.

\facility{\emph{Facilities: HST} (ACS, WFC3)}

\bibliography{lars}
\clearpage

\appendix

\section{Ly$\alpha$ and H$\alpha$ Emission Maps and Scattering Models}
We present here the Ly$\alpha$ and H$\alpha$ emission maps as well as the corresponding Ly$\alpha$/H$\alpha$ and H$\alpha$/H$\beta$ ratios for the LARS galaxies, with their apertures applied. Additionally, we include the H$\alpha$ vs. Ly$\alpha$ surface brightness distributions and trace the pixels corresponding to clumpy dust or overbright Ly$\alpha$ to their origins within each galaxy. Finally, we show the observed Ly$\alpha$ versus H$\alpha$ surface brightness distributions and compare them to the empirical Ly$\alpha$ scattering model results.

\begin{figure*}
\centering
\scalebox{0.5}
{\includegraphics{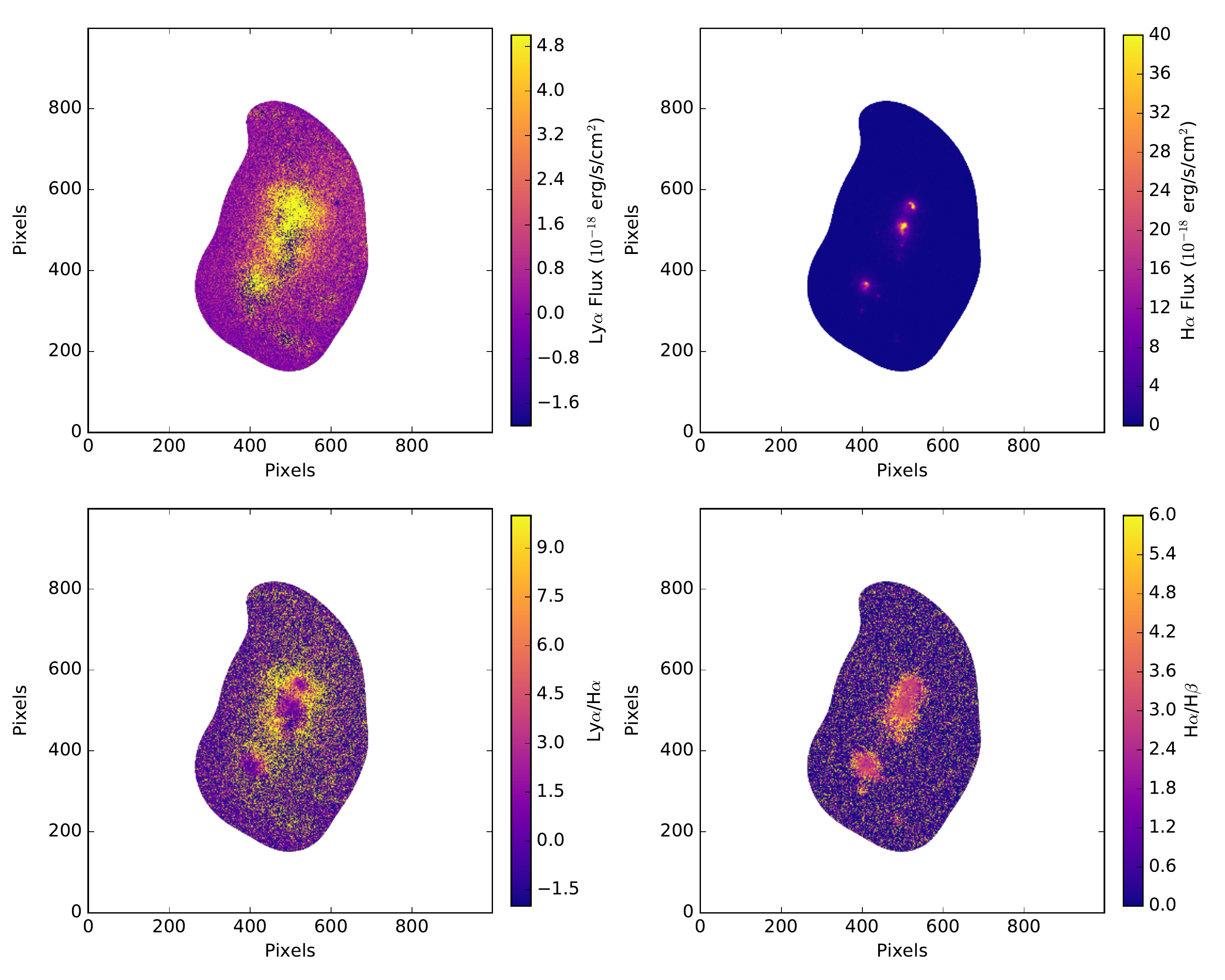}}
\caption{Images of LARS02. The calculated aperture has been applied. \emph{Top left}: The Ly$\alpha$ emission map. \emph{Top right}: The H$\alpha$ emission map. \emph{Bottom left}: The Ly$\alpha$/H$\alpha$ ratio, with an average uncertainty of 0.01. \emph{Bottom right}: The H$\alpha$/H$\beta$ ratio, with an average uncertainty of 0.001.}
\label{images02}
\end{figure*}
\begin{figure*}
\centering
\scalebox{0.32}
{\includegraphics{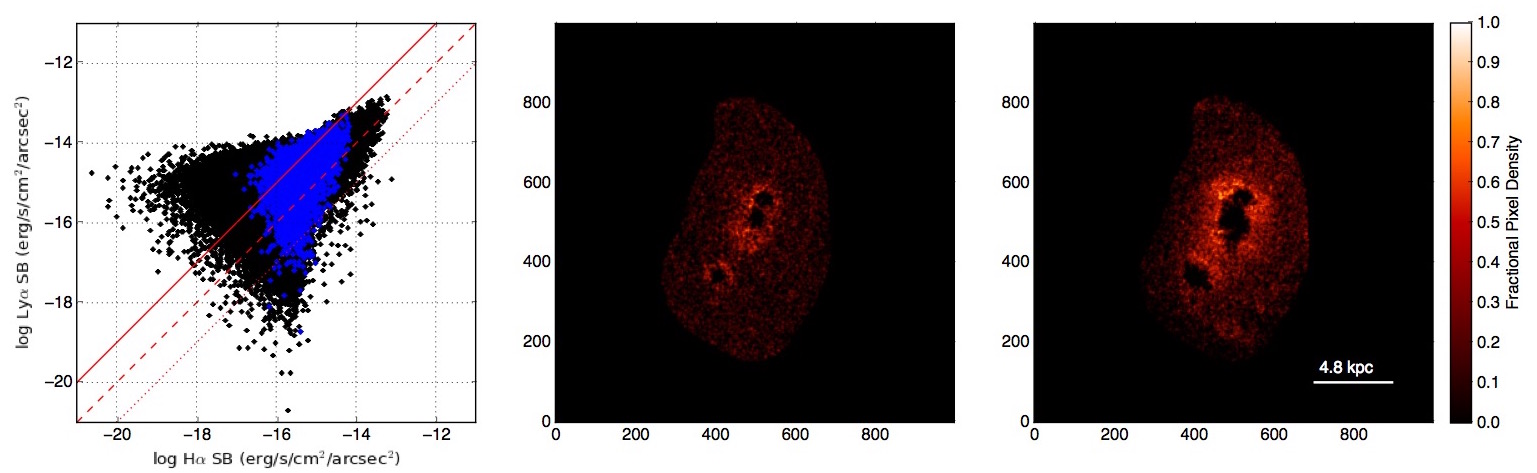}}
\caption{Same as Figure~\ref{sbplotcardscatt01} but for LARS02.}
\label{sbplotcardscatt02}
\end{figure*}
\begin{figure*}
\centering
\scalebox{0.7}
{\includegraphics{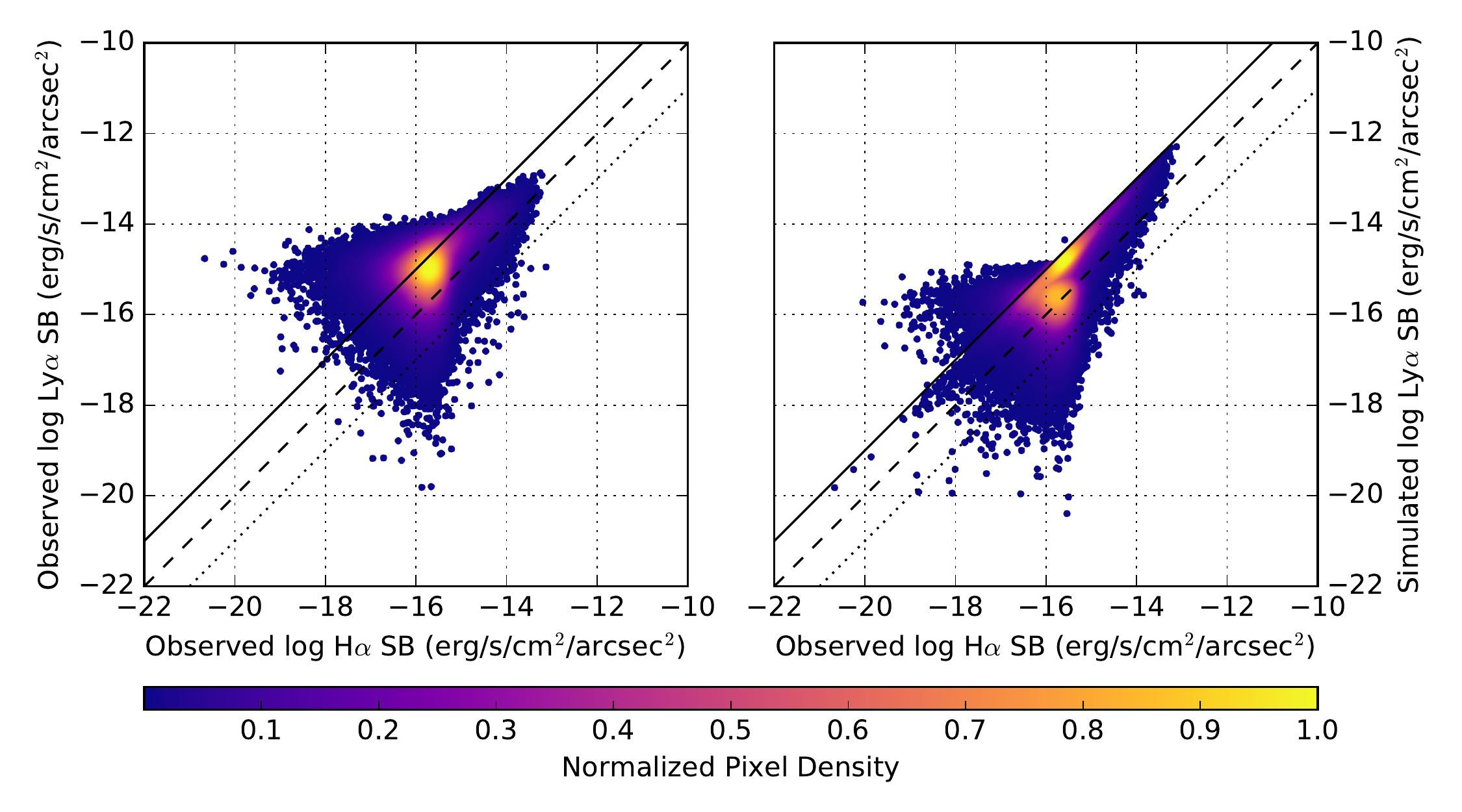}}
\caption{Same as Figure~\ref{sbplot01}, but for LARS02.}
\label{sbplot02}
\end{figure*}

\begin{figure*}
\centering
\scalebox{0.5}
{\includegraphics{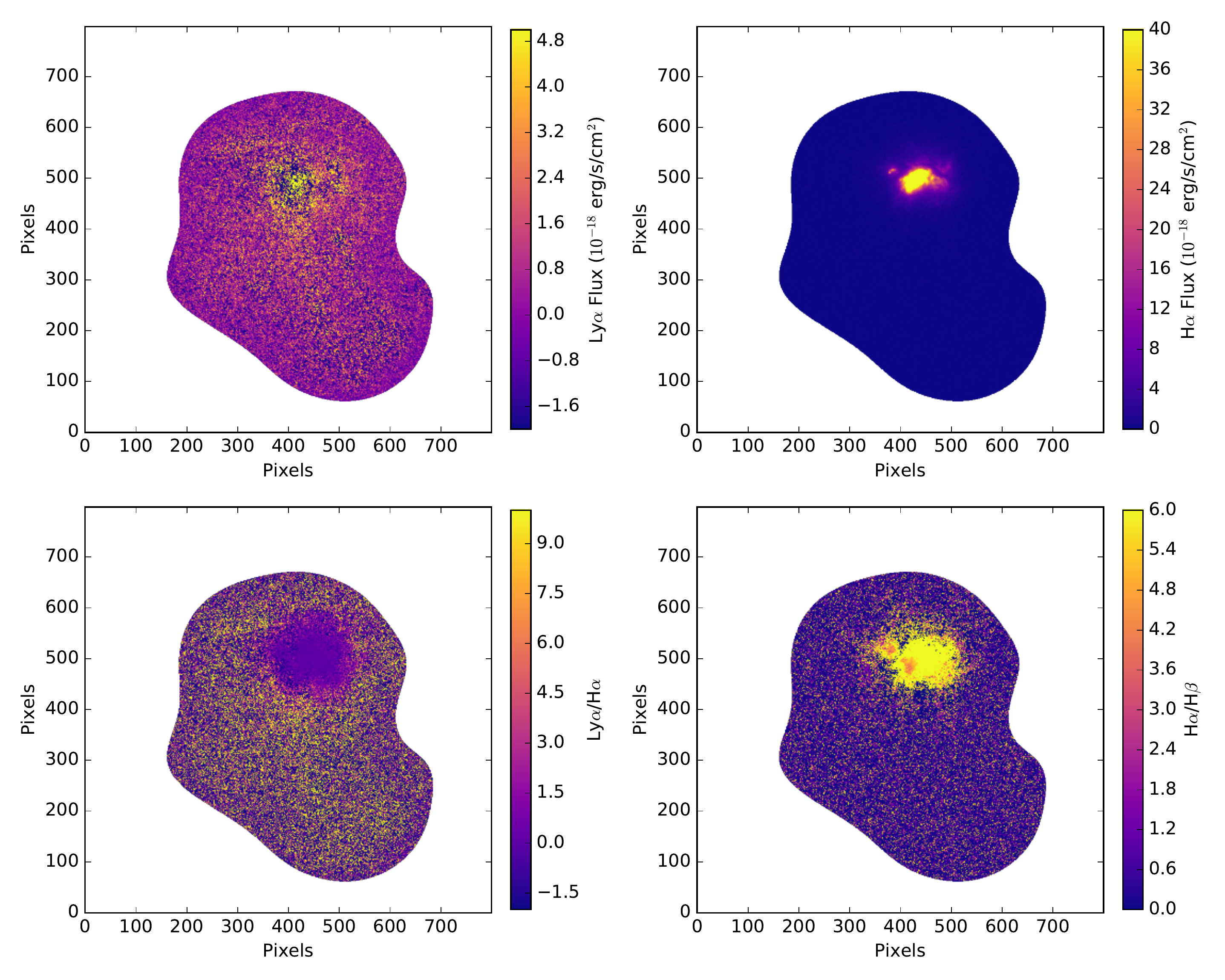}}
\caption{Images of LARS03. The calculated aperture has been applied. \emph{Top left}: The Ly$\alpha$ emission map. \emph{Top right}: The H$\alpha$ emission map. \emph{Bottom left}: The Ly$\alpha$/H$\alpha$ ratio, with an average uncertainty of 0.003. \emph{Bottom right}: The H$\alpha$/H$\beta$ ratio, with an average uncertainty of 0.009.}
\label{images03}
\end{figure*}
\begin{figure*}
\centering
\scalebox{0.32}
{\includegraphics{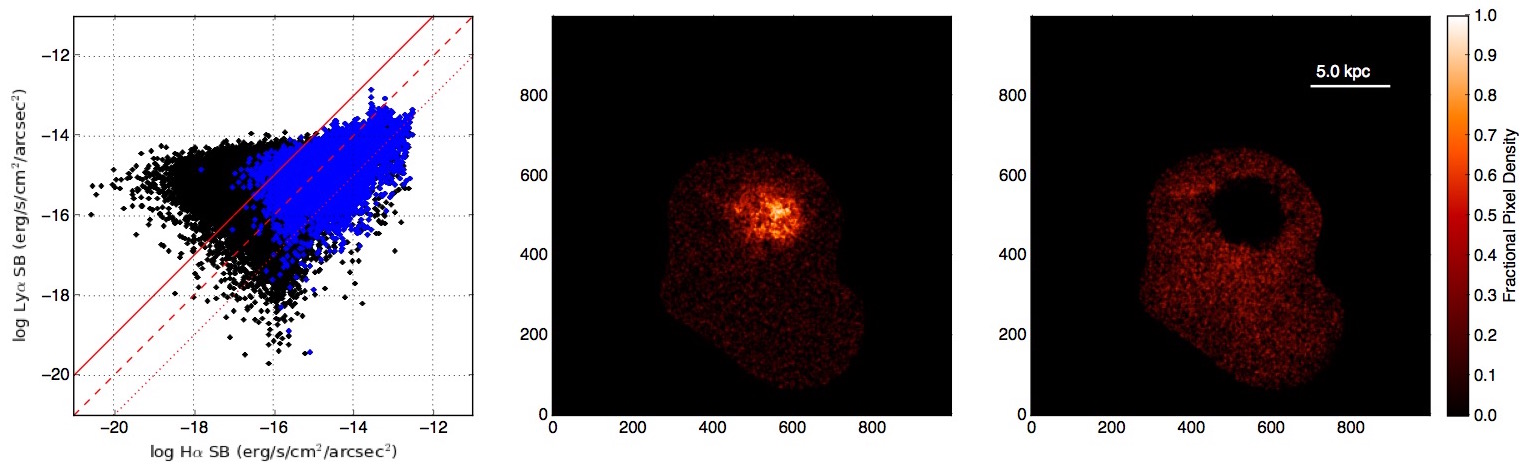}}
\caption{Same as Figure~\ref{sbplotcardscatt01} but for LARS03.}
\label{sbplotcardscatt03}
\end{figure*}
\begin{figure*}
\centering
\scalebox{0.7}
{\includegraphics{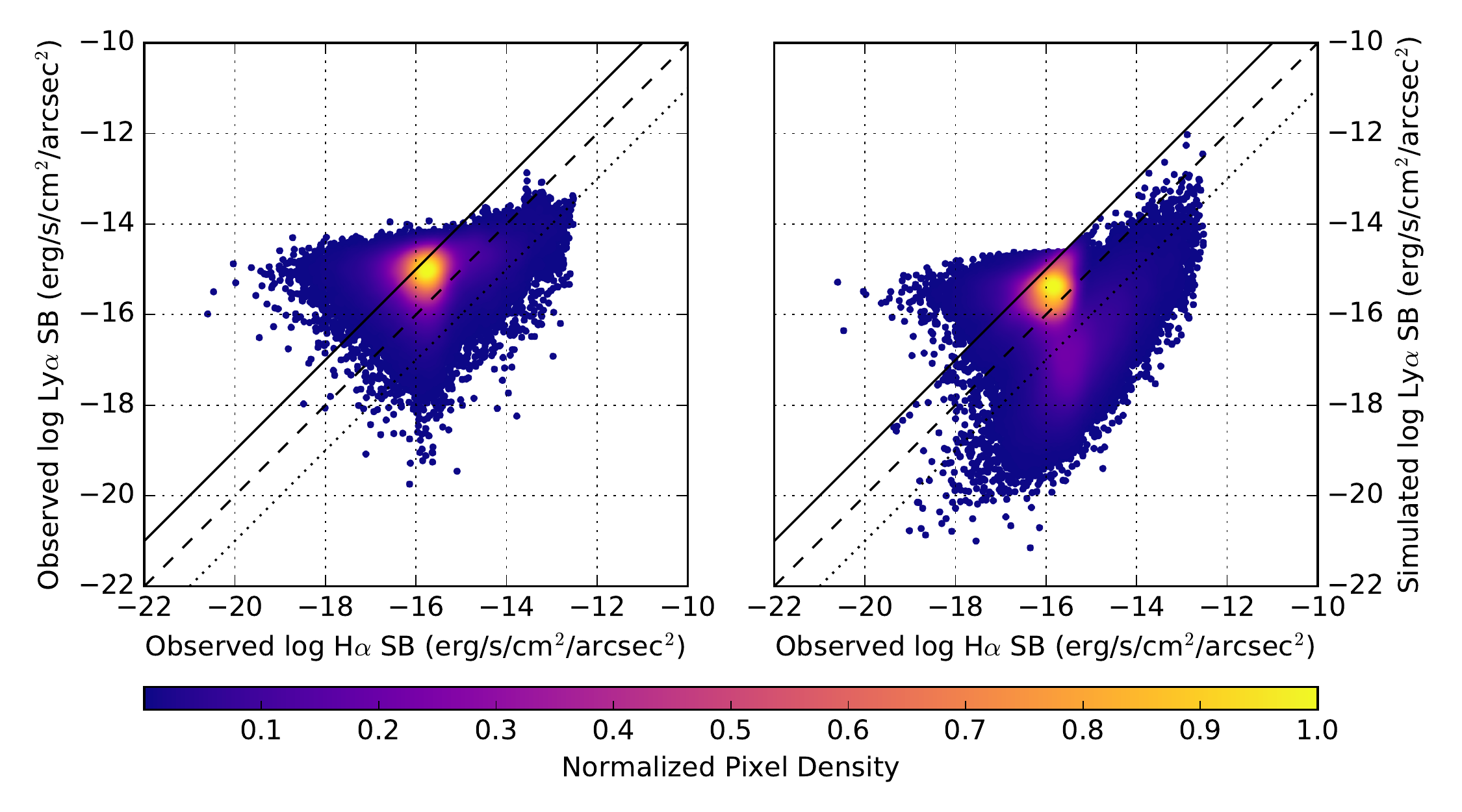}}
\caption{Same as Figure~\ref{sbplot01}, but for LARS03.}
\label{sbplot03}
\end{figure*}

\begin{figure*}
\centering
\scalebox{0.5}
{\includegraphics{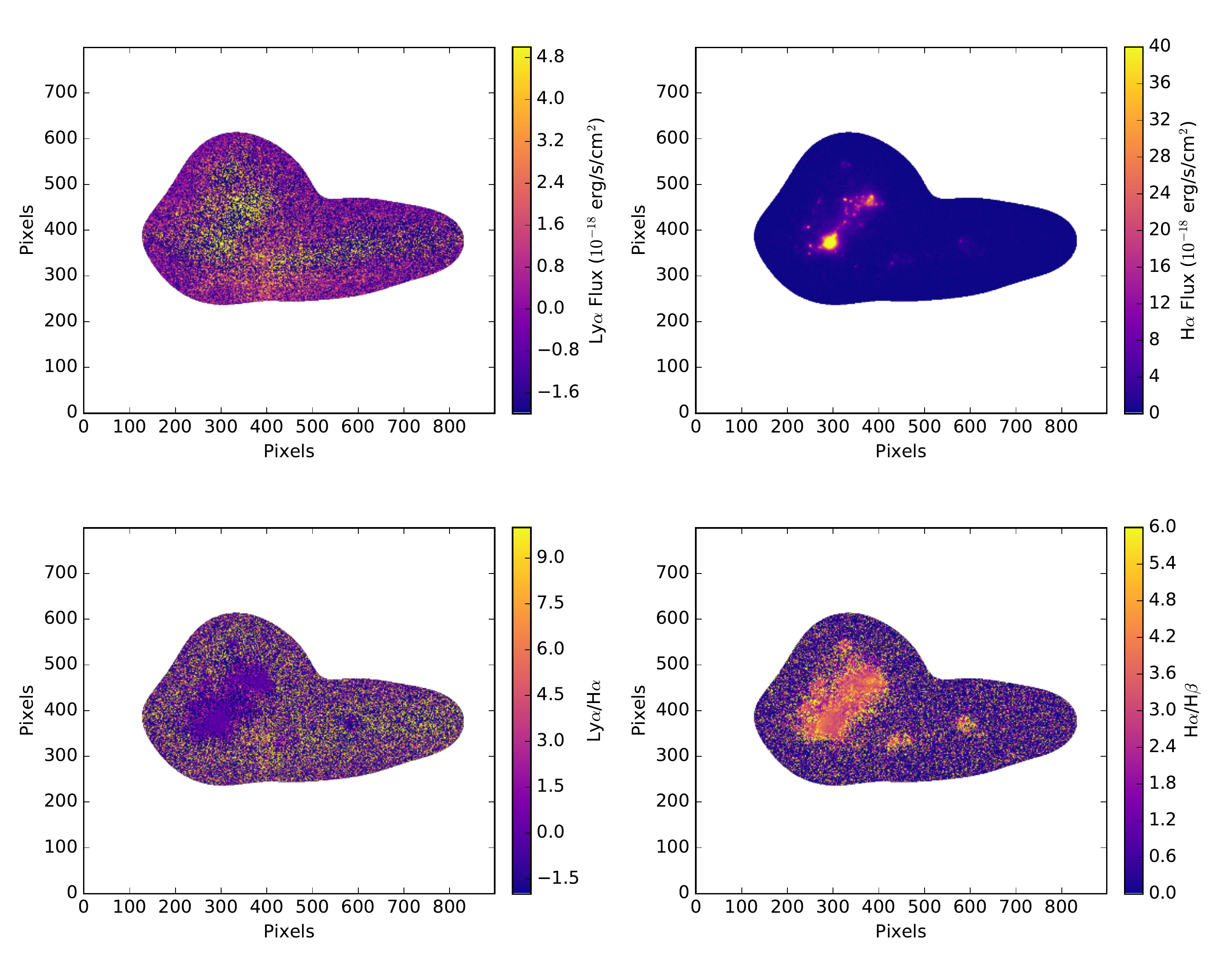}}
\caption{Images of LARS04. The calculated aperture has been applied. \emph{Top left}: The Ly$\alpha$ emission map. \emph{Top right}: The H$\alpha$ emission map. \emph{Bottom left}: The Ly$\alpha$/H$\alpha$ ratio, with an average uncertainty of 0.007. \emph{Bottom right}: The H$\alpha$/H$\beta$ ratio, 0.002.}
\label{images04}
\end{figure*}
\begin{figure*}
\centering
\scalebox{0.32}
{\includegraphics{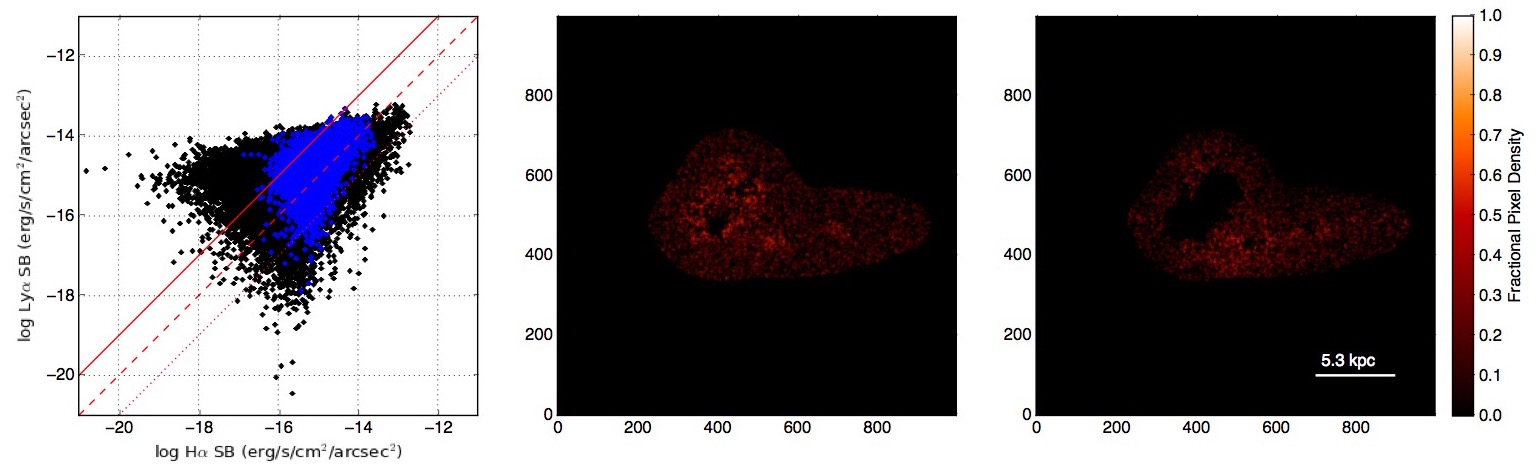}}
\caption{Same as Figure~\ref{sbplotcardscatt01} but for LARS04.}
\label{sbplotcardscatt04}
\end{figure*}
\begin{figure*}
\centering
\scalebox{0.7}
{\includegraphics{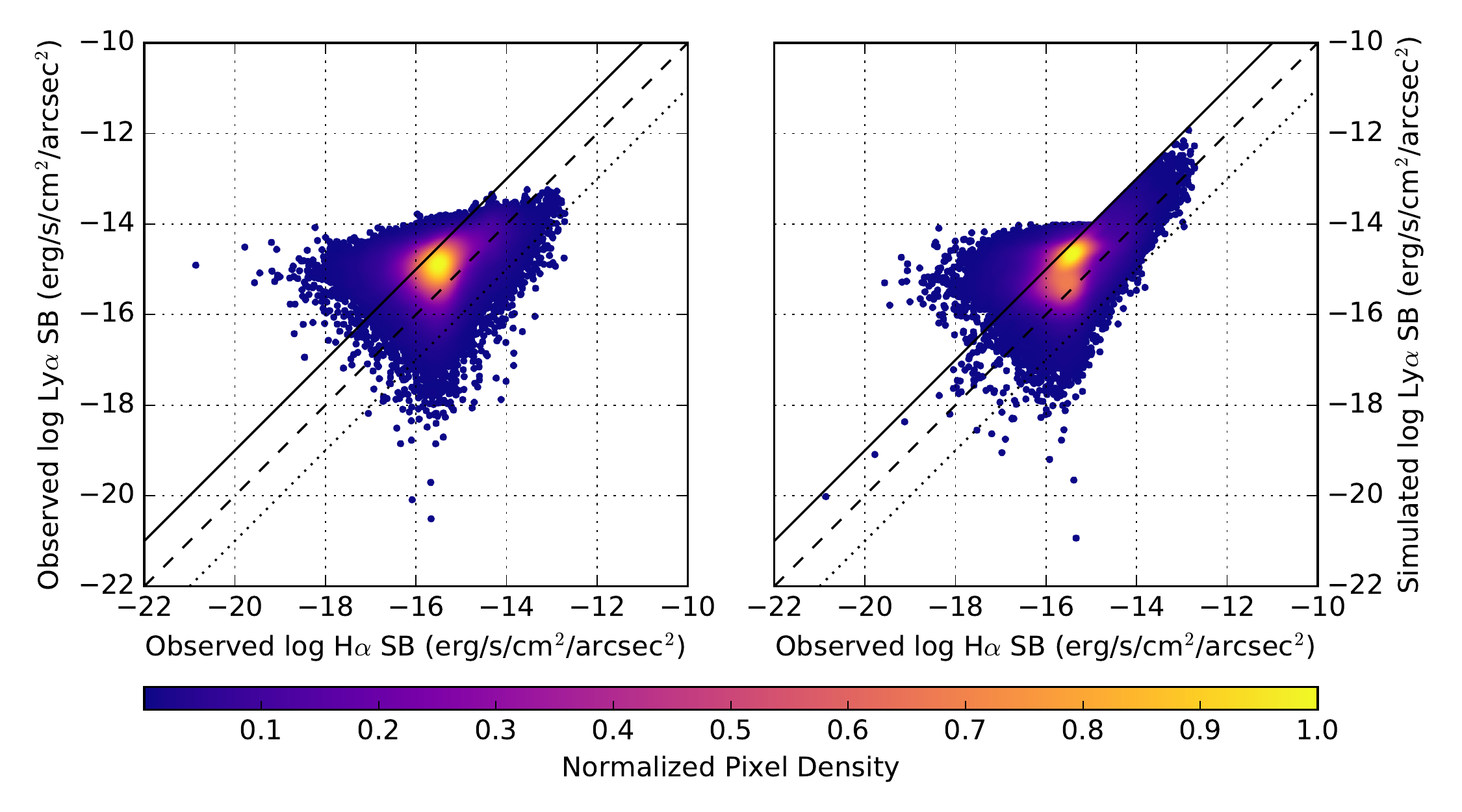}}
\caption{Same as Figure~\ref{sbplot01}, but for LARS04.}
\label{sbplot04}
\end{figure*}

\begin{figure*}
\centering
\scalebox{0.5}
{\includegraphics{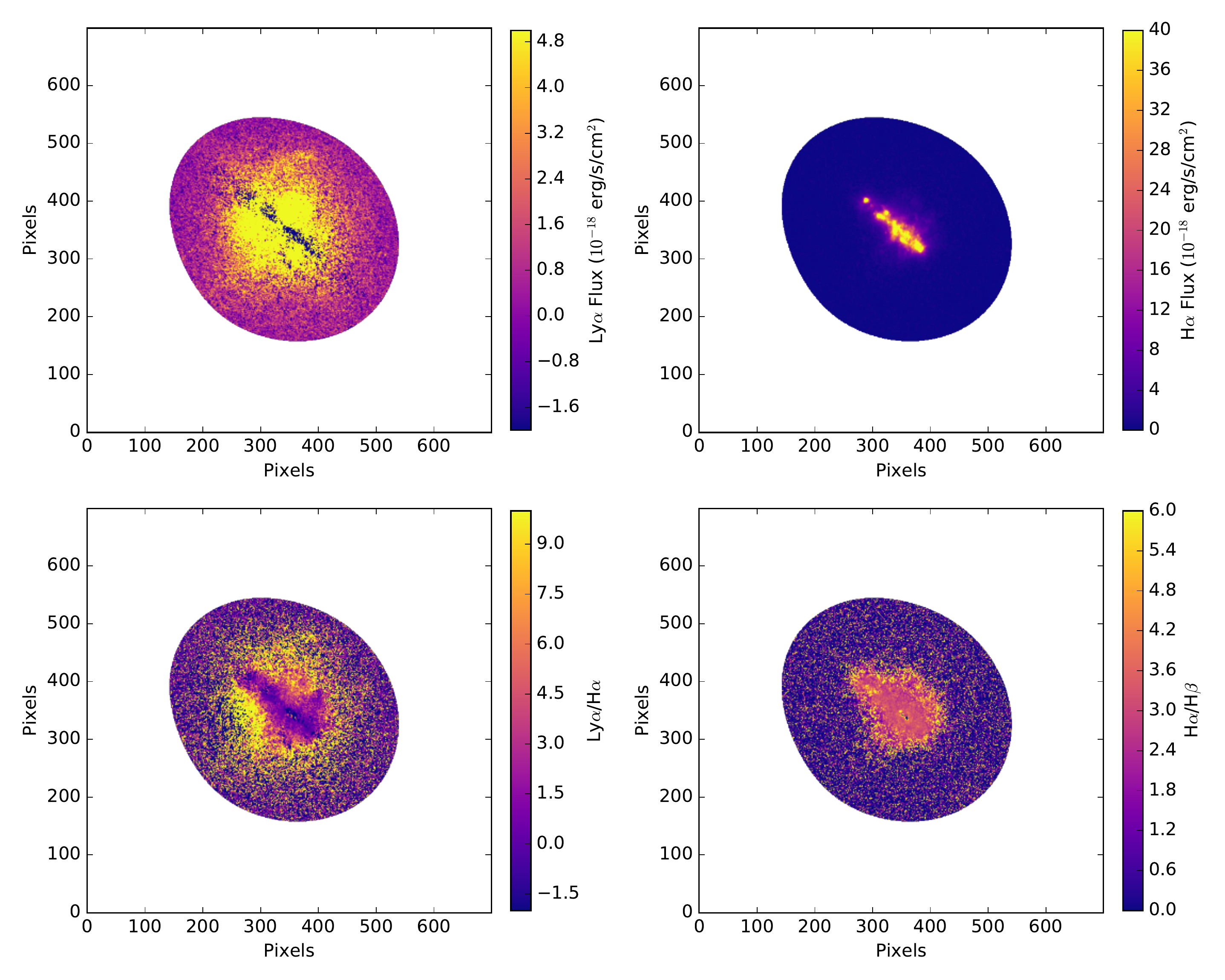}}
\caption{Images of LARS05. The calculated aperture has been applied. \emph{Top left}: The Ly$\alpha$ emission map. \emph{Top right}: The H$\alpha$ emission map. \emph{Bottom left}: The Ly$\alpha$/H$\alpha$ ratio, with an average uncertainty of 0.01. \emph{Bottom right}: The H$\alpha$/H$\beta$ ratio, with an average uncertainty of 0.002.}
\label{images05}
\end{figure*}
\begin{figure*}
\centering
\scalebox{0.32}
{\includegraphics{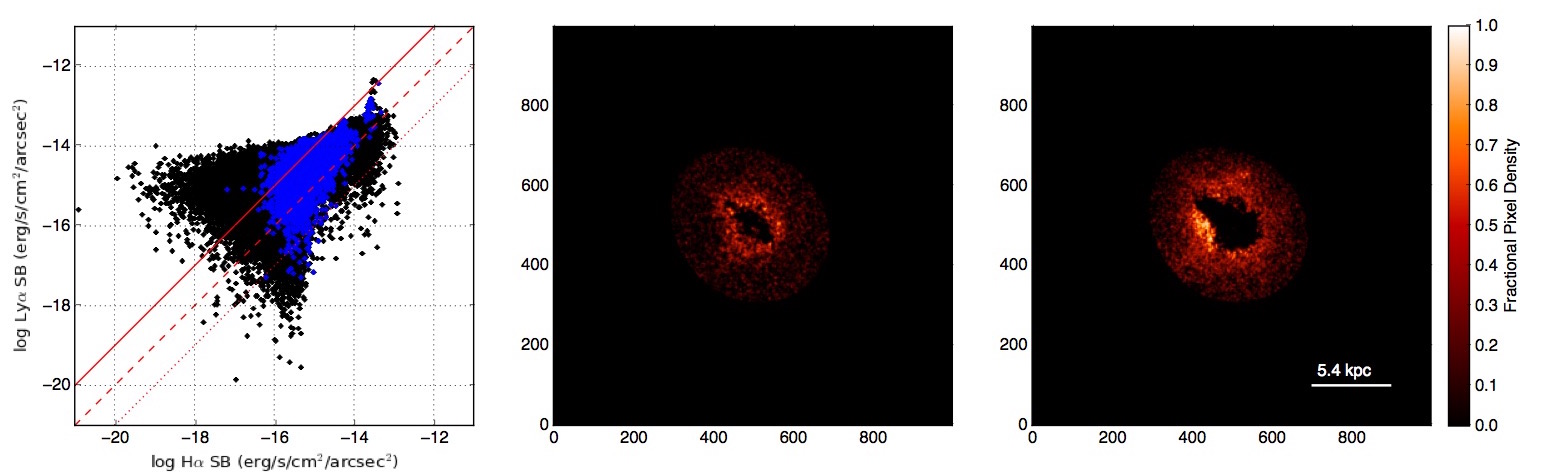}}
\caption{Same as Figure~\ref{sbplotcardscatt01} but for LARS05.}
\label{sbplotcardscatt05}
\end{figure*}
\begin{figure*}
\centering
\scalebox{0.7}
{\includegraphics{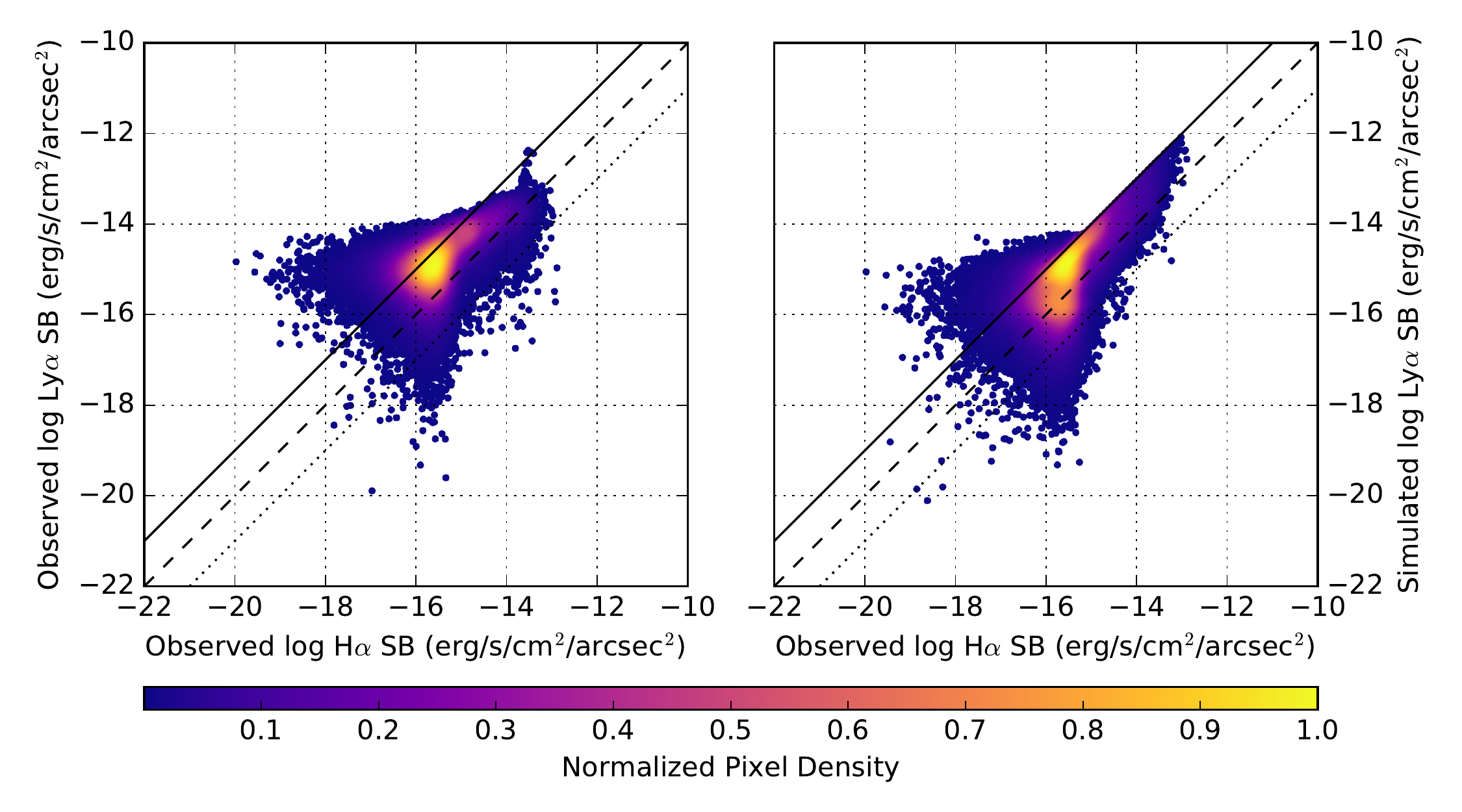}}
\caption{Same as Figure~\ref{sbplot01}, but for LARS05.}
\label{sbplot05}
\end{figure*}

\begin{figure*}
\centering
\scalebox{0.5}
{\includegraphics{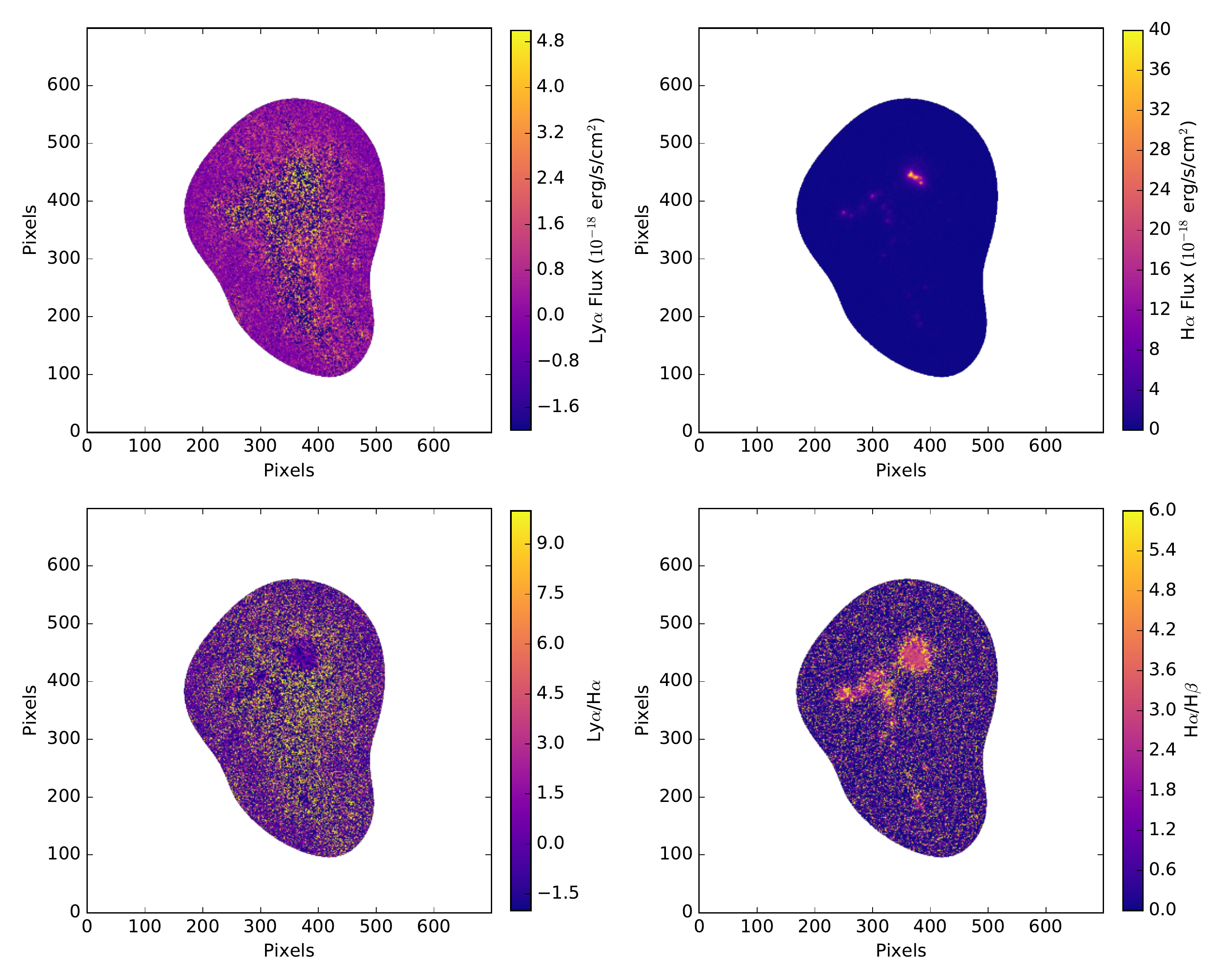}}
\caption{Images of LARS06. The calculated aperture has been applied. \emph{Top left}: The Ly$\alpha$ emission map. \emph{Top right}: The H$\alpha$ emission map. \emph{Bottom left}: The Ly$\alpha$/H$\alpha$ ratio, with an average uncertainty of 0.02. \emph{Bottom right}: The H$\alpha$/H$\beta$ ratio, with an average uncertainty of 0.002.}
\label{images06}
\end{figure*}
\begin{figure*}
\centering
\scalebox{0.32}
{\includegraphics{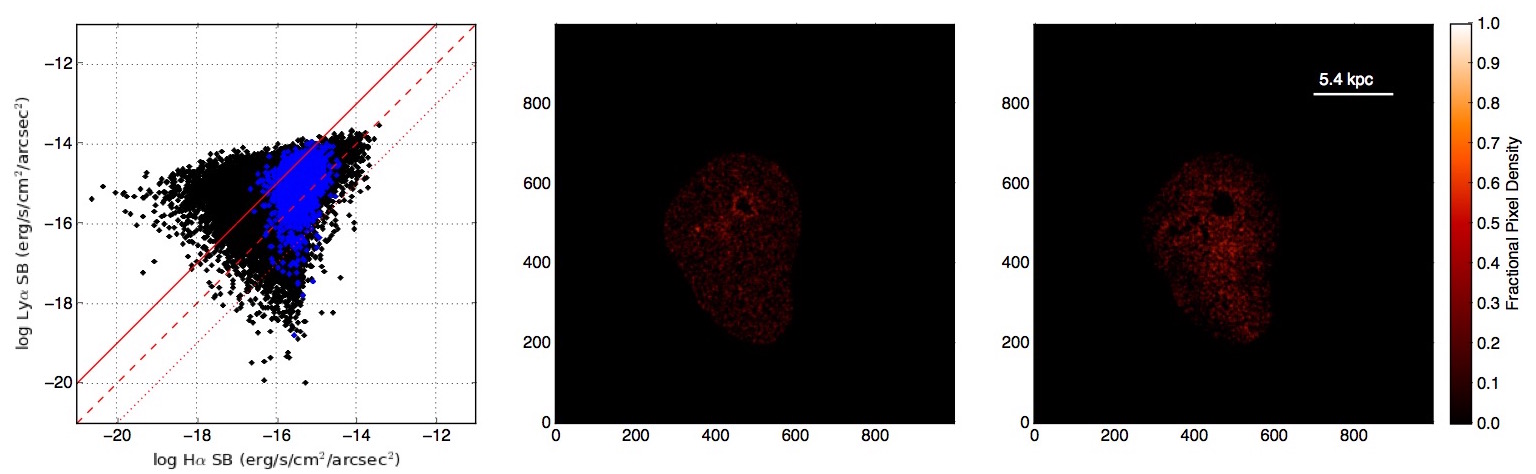}}
\caption{Same as Figure~\ref{sbplotcardscatt01} but for LARS06.}
\label{sbplotcardscatt06}
\end{figure*}
\clearpage
\begin{figure*}
\centering
\scalebox{0.7}
{\includegraphics{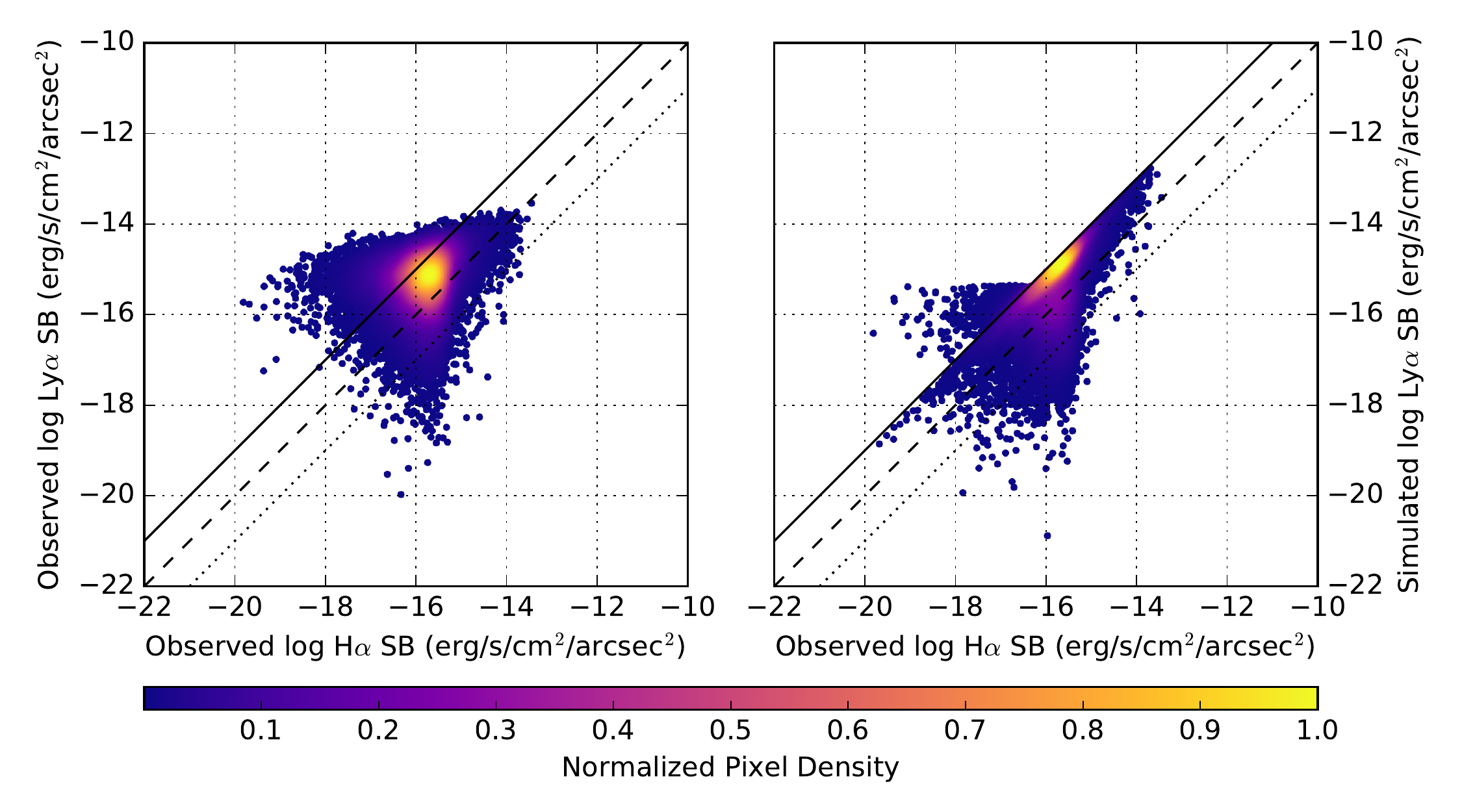}}
\caption{Same as Figure~\ref{sbplot01}, but for LARS06.}
\label{sbplot06}
\end{figure*}

\begin{figure*}
\centering
\scalebox{0.5}
{\includegraphics{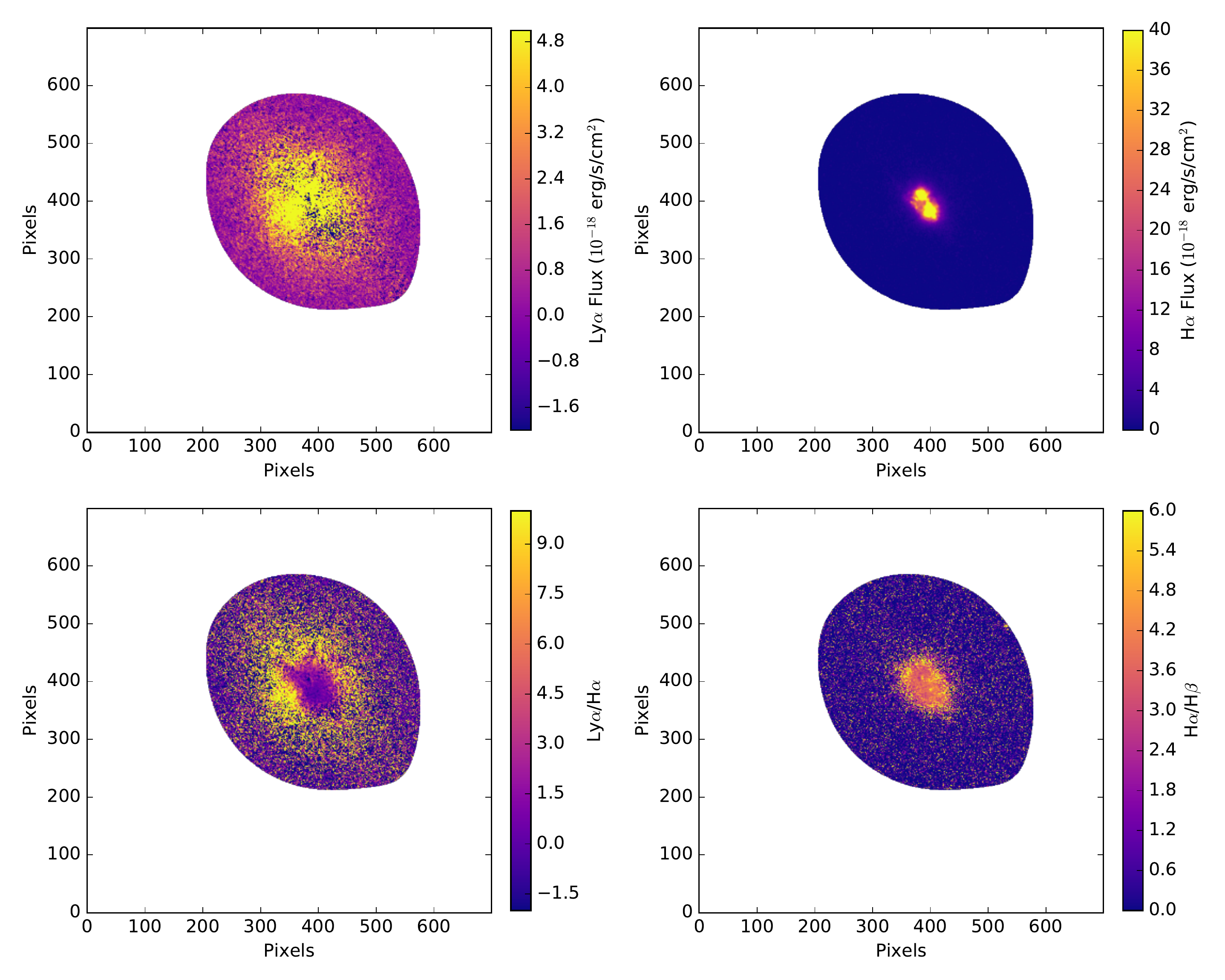}}
\caption{Images of LARS07. The calculated aperture has been applied. \emph{Top left}: The Ly$\alpha$ emission map. \emph{Top right}: The H$\alpha$ emission map. \emph{Bottom left}: The Ly$\alpha$/H$\alpha$ ratio, with an average uncertainty of 0.02. \emph{Bottom right}: The H$\alpha$/H$\beta$ ratio, with an average uncertainty of 0.003.}
\label{images07}
\end{figure*}
\begin{figure*}
\centering
\scalebox{0.32}
{\includegraphics{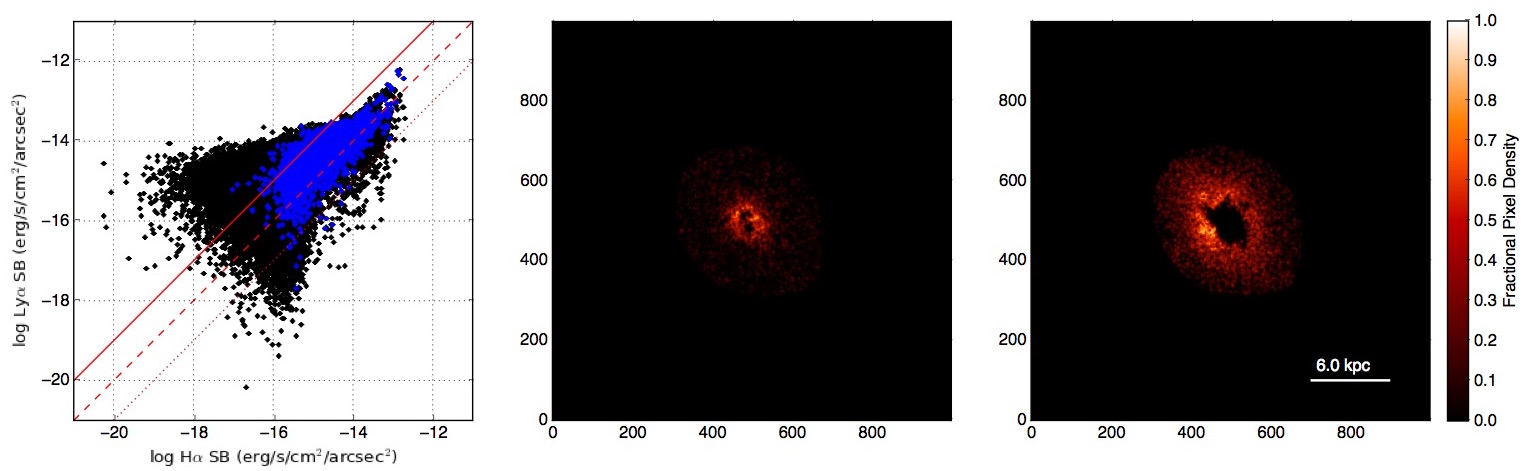}}
\caption{Same as Figure~\ref{sbplotcardscatt01} but for LARS07.}
\label{sbplotcardscatt07}
\end{figure*}
\begin{figure*}
\centering
\scalebox{0.7}
{\includegraphics{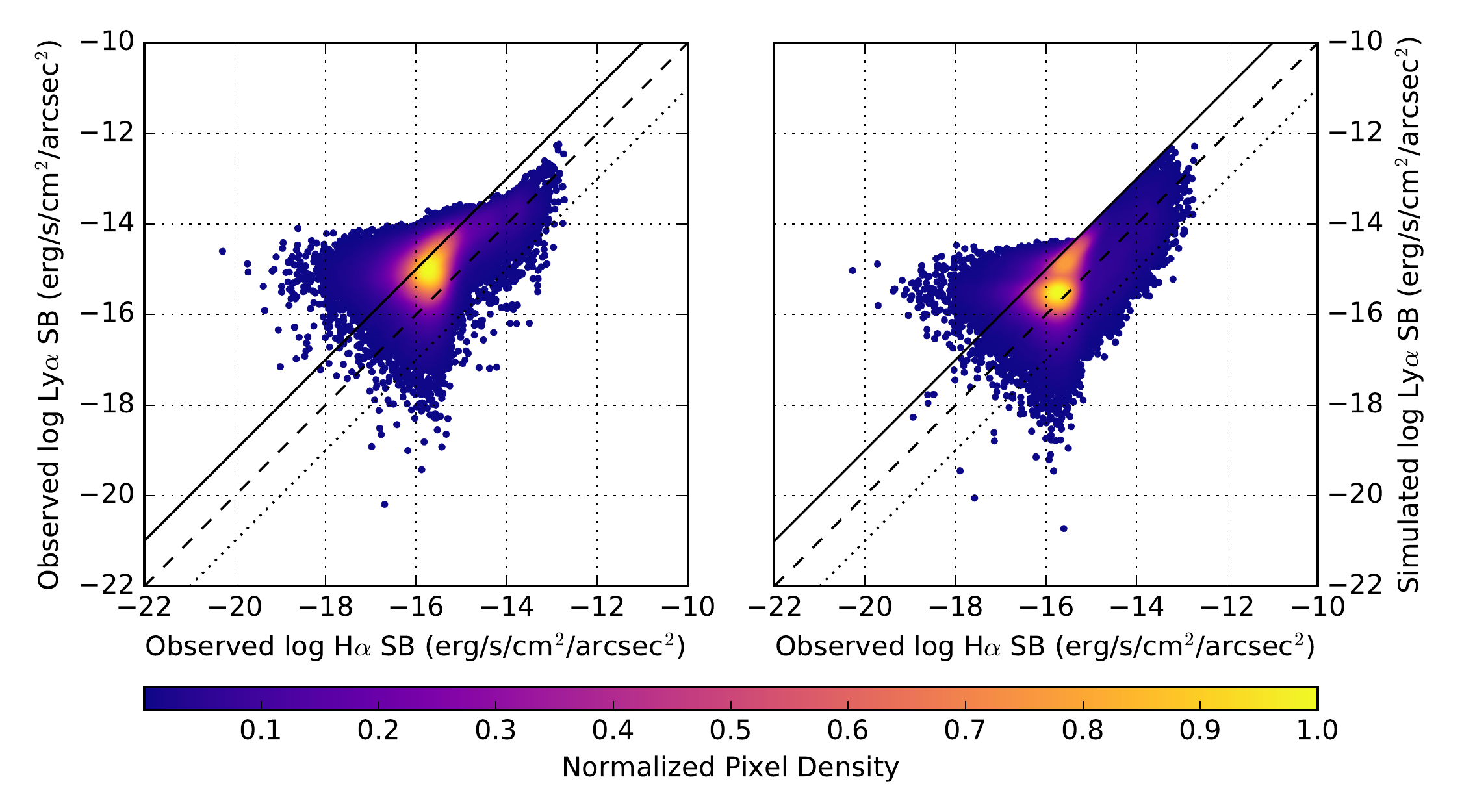}}
\caption{Same as Figure~\ref{sbplot01}, but for LARS07.}
\label{sbplot07}
\end{figure*}

\begin{figure*}
\centering
\scalebox{0.5}
{\includegraphics{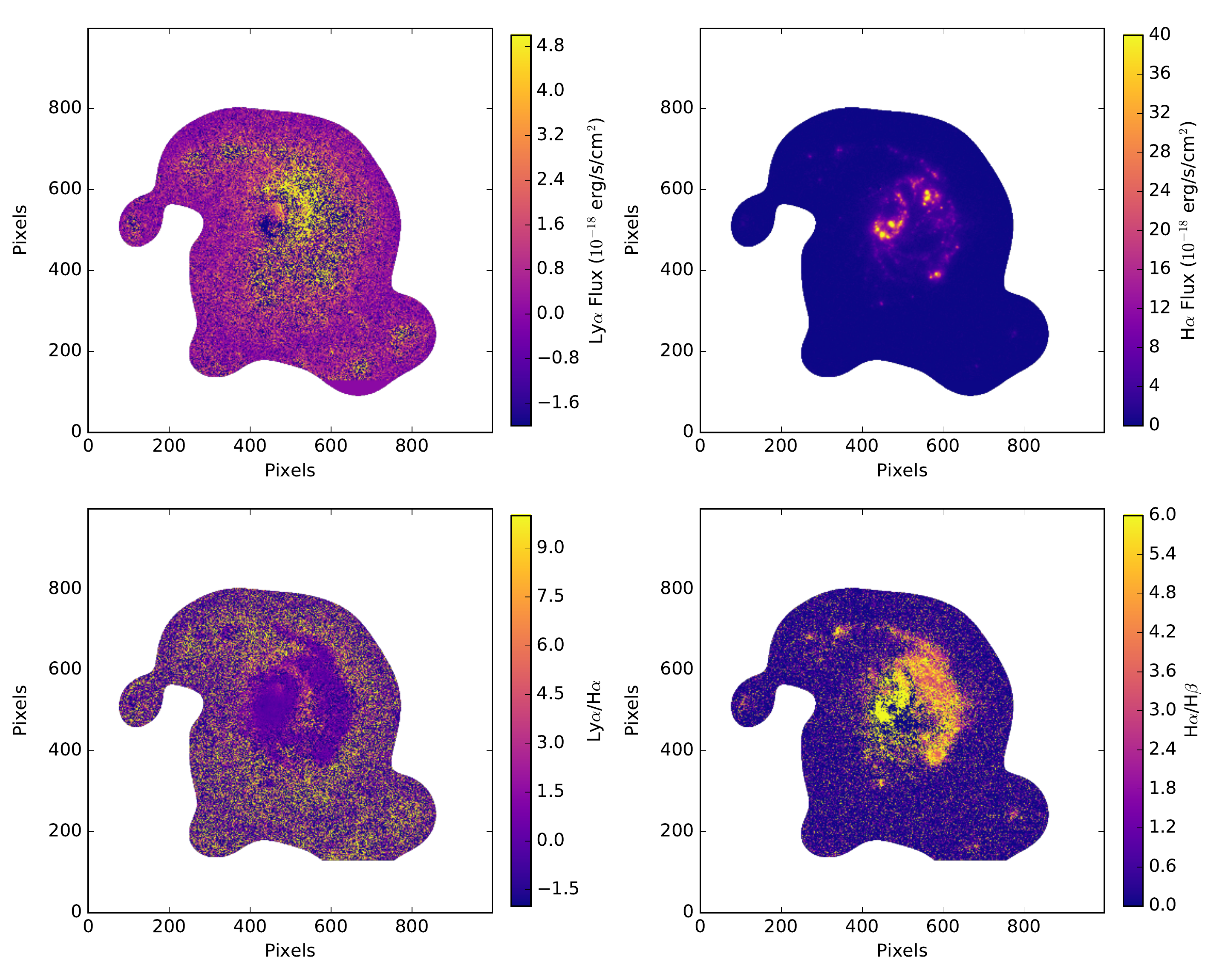}}
\caption{Images of LARS08. The calculated aperture has been applied. \emph{Top left}: The Ly$\alpha$ emission map. \emph{Top right}: The H$\alpha$ emission map. \emph{Bottom left}: The Ly$\alpha$/H$\alpha$ ratio, with an average uncertainty of 0.003. \emph{Bottom right}: The H$\alpha$/H$\beta$ ratio, with an average uncertainty of 0.007.}
\label{images08}
\end{figure*}
\begin{figure*}
\centering
\scalebox{0.32}
{\includegraphics{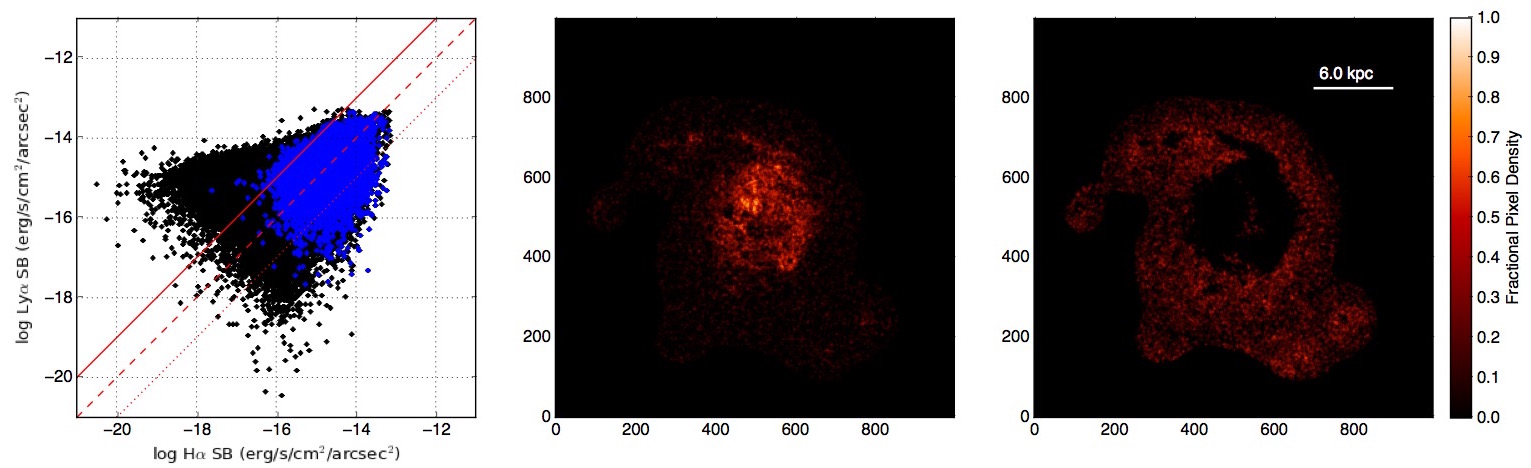}}
\caption{Same as Figure~\ref{sbplotcardscatt01} but for LARS08.}
\label{sbplotcardscatt08}
\end{figure*}
\begin{figure*}
\centering
\scalebox{0.7}
{\includegraphics{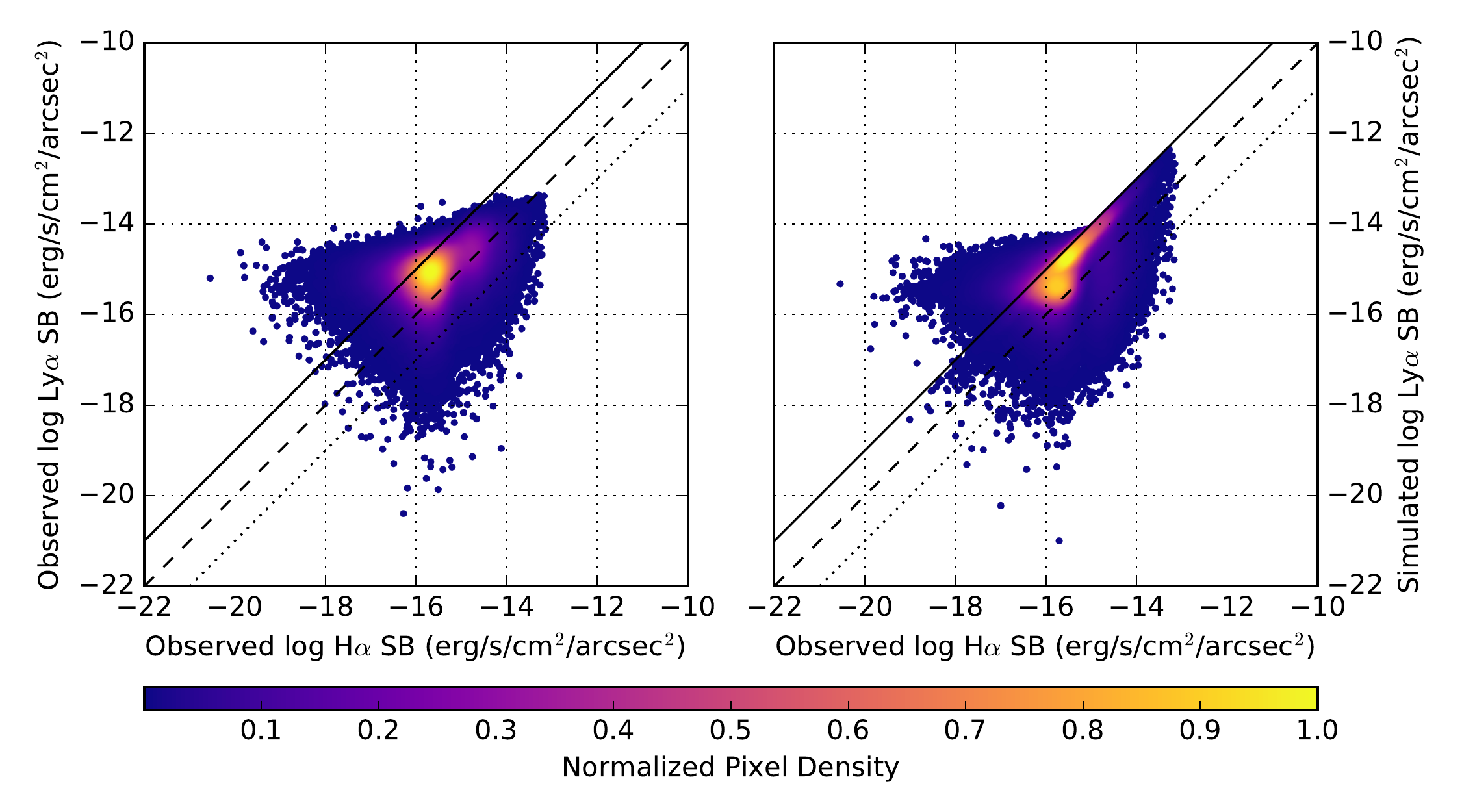}}
\caption{Same as Figure~\ref{sbplot01}, but for LARS08.}
\label{sbplot08}
\end{figure*}

\begin{figure*}
\centering
\scalebox{0.5}
{\includegraphics{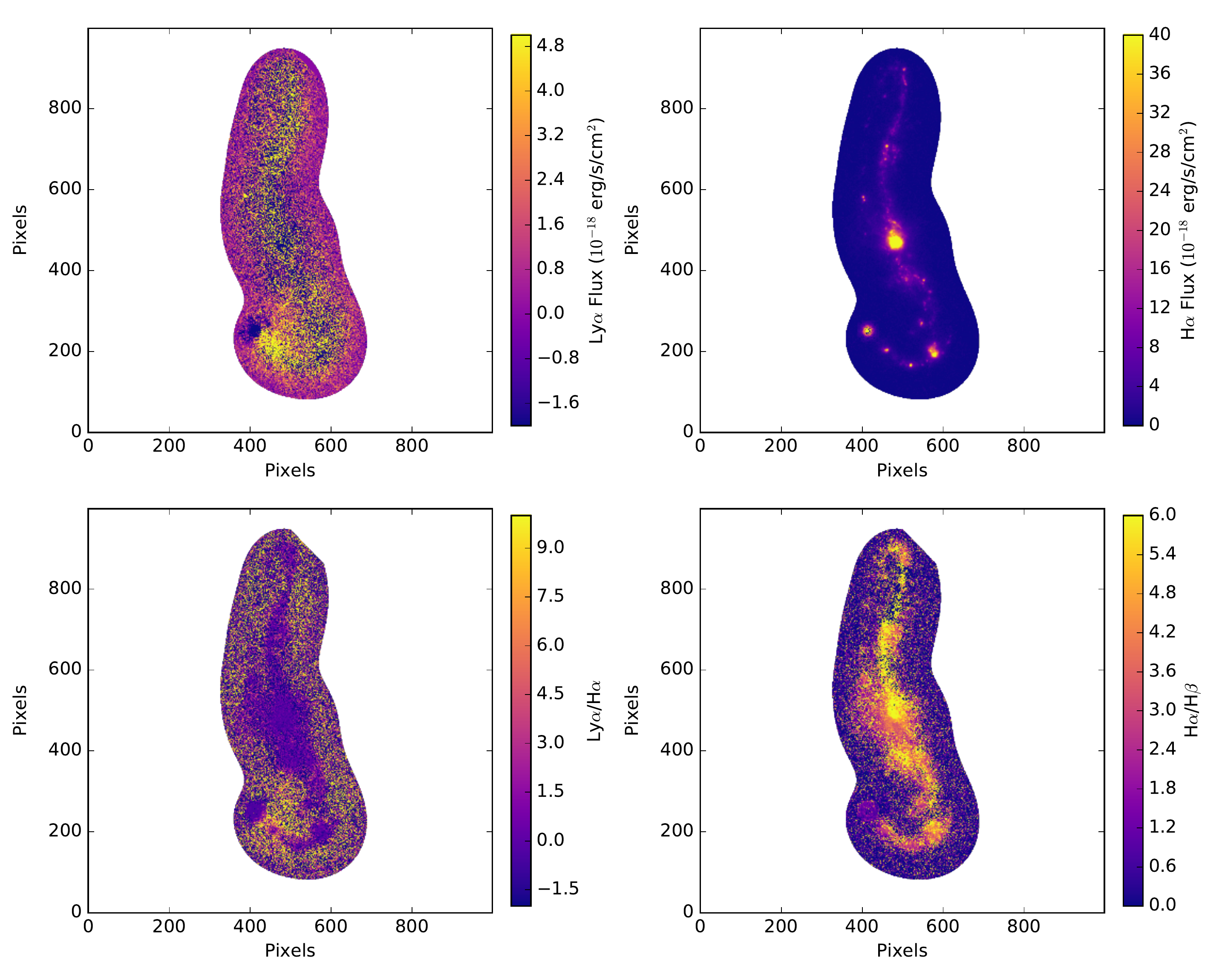}}
\caption{Images of LARS09. The calculated aperture has been applied. \emph{Top left}: The Ly$\alpha$ emission map. \emph{Top right}: The H$\alpha$ emission map. \emph{Bottom left}: The Ly$\alpha$/H$\alpha$ ratio, with an average uncertainty of 0.004. \emph{Bottom right}: The H$\alpha$/H$\beta$ ratio, with an average uncertainty of 0.002.}
\label{images09}
\end{figure*}
\begin{figure*}
\centering
\scalebox{0.32}
{\includegraphics{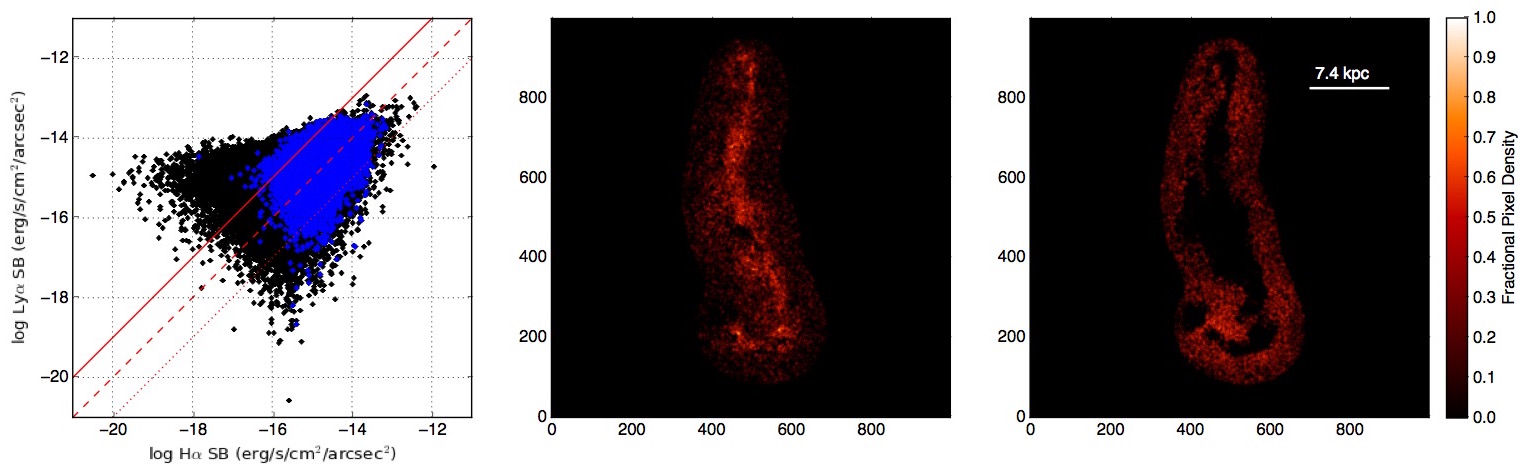}}
\caption{Same as Figure~\ref{sbplotcardscatt01} but for LARS09.}
\label{sbplotcardscatt09}
\end{figure*}
\begin{figure*}
\centering
\scalebox{0.7}
{\includegraphics{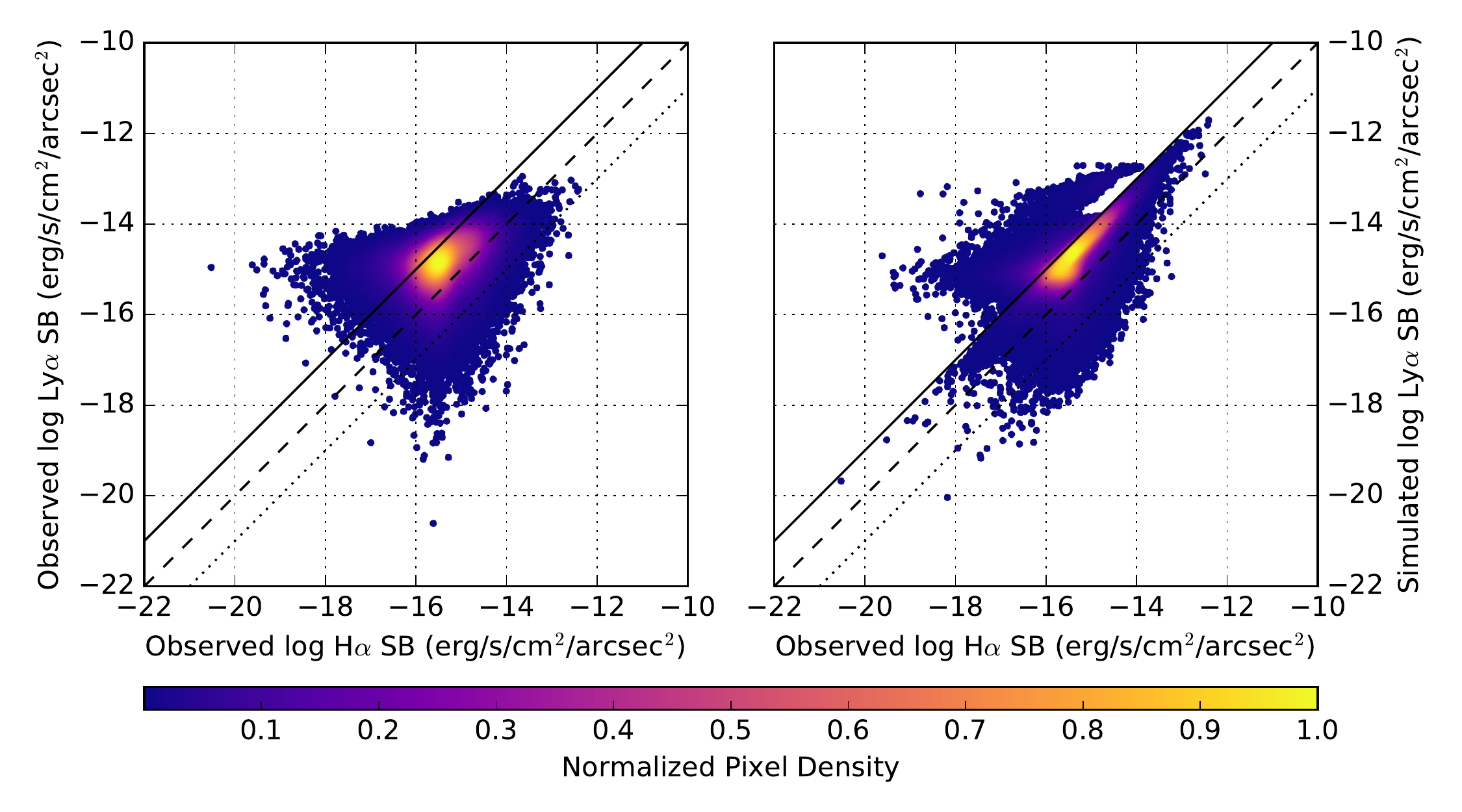}}
\caption{Same as Figure~\ref{sbplot01}, but for LARS09.}
\label{sbplot09}
\end{figure*}

\begin{figure*}
\centering
\scalebox{0.5}
{\includegraphics{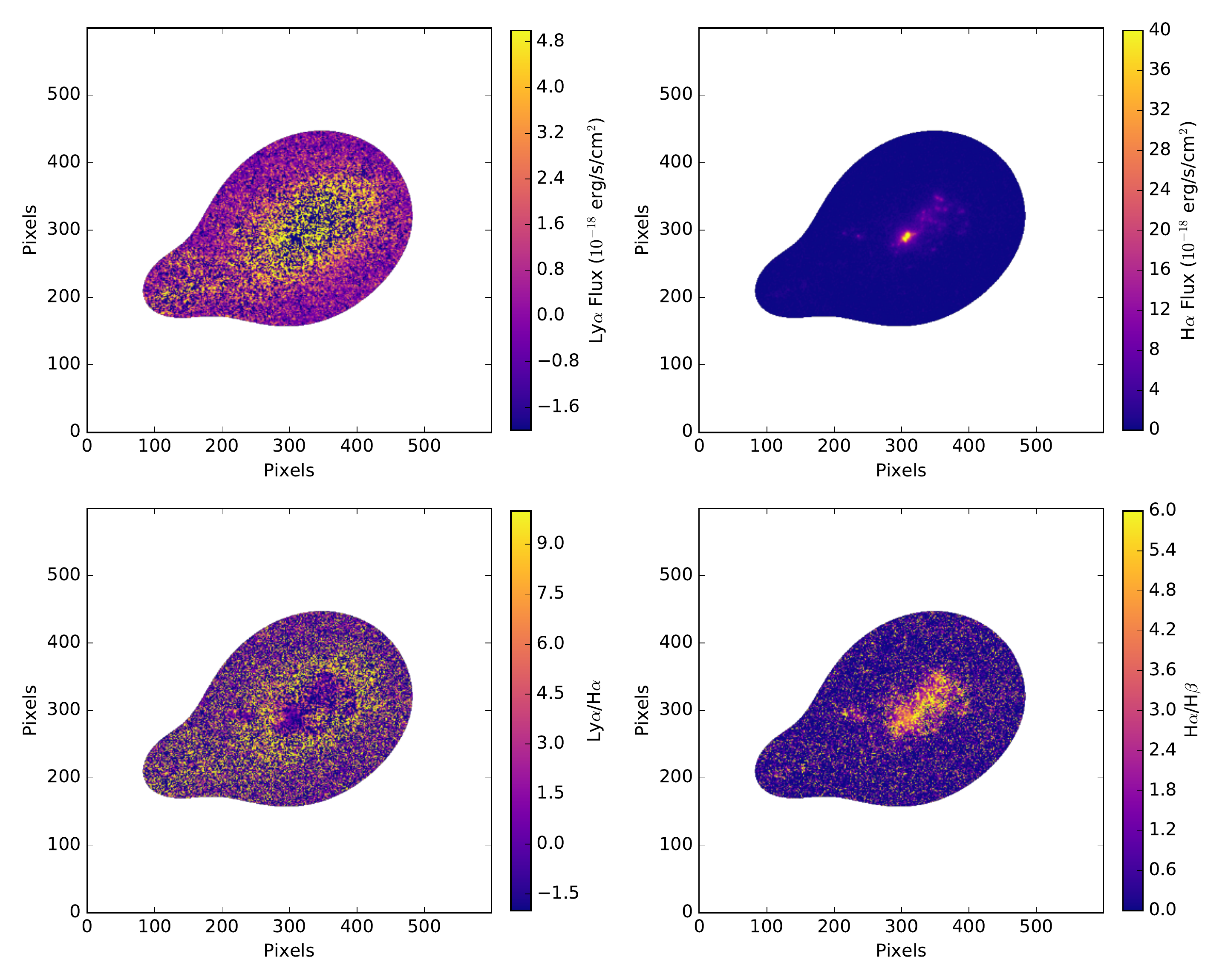}}
\caption{Images of LARS10. The calculated aperture has been applied. \emph{Top left}: The Ly$\alpha$ emission map. \emph{Top right}: The H$\alpha$ emission map. \emph{Bottom left}: The Ly$\alpha$/H$\alpha$ ratio, with an average uncertainty of 0.01. \emph{Bottom right}: The H$\alpha$/H$\beta$ ratio, with an average uncertainty of 0.003.}
\label{images10}
\end{figure*}
\begin{figure*}
\centering
\scalebox{0.32}
{\includegraphics{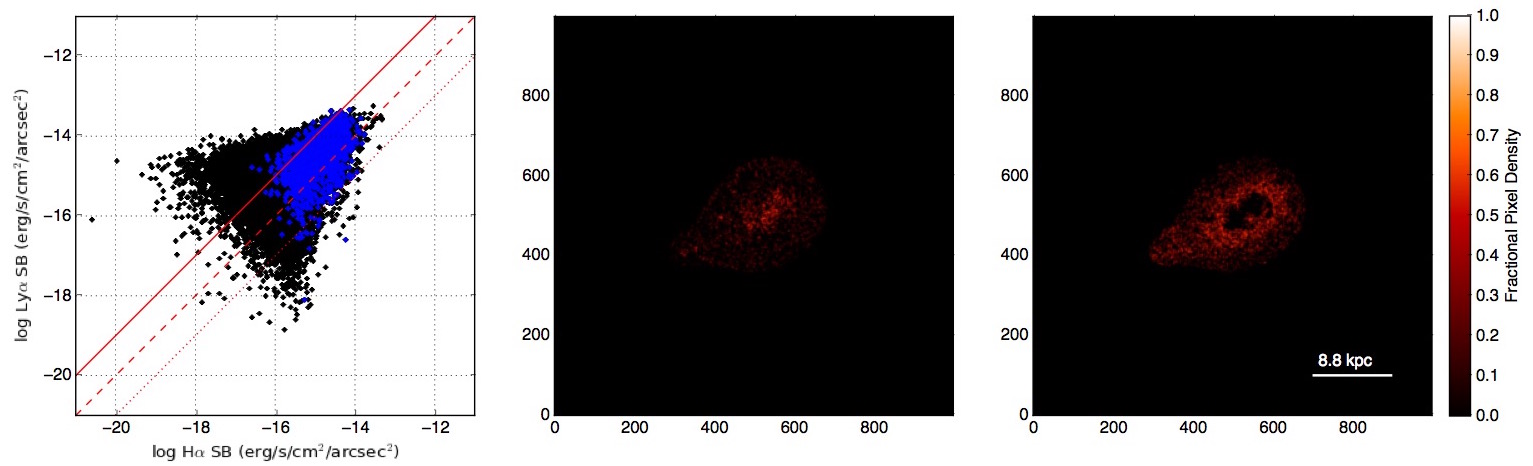}}
\caption{Same as Figure~\ref{sbplotcardscatt01} but for LARS10.}
\label{sbplotcardscatt10}
\end{figure*}
\clearpage
\begin{figure*}
\centering
\scalebox{0.7}
{\includegraphics{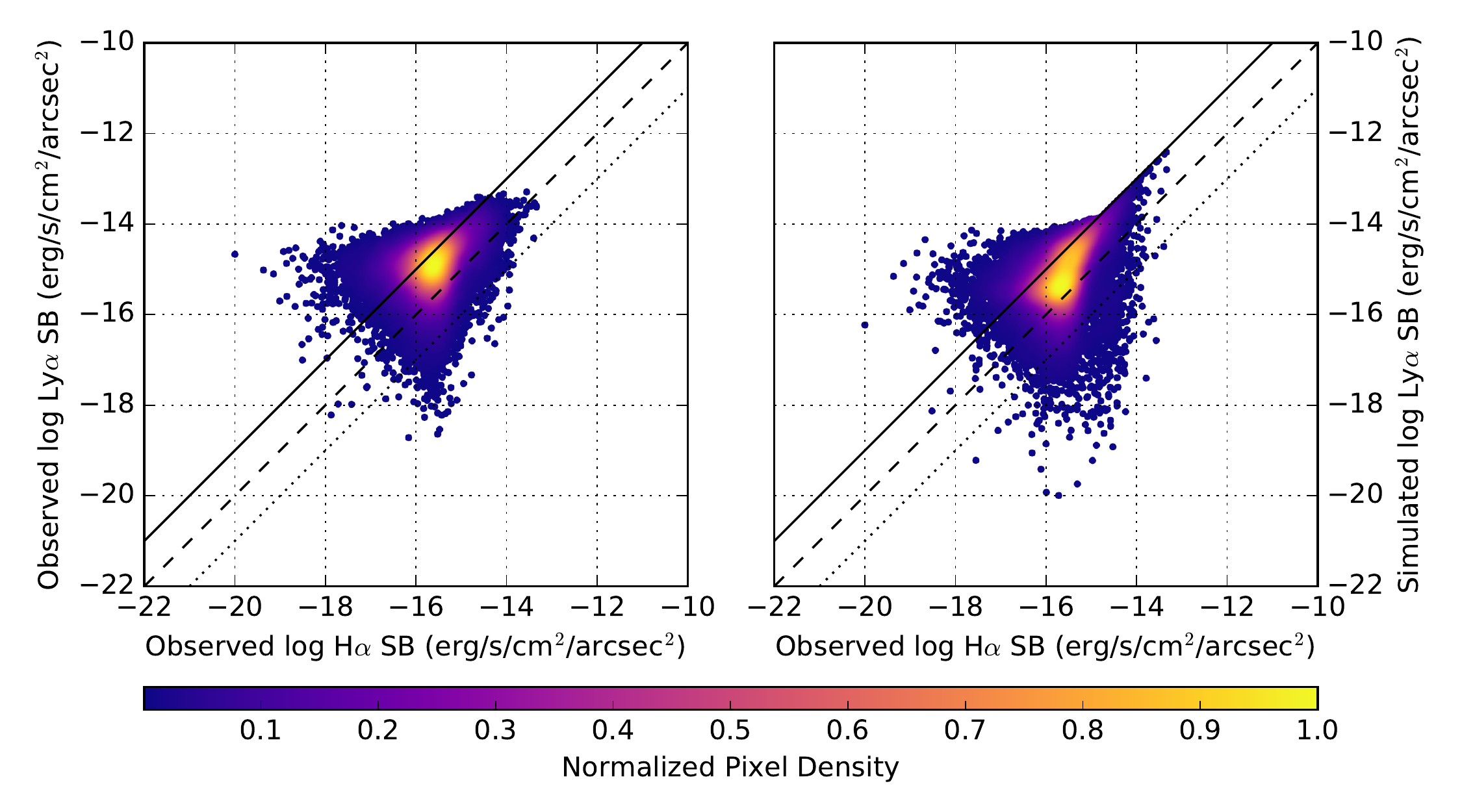}}
\caption{Same as Figure~\ref{sbplot01}, but for LARS10.}
\label{sbplot10}
\end{figure*}

\begin{figure*}
\centering
\scalebox{0.5}
{\includegraphics{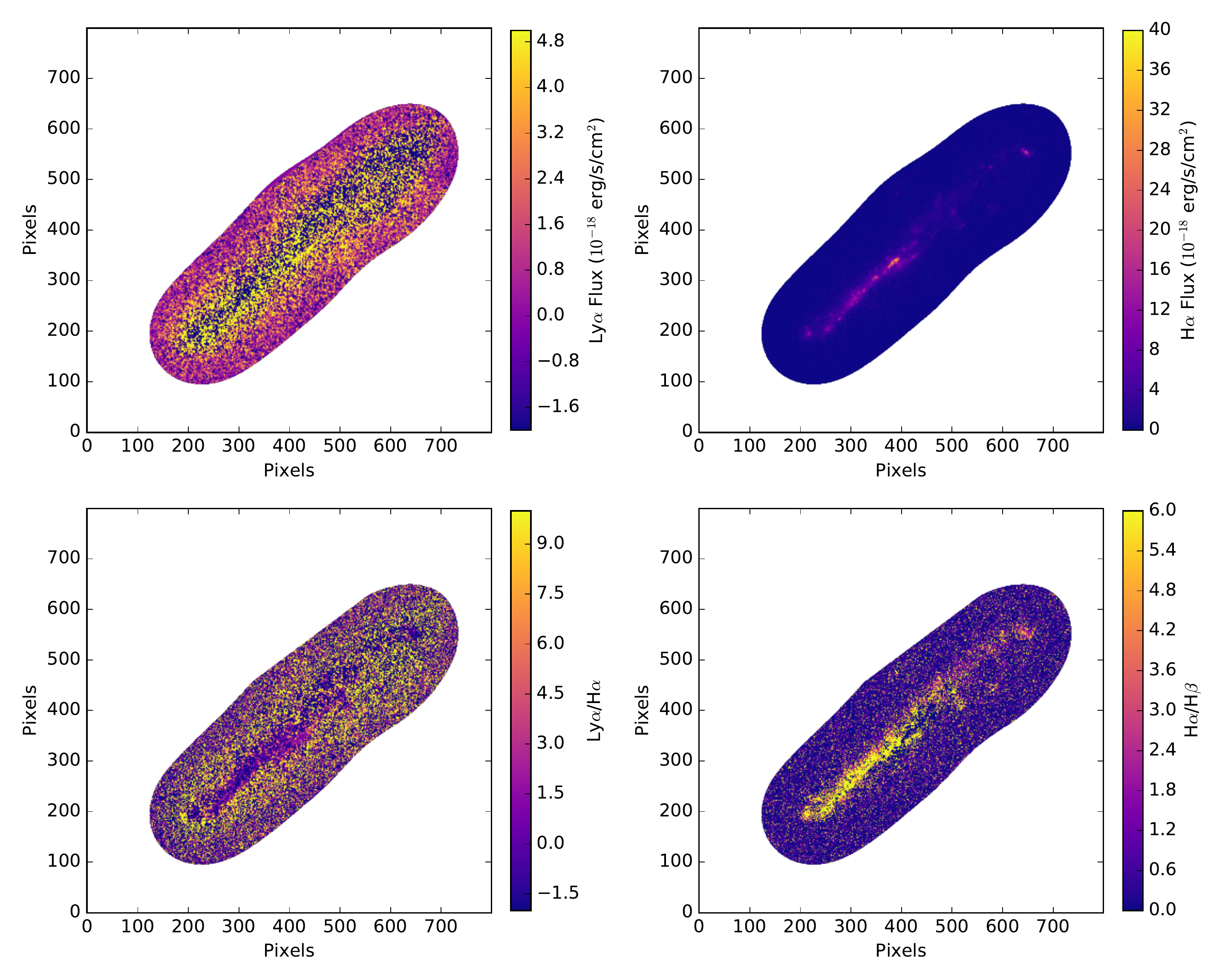}}
\caption{Images of LARS11. The calculated aperture has been applied. \emph{Top left}: The Ly$\alpha$ emission map. \emph{Top right}: The H$\alpha$ emission map. \emph{Bottom left}: The Ly$\alpha$/H$\alpha$ ratio, with an average uncertainty of 0.009. \emph{Bottom right}: The H$\alpha$/H$\beta$ ratio, with an average uncertainty of 0.01.}
\label{images11}
\end{figure*}
\begin{figure*}
\centering
\scalebox{0.32}
{\includegraphics{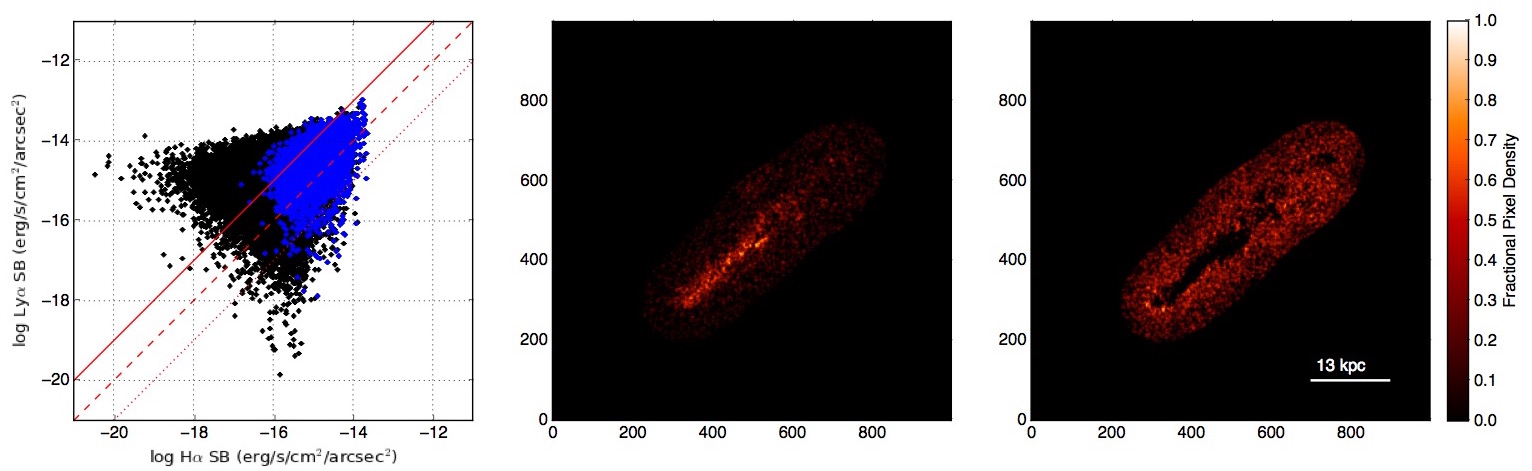}}
\caption{Same as Figure~\ref{sbplotcardscatt01} but for LARS11.}
\label{sbplotcardscatt11}
\end{figure*}
\begin{figure*}
\centering
\scalebox{0.7}
{\includegraphics{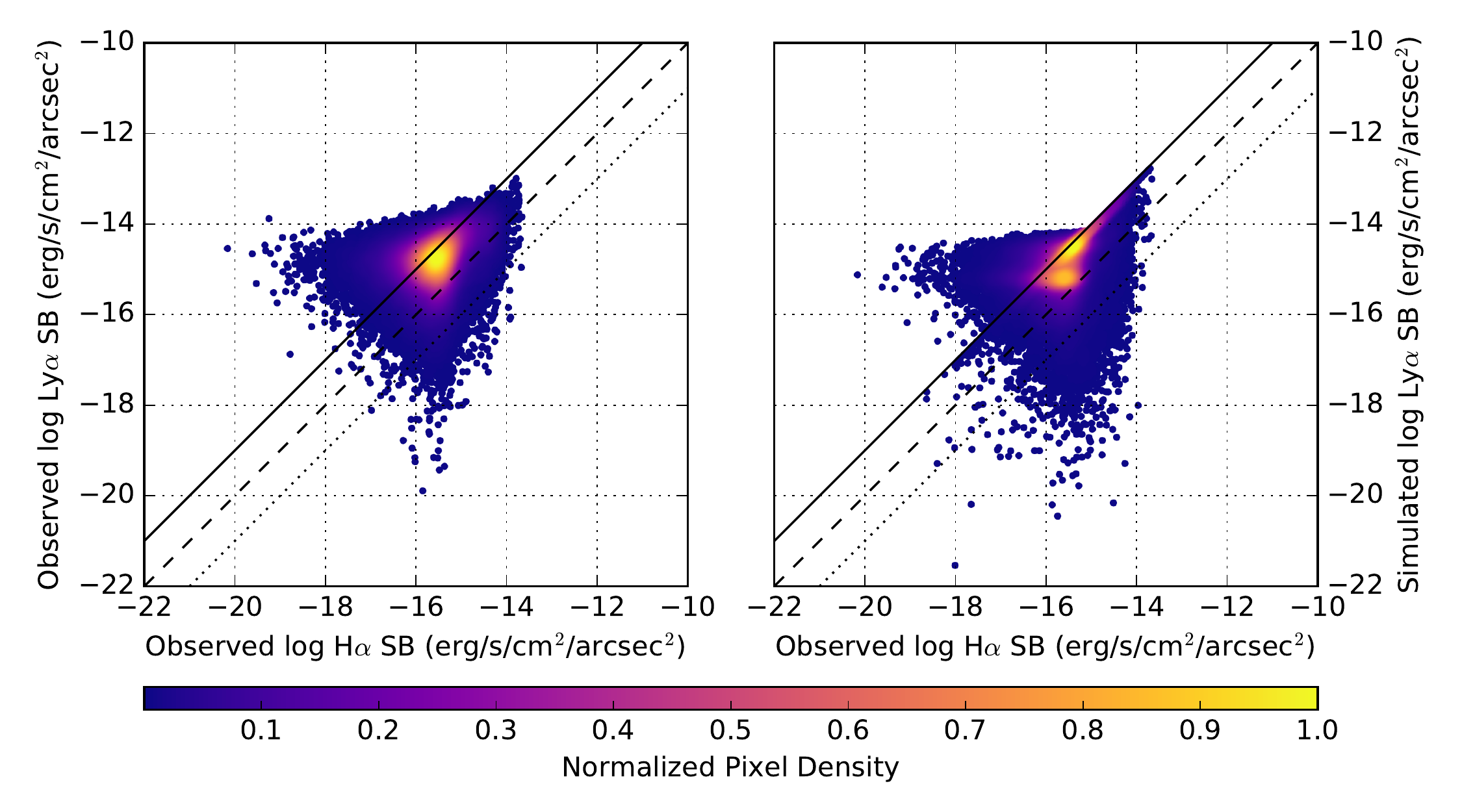}}
\caption{Same as Figure~\ref{sbplot01}, but for LARS11.}
\label{sbplot11}
\end{figure*}

\begin{figure*}
\centering
\scalebox{0.5}
{\includegraphics{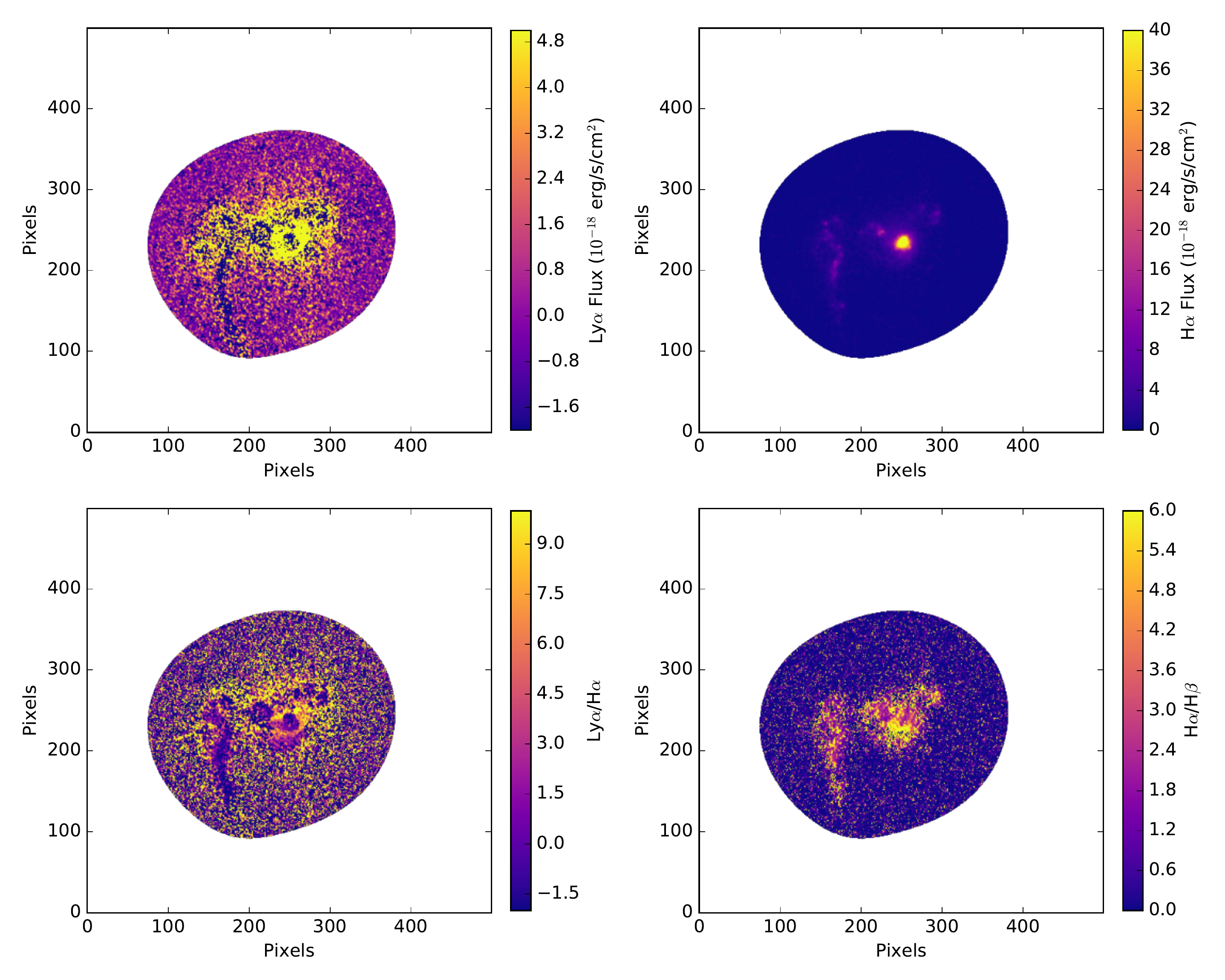}}
\caption{Images of LARS13. The calculated aperture has been applied. \emph{Top left}: The Ly$\alpha$ emission map. \emph{Top right}: The H$\alpha$ emission map. \emph{Bottom left}: The Ly$\alpha$/H$\alpha$ ratio, with an average uncertainty of 0.07. \emph{Bottom right}: The H$\alpha$/H$\beta$ ratio, with an average uncertainty of 0.004.}
\label{images13}
\end{figure*}
\begin{figure*}
\centering
\scalebox{0.32}
{\includegraphics{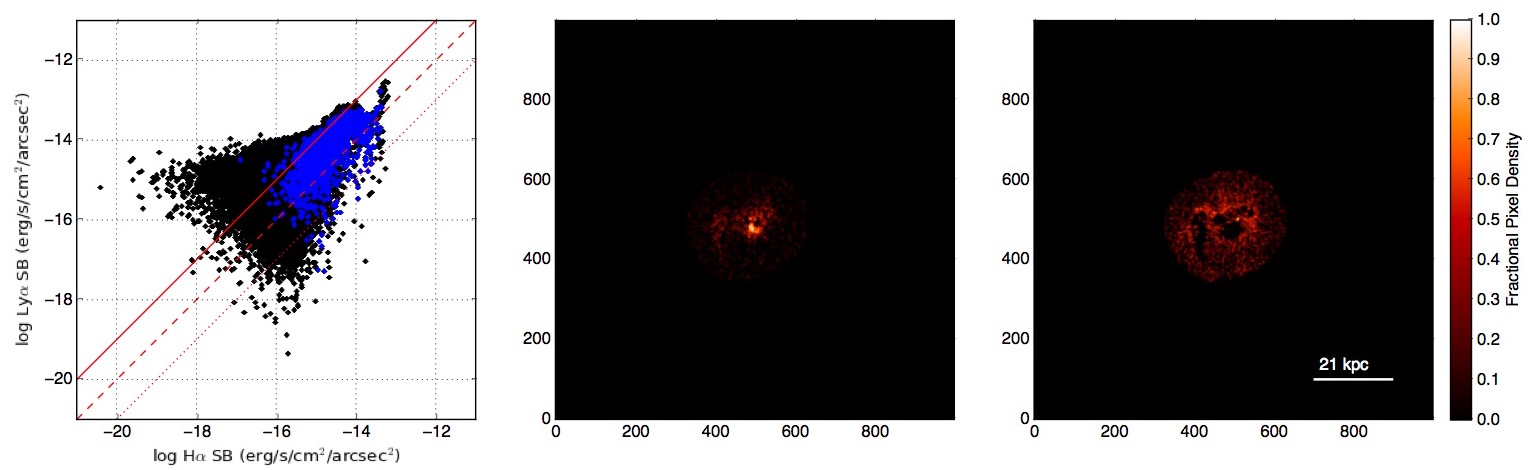}}
\caption{Same as Figure~\ref{sbplotcardscatt01} but for LARS13.}
\label{sbplotcardscatt13}
\end{figure*}
\begin{figure*}
\centering
\scalebox{0.7}
{\includegraphics{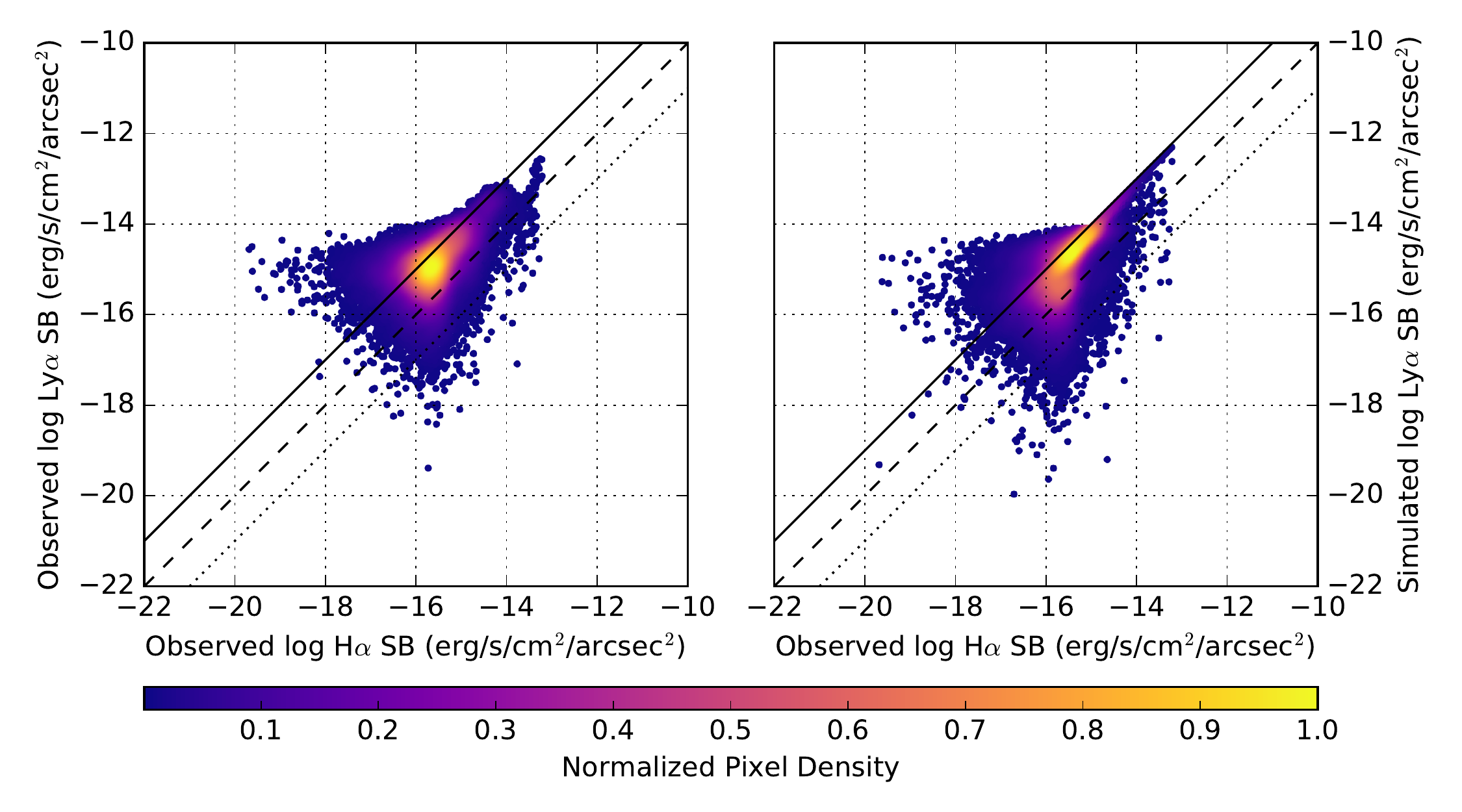}}
\caption{Same as Figure~\ref{sbplot01}, but for LARS13.}
\label{sbplot13}
\end{figure*}

\begin{figure*}
\centering
\scalebox{0.5}
{\includegraphics{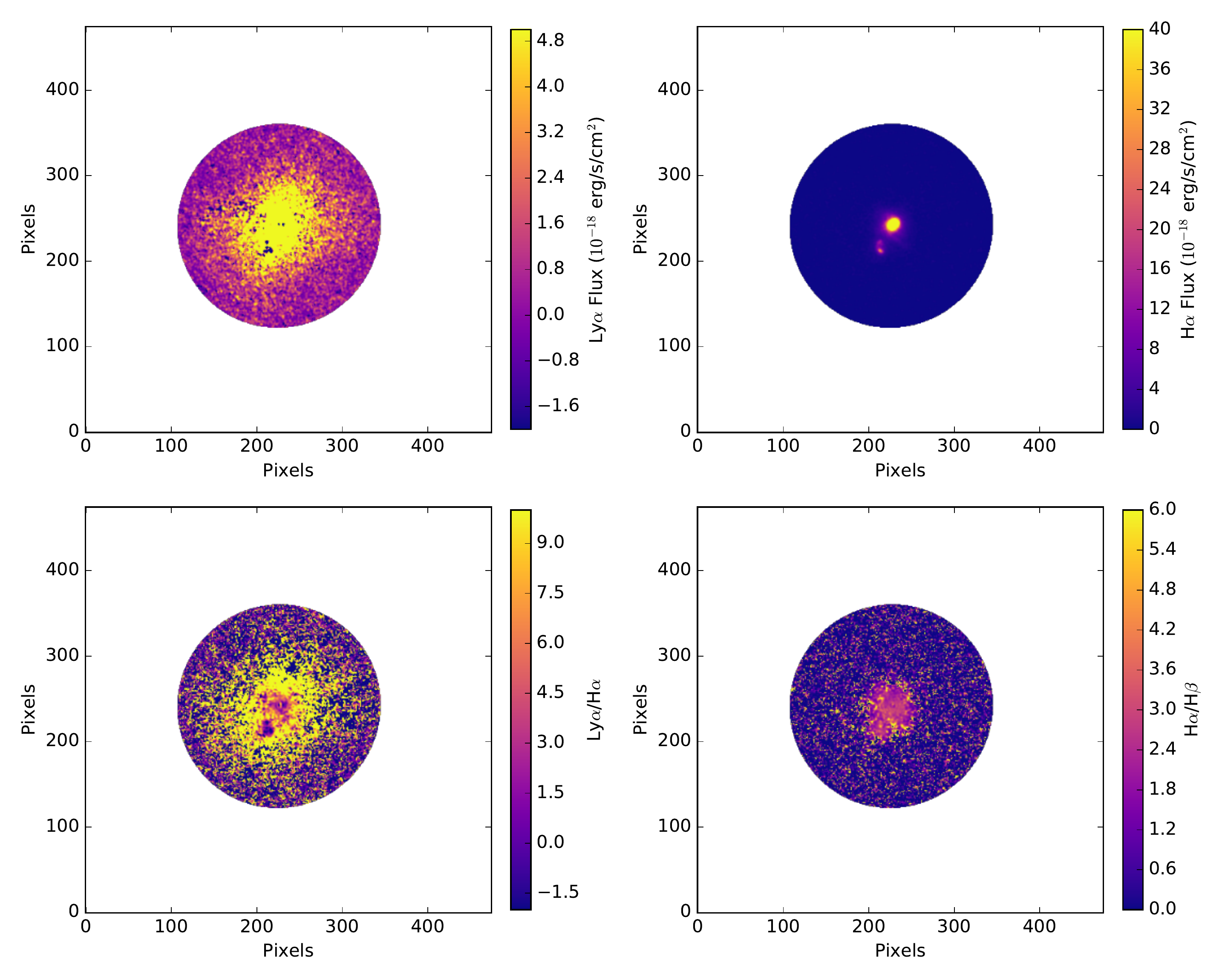}}
\caption{Images of LARS14. The calculated aperture has been applied. \emph{Top left}: The Ly$\alpha$ emission map. \emph{Top right}: The H$\alpha$ emission map. \emph{Bottom left}: The Ly$\alpha$/H$\alpha$ ratio, with an average uncertainty of 0.02. \emph{Bottom right}: The H$\alpha$/H$\beta$ ratio, with an average uncertainty of 0.002.}
\label{images14}
\end{figure*}
\begin{figure*}
\centering
\scalebox{0.32}
{\includegraphics{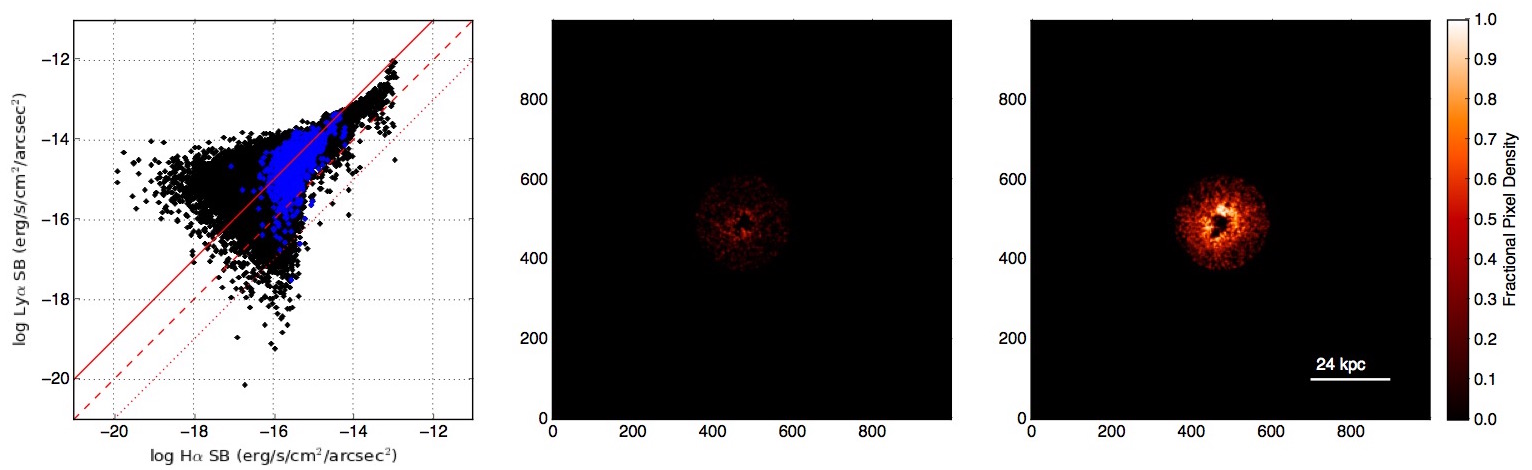}}
\caption{Same as Figure~\ref{sbplotcardscatt01} but for LARS14.}
\label{sbplotcardscatt14}
\end{figure*}
\begin{figure*}
\centering
\scalebox{0.7}
{\includegraphics{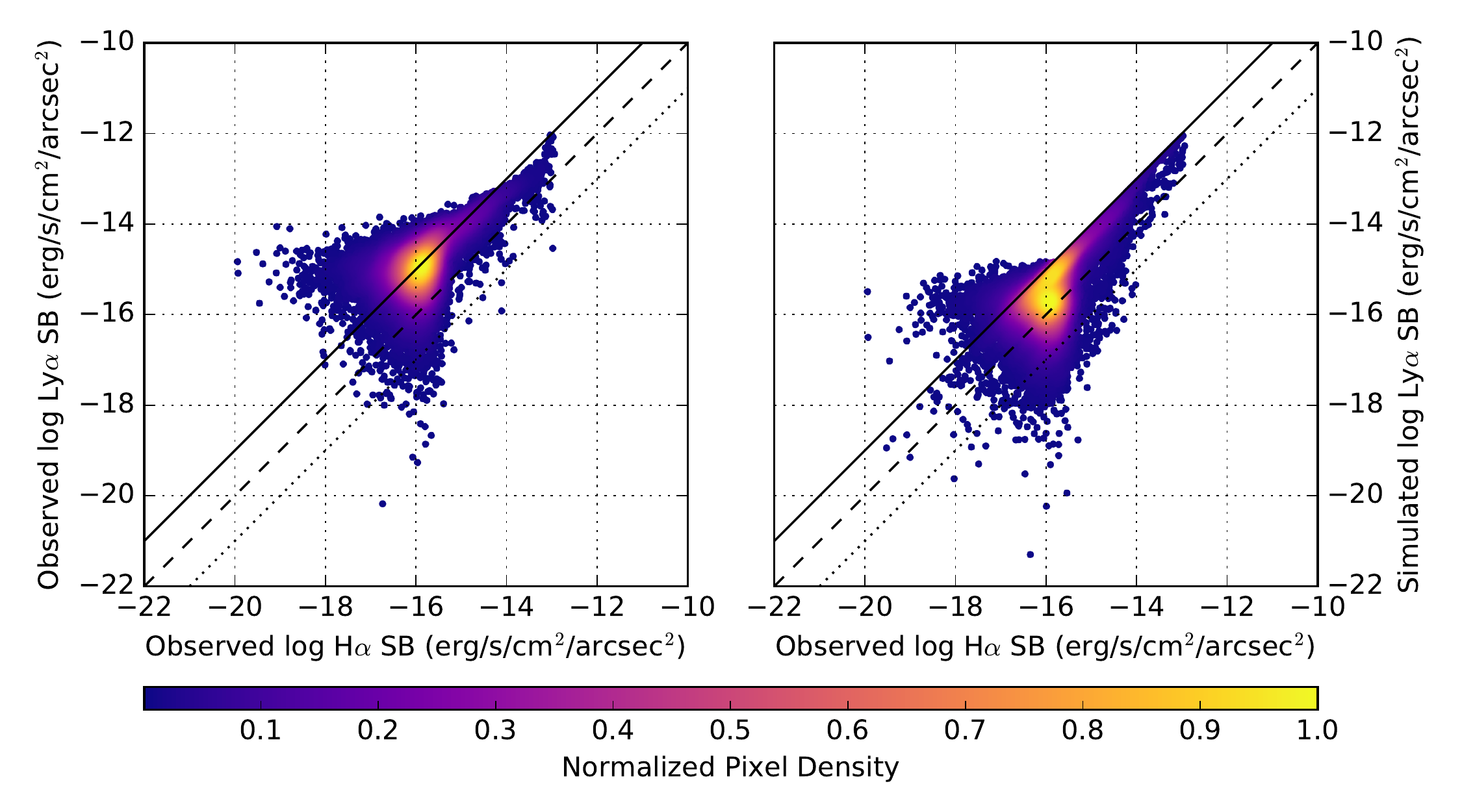}}
\caption{Same as Figure~\ref{sbplot01}, but for LARS14.}
\label{sbplot14}
\end{figure*}

\end{document}